\documentclass[]{aa} 

\usepackage{graphicx}
\usepackage{lscape}
\usepackage{txfonts}
\usepackage{natbib}
\usepackage{url}
\usepackage[T1]{fontenc}

\usepackage[bookmarks=false]{hyperref}

\newcommand{\flux}{erg\,s$^{-1}$\,cm$^{-2}$}

\newcommand{\mz}{{\small $\mathcal{M}$-Z}}
\newcommand{\Sz}{{\small {\large $\Sigma$}-Z}}

\DeclareRobustCommand{\ion}[2]{%
\relax\ifmmode
\ifx\testbx\f@series
{\mathbf{#1\,\mathsc{#2}}}\else
{\mathrm{#1\,\mathsc{#2}}}\fi
\else\textup{#1\,{\mdseries\textsc{#2}}}%
\fi}

\newcommand{\HII}{\ion{H}{ii}~}

\begin{document}
   \title{Imprints of galaxy evolution on \HII regions}
   \subtitle{Memory of the past uncovered by the CALIFA survey}

   \author{
S.\,F. S\'anchez\inst{\ref{unam}}
\and
E. P\'erez \inst{\ref{iaa}}
\and
F.\,F. Rosales-Ortega\inst{\ref{inaoe}}
\and
D.\,Miralles-Caballero\inst{\ref{uam}}
\and
A. R. L\'{o}pez-S\'{a}nchez \inst{\ref{aao},\ref{mac}}
\and
J. Iglesias-P\'aramo\inst{\ref{iaa},\ref{caha}}
\and
R.\,A. Marino\inst{\ref{ucm}}
\and
L.\,S\'anchez-Menguiano\inst{\ref{iaa},\ref{ugr}}
\and
R. Garc\'\i a-Benito \inst{\ref{iaa}}
\and
D. Mast\inst{\ref{Jan}}
\and
M.A. Mendoza\inst{\ref{iaa}}
\and
P. Papaderos \inst{\ref{porto}}
\and
S. Ellis\inst{\ref{sydney},\ref{aao}}
\and
L. Galbany\inst{\ref{mil},\ref{chile}}
\and
C. Kehrig\inst{\ref{iaa}} 
\and
A. Monreal-Ibero\inst{\ref{paris}}
\and
R. Gonz\'alez Delgado\inst{\ref{iaa}}
\and
M. Moll\'a\inst{\ref{ciemat}}
 \and
B. Ziegler\inst{\ref{viena}}
\and
A. de Lorenzo-C\'aceres\inst{\ref{StAnd}}
\and
J. Mendez-Abreu \inst{\ref{StAnd}}
\and
J. Bland-Hawthorn \inst{\ref{sydney}}
\and 
S. Bekerait\.{e} \inst{\ref{aip}}
\and 
M. M. Roth \inst{\ref{aip}}
\and
A. Pasquali\inst{\ref{hei}}
\and
A.\, D\'\i az\inst{\ref{uam}}
\and
D. Bomans\inst{\ref{bochum}}
\and
G. van de Ven \inst{\ref{mpia}}
\and 
L. Wisotzki \inst{\ref{aip}}
\and
The CALIFA collaboration
}

   \institute{
     \label{unam}Instituto de Astronom\'\i a,Universidad Nacional Auton\'oma de Mexico, A.P. 70-264, 04510, M\'exico,D.F.
       \and
     \label{iaa}Instituto de Astrof\'{\i}sica de Andaluc\'{\i}a (CSIC), Glorieta de la Astronom\'\i a s/n, Aptdo. 3004, E18080-Granada, Spain\\ \email{sanchez@iaa.es}.
     \and
     \label{inaoe} Instituto Nacional de Astrof{\'i}sica, {\'O}ptica y Electr{\'o}nica, Luis E. Erro 1, 72840 Tonantzintla, Puebla, Mexico. 
     \and
     \label{uam}Departamento de F\'isica Te\'orica, Universidad Aut\'onoma de Madrid, 28049 Madrid, Spain.
     \and
     \label{aao} Australian Astronomical Observatory, PO Box 915, North Ryde, NSW 1670, Australia
     \and
         \label{mac} Department of Physics and Astronomy, Macquarie University, NSW 2109, Australia
     \and
         \label{caha} Centro Astron\'omico Hispano Alem\'an, Calar Al
to, (CSIC-MPG),
        C/Jes\'{u}s Durb\'{a}n Rem\'{o}n 2-2, E-04004 Almer\'{\i}a, Spain 
     \and
     \label{ucm}CEI Campus Moncloa, UCM-UPM, Departamento de Astrof\'{i}sica y CC$.$ de la Atm\'{o}sfera, Facultad de CC$.$ F\'{i}sicas, Universidad Complutense de Madrid, Avda.\,Complutense s/n, 28040 Madrid, Spain.
     \and
     \label{ugr}Dpto. de F\'\i sica Te\'orica y del Cosmos, University of Granada, Facultad de Ciencias (Edificio Mecenas), 18071 Granada, Spain 
     \and
     \label{Jan} Instituto de Cosmologia, Relatividade e Astrof\'{i}sica - ICRA, Centro Brasileiro de Pesquisas F\'{i}sicas, Rua Dr.Xavier Sigaud 150, CEP 22290-180, Rio de Janeiro, RJ, Brazil    
     \and
     \label{porto}Centro de Astrof\'\i sica and Faculdade de Ciencias,
     Universidade do Porto, Rua das Estrelas, 4150-762 Porto, Portugal.
     \and
     \label{sydney}Sydney Institute for Astronomy, School of Physics A28, University of Sydney, NSW 2006, Australia.
     \and
     \label{mil} Millennium Institute of Astrophysics, Universidad de Chile, Casilla 36-D, Santiago, Chile
     \and
     \label{chile} Departamento de Astronom\'\i a, Universidad de Chile, Casilla 36-D, Santiago, Chile
     \and
     \label{paris} GEPI, Observatoire de Paris, CNRS UMR8111, Universit\'e Paris Diderot, Place Jules Janssen, 92190 Meudon, France 
     \and
     \label{ciemat}Departamento de Investigaci\'on B\'asica, CIEMAT, Avda. Complutense 40 E-28040 Madrid, Spain.
     \and
     \label{viena}University of Vienna, T\"urkenschanzstrasse 17, 1180 Vienna, Austria
\and
\label{StAnd}School of Physics and Astronomy, University of St Andrews, North Haugh, St Andrews, KY16 9SS, U.K. (SUPA)
\and
 \label{aip}Leibniz-Institut f\"ur Astrophysik Potsdam (AIP), An der Sternwarte 16, D-14482 Potsdam, Germany.
\and
\label{hei} Universitaet Heidelberg, Zentrum fuer Astronomie, Astronomisches Rechen Institut, Moenchhofstr. 12 - 14, 69120 Heidelberg, Germany.
\and
\label{bochum}  Astronomical Institute of the Ruhr-University Bochum Universitaetsstr. 150, 44801 Bochum, Germany.
 \and
\label{mpia}Max-Planck-Institut f\"ur Astronomie, Heidelberg, Germany.
              }

   \date{Received ----- ; accepted ---- }


\abstract{ \HII regions in galaxies are the sites of star formation and thus particular places
to understand the build-up of stellar mass in the universe. The line ratios of this ionized gas are frequently
  used to characterize the ionization conditions. In particular
  the oxygen abundances are assumed to trace the chemical
  enrichment of galaxies.}{ We
  explore here the connections between the ionization conditions and
  the properties of the overall underlying stellar population (ionizing or
  not-ionizing) in \HII regions, in order to uncover the real physical connection
  between them.}{ We use the \HII
  regions catalogue from the CALIFA survey, the largest in existence
  with more than 5000 \HII regions, to explore their distribution across the classical
  [\ion{O}{iii}]~$\lambda$5007/H$\beta$
  vs. [\ion{N}{ii}]~$\lambda$6583/H$\alpha$ diagnostic diagram, and how it
depends on the oxygen abundance, ionization parameter, electron density, and dust
  attenuation. The location of \HII regions within this diagram is compared
  with predictions from photoionization models. Finally, we explore
  the dependence of the location within the diagnostic diagram on the
  properties of the host galaxies, the galactocentric distances and
  the properties of the underlying stellar population.}{ We found that
  the \HII regions with weaker ionization strengths and more metal
  rich are located in the bottom-right area of the diagram. On the
  contrary, those regions with stronger ionization strengths and more
  metal poor are located in the upper-left end of the diagram.
  Photoionization models per se do not predict these correlations between the
  parameters and the line ratios. The \HII regions located in earlier-type
  galaxies, closer to the center, and formed in older and more metal
  rich regions of the galaxies are located in the bottom-right area of
  the diagram. On the other hand, those regions located in
  late-type galaxies, in the outer regions of the disks, and formed on
  younger and more metal poor regions lie in the top-left area
  of the diagram. The two explored line ratios show strong
  correlations with the age and metallicity of the underlying stellar
  population.}{ These results indicate that although \HII regions are
  short lived events, they are affected by the total underlying
  stellar population. One may say that \HII regions keep a memory of
  the stellar evolution and chemical enrichment that have left an
  imprint on the both the ionizing stellar population and the ionized
  gas.}

   \maketitle


\section{Introduction}

Classical \HII regions are large, low-density clouds of partially
ionized gas in which star formation has recently taken place ($<$15
Myr). The short-lived blue stars forged in these regions emit large
amounts of ultraviolet light that ionizes the surrounding gas. They
show very different physical scales, from few parsecs, like Orion
nebula (D$\sim$8 pc), or even smaller \citep{ander14}, to hundreds of parsecs, such as 30 Doradus
(D$\sim$200 pc), NGC\,604 (D$\sim$460 pc) or NGC\,5471 (D$\sim$1 kpc)
as reported by \cite{oey03} and \cite{rgb2011}. These latter ones are
the prototypes of the extragalactic giant \HII regions found
frequently in the disks of spiral galaxies
\citep[e.g.][]{hodg83,dott87,dott89,knap98}, or starburst and blue compact galaxies
\citep[e.g.][]{kehrig08,angel09,cairos12}.

\cite{baldwin81} first proposed the
     [\ion{O}{iii}]~$\lambda$5007/H$\beta$ versus
     [\ion{N}{ii}]~$\lambda$6583/H$\alpha$ diagnostic diagram (now
     known as the BPT diagram) to separate emission-line objects
     according to the main excitation mechanism: normal \HII regions,
     planetary nebulae, and objects photoionized by a harder radiation
     field. A hard radiation field can be produced by either a
     power-law continuum from an AGN, shock excitation, planetary nebulae or even
     post-AGB stars
     \citep[e.g.][]{binn94,binn09,mori09,flores11,kehrig12,sign13,papa13}.
     \cite{veilleux87} and \cite{osterbrock89} extended and refined
     this classification scheme, incorporating new diagnostic
     diagrams. \cite{osterbrock89} used theoretical photoionization
     models to infer the shape of the classification line between
     star-forming (SF) and AGN galaxies, and they added two new
     diagnostics diagrams that exploit the [\ion{O}{i}]/H$\beta$ versus
     [\ion{S}{ii}]/H$\alpha$ line ratios. \cite{dopita00} and
     \cite{kewley01} combined stellar population synthesis and
     photoionization models to build the first purely theoretical
     classification scheme for separating pure AGN from galaxies
     containing star formation, and \cite{kauffmann03} used SDSS data
     to observationaly constrains these classifications.

In essence, these models assume that the main factors that control the
emission line spectrum are the chemical abundances of the heavy
elements in the gas phase of an \HII region (oxygen being the most
important), the shape of the ionizing radiation spectrum, and the
geometrical distribution of gas with respect to the ionizing
sources. Generally speaking, all the geometrical factors are subsumed
into a single factor, the ionization parameter $q$ (with dimensions
$cm\ s^{-1}$) or the (dimensionless) ionization parameter $u =
q/c$. They also assume  {\it a priori} that these parameters are
independent, and thus these models are presented as grids of oxygen
abundance, ionization parameter, and electron densities for a given
ionizing population.

Different demarcation lines have been proposed for this diagram. The
most popular ones are the \cite{kauffmann03} and \cite{kewley01}
curves. They are usually invoked to distinguish between star-forming
regions (below the Kauffmann et al. 2013 curve) and AGNs (above the
Kewley et al. 2001 curve). The location between both curves is
normally assigned to a mixture of different sources of
ionization. Additional demarcation lines have been proposed for the
region above the \cite{kewley01} curve to segregate between Seyfert
and LINERs \citep[e.g.,][]{kewley06,cid-fernandes10}. { More
  recently it has been explored the use of 3D diagnostic diagrams to
  segregate the different physical ionizaion sources in a cleaner way
  \citep[e.g.][]{vogt14}}.

There are well known trends of the ionization conditions across the
BPT diagram. For example, as the gas-phase chemical abundances are
increased, the primary effect on the \HII region is to decrease the
electron temperature, since there is more cooling per
photoionization. This affects the temperature sensitive line ratios
such as [\ion{O}{iii}]$\lambda$4959,$\lambda$5007/$\lambda$4363, but
also many others \citep[e.g.][]{marino13}, and in particular the two
line ratios used in the BPT diagram:
[\ion{O}{iii}]~$\lambda$5007/H$\beta$ versus
[\ion{N}{ii}]~$\lambda$6583/H$\alpha$. Indeed, early photographic
spectra revealed that, in disk galaxies, \HII regions located in the
outer spiral arms displayed large
[\ion{O}{iii}]~$\lambda$5007/H$\beta$ ratios, while \HII\ regions
located in the inner spiral arms show relatively large
[\ion{N}{ii}]~$\lambda$6583/H$\alpha$ ratios
\citep[e.g.][]{sand69,sear71}. It was soon realized that this is due
to the existence of a global abundance gradient across the galaxy,
connecting for the first time the location of \HII regions within the
BPT diagram with the chemical evolution in disk galaxies.

Since these early studies, many others have unveiled some aspects of
the complex processes at play between the chemical evolution of
galaxies and the ionization conditions in \HII regions. These studies
have been successful in determining important relationships, scaling
laws and systematic patterns: (i) luminosity-metallicity,
mass-metallicity, and surface brightness vs. metallicity relations
\citep{leque79,Skillman:1989p1592,VilaCostas:1992p322,zaritsky94,tremonti04};
(ii) effective yield vs. luminosity and circular velocity relations
\citep{Garnett:2002p339}; (iii) abundance gradients and the effective
radius of disks \citep{diaz89}; (iv) systematic differences in the
gas-phase abundance gradients between normal and barred spirals
\citep{zaritsky94,Martin:1994p1602}; (v) characteristic vs. integrated
abundances \citep{moustakas06}; (vi) galactocentric abundance gradients
\cite[e.g.][]{bresolin09,yoac10,rosales11,mari11,bresolin12}, etc.

{ However, many of these studies, in particular the older ones,
  present evident selection effects. In most of the cases the
  spectroscopic surveys of \HII regions rely on previous imaging
  surveys using narrow-band images
  \citep[e.g.][]{Martin:1994p1602}. Those images are then processed to
  provide with a catalogue of \HII regions suitable to be observed
  using long-slit spectroscopy. In this selection the signal-to-noise
  of the detected emission line in the narrow-band images plays an
  important role. Therefore most of these surveys are biased towards
  high contrast (equivalent width of emission lines) \HII regions,
  located in the outer disks of galaxies. They tend to be focused on
  late-type spirals (Sc and Sd), which offer the best constrast, due
  to their lower surface-brightness in the continuum. This may has
  biased our understanding of the properties of \HII regions, since in
  most of the cases the studies were restricted to regions ionized by
  younger and/or massive stellar clusters . Indeed, \cite{kennicutt89}
  recognized that \ion{H}{ii} regions in the center of galaxies
  distinguish themselves spectroscopically from those in the disk by
  their stronger low-ionization forbidden emission. The nature of this
  difference was not clear. It may be due to contamination by an extra
  source of ionization, such as diffuse emission or the presence of an
  AGN.  However, other stellar processes, such as nitrogen enhancement
  due to a natural aging process of \ion{H}{ii} regions and the
  surrounding ISM, can produce the same effect. These early results
  were confirmed by \cite{ho97}, who demonstrated that inner
  star-forming regions may populate the right branch of the BPT
  diagram at a location above the demarcation line defined later by
  \cite{kauffmann03}.  }

{ Another bias that may be introduced by classical
  slit-spectroscopy surveys of \HII regions is the aperture
  effects. As already noticed by \cite{dopita14}, Long-slit
  observations do not integrate over the whole of an \HII region,
  particularly in the case of observations of nearby galaxies. By
  picking out the brightest regions of such \HII regions, we bias the
  data toward the high-excitation regions and the spectrum is not
  representative of the whole \HII region. This is clearly illustrated
  when the integrated and spatial resolved properties of \HII regions are compared
  \citep[e.g.][]{sanchez07c,rela10,monreal11}. }

The advent of Multi-Object Spectrometers and Integral Field
Spectroscopy (IFS) instruments with large fields of view offers us the
opportunity to undertake a new generation of emission-line surveys,
based on samples of hundreds of \ion{H}{ii} regions and full
two-dimensional (2D) coverage of the disks of nearby spiral galaxies
\citep[e.g.][]{rosales-ortega10,sanchez12a,sanchez14}.  Large catalogs
of \HII regions have been produced, with thousands of objects, over
statistical samples of galaxies covering a wide range of morphological
types \citep{sanchez12b,sanchez13}. 
These catalogs led to the discovery of  a clear correlation between the oxygen abundance of
individual \HII regions and the stellar mass densities of the
underlying stellar population \citep{rosales12,sanchez13}, a radial
gradient of different ionization conditions \citep{sanchez12b}, and the
existence of a common abundance gradient in all disk-dominated
galaxies \citep{sanchez14}. { Interestingly, these results confirmed the
previous ones presented by \cite{kennicutt89} and \cite{ho97}, regarding the presence of \HII
regions above the demarcation line defined by \cite{kauffmann03}. Those
\HII regions are not restricted to the
central regions and can be found at any galactocentric distance,
even at more than 2 $r_e$, which excludes/minimizes a possible
contamination by a central source of ionization. A more plausible explanation
is that we are observing more aged \HII regions as indicated before.}

All these relations point towards a connection between the ionization
conditions in \HII regions and the properties of the underlying
stellar population, and therefore, the overall evolution of the host
galaxies. This connection is not taken into account by
photo-ionization models. { Evolutionary processes of the
ionizing cluster \citep[e.g.][]{dopita06a} and/or constrainsts impossed by the evolution
of the host galaxies \citep[e.g.][]{dopita06b} are required to exaplain these relations.}

In the current study we explore the distribution of \HII regions
across the BPT diagram to try to understand the nature of this
connection. To do so, we use the largest catalog of \HII regions
with spectroscopic information currently available. This catalog was
extracted from the CALIFA dataset
\citep{sanchez12a}, and comprises more than 5000 \HII regions
\citep{sanchez14}. The structure of this article is as follows:
Section \ref{sample} presents the sample of galaxies and \HII regions
used along this study. Section \ref{ana} summarizes the analysis
of the data. Section \ref{models} comprises the comparison
between the distribution across the BPT diagram with photo-ionization
models.
 In Sec. \ref{empirical} we explore how the
parameters that define the ionization conditions actually change  across the BPT 
diagram. Sections \ref{morph} and \ref{dist} explore the change in the
distribution of \HII regions as a function of the morphology of the host
galaxy and their galactocentric distance. Finally, 
Sec. \ref{age_met} explores the dependency of the location within
the BPT diagram with the properties of the underlying stellar
population. Section \ref{dis} discusses the results of the analysis,
and the conclusions are summarized in Sec. \ref{sum}.

\section{Sample of galaxies and dataset}\label{sample}

The galaxies were selected from the CALIFA observed sample. Since
CALIFA is an ongoing survey, whose observations are scheduled on a
monthly basis (i.e., dark nights), the list of objects increases
regularly. The current results are based on the 306 galaxies observed
before May 2013. Their main characteristics were already described in
\cite{sanchez14}.

The details of the survey, sample, observational strategy, and
reduction are explained in \cite{sanchez12a}. All galaxies were
observed using PMAS \citep{roth05} in the PPAK configuration
\citep{kelz06}, covering a hexagonal field-of-view (FoV) of
74$\arcsec$$\times$64$\arcsec$, sufficient to map the full optical
extent of the galaxies up to two to three disk effective radii. This is
possible because of the diameter selection of the sample (Walcher et
al., 2014). The observing strategy guarantees  complete coverage
of the FoV, with a final spatial resolution of FWHM$\sim$3$\arcsec$,
corresponding to $\sim$1 kpc at the average redshift of the
survey. The sampled wavelength range and spectroscopic resolution
(3745-7500 \AA, $\lambda/\Delta\lambda\sim$850, for the low-resolution
setup) are more than sufficient to explore the most prominent ionized
gas emission lines, from [\ion{O}{ii}]$\lambda$3727 
to [\ion{S}{ii}]$\lambda$6731 at the redshift of our targets, on
one hand, and to deblend and subtract the underlying stellar
population, on the other \citep[e.g.,][]{sanchez12a,kehrig12,cid-fernandes13,cid-fernandes14}. The dataset was
reduced using version 1.3c of the CALIFA pipeline, whose modifications
with respect to the one presented in \cite{sanchez12a} are described
in detail in \cite{husemann13}. In summary, the data fulfil the
predicted quality-control requirements with a spectrophotometric
accuracy that is better than 15\%\ everywhere within the wavelength range,
both absolute and relative with a depth that allows us to detect
emission lines in individual \HII regions as faint as
$\sim$10$^{-17}$\flux, and with a signal-to-noise ratio of
S/N$\sim$3-5. For the emission lines considered in the current study,
the S/N is well above this limit, and the measurement errors are
negligible in most of the cases. In any case, they have been
propagated and included in the final error budget.

The final product of the data reduction is a regular-grid datacube,
with $x$ and $y$ coordinates that indicate the right ascension and
declination of the target and $z$ a common step in
wavelength. The CALIFA pipeline also provides  the propagated error
cube, a proper mask cube of bad pixels, and a prescription of how to
handle the errors when performing spatial binning (due to covariance
between adjacent pixels after image reconstruction). These datacubes,
together with the ancillary data described in Walcher et al. (2014), are the basic starting points of our analysis.

\subsection{\HII detection and extraction}\label{HIIreg}

The segregation of \ion{H}{ii} regions and the extraction of the
corresponding spectra is performed using a semi-automatic procedure
named {\sc HIIexplorer}\footnote{\url{http://www.caha.es/sanchez/HII_explorer/}}. The
procedure is based on some basic assumptions: (a) \ion{H}{ii} regions
are peaky and isolated structures with a strong ionized gas emission, which is
significantly above the stellar continuum emission and the average ionized gas
emission across the galaxy. This is particularly true for 
H$\alpha$ because (b) \ion{H}{ii} regions have a typical physical
size of about a hundred or a few hundred parsecs
\citep[e.g.,][]{rosa97,lopez2011,oey03}, which corresponds to a typical
projected size of a few arcsec at the  distance  of the galaxies.

These basic assumptions are based on the fact that most of the H$\alpha$
luminosity observed in spiral and irregular galaxies is a direct
tracer of the ionization of the interstellar medium (ISM) by the
ultraviolet (UV) radiation produced by young high-mass OB
stars. Since only high-mass, short-lived stars contribute
significantly to the integrated ionizing flux, this luminosity is a
direct tracer of the current star-formation rate (SFR), independent of the
previous star-formation history. Therefore, clumpy structures detected in
the H$\alpha$ intensity maps are most probably
associated with classical \ion{H}{ii} regions (i.e., those regions for which the oxygen
abundances have been calibrated).

The details of {\sc HIIexplorer} are given in \cite{sanchez12b} and
\cite{rosales12}. In summary we create a narrow-band image centred on
the wavelength of H$\alpha$ at the redshift of the object. Then we run
{\sc HIIexplorer} to detect and extract the spectra of each individual
\HII region, adopting the parameters presented in
\cite{sanchez14}. { The algorithm starts looking for the brightest
  pixel in the map. Then, the code aggregates the adjacent pixels until all
  pixels with flux larger than 10\% of the peak flux of the region and
  within 500 pc or 3.5 spaxels from the center have been
  accumulated. The distance limit take into account the typical size
  of \HII regions of a few hundreds of parsecs
  \citep[e.g.][]{rosa97,lopez2011}. Then, the selected region is
  masked and the code keeps iterating until no peak with flux ex-
  ceeding the median H$\alpha$ emission flux of the galaxy is left.
  \cite{mast14} studied the loss of resolution in IFS using nearby
  galaxies observed by PINGS \citep{rosales-ortega10}. Some of these
  galaxies were simulated at higher redshifts to match the
  characteristics and resolution the galaxies observed by the CALIFA
  survey. Regarding the \HII region selection, the authors conclude
  that at $z\sim$0.02 the \HII clumps can contain on average from 1 to
  6 of the \HII regions obtained from the original data at
  $z\sim0.001$. Another caveat is that this procedure tends to select
  regions with similar sizes although real \HII regions actually have
  different sizes. However, the actual adopted size is of the order of
  the FWHM of the CALIFA data for the used version of the data
  reduction \citep{husemann13}.}

Then, for each individual extracted
spectrum we modelled  the stellar continuum using
{\tt FIT3D}\footnote{\url{http://www.caha.es/sanchez/FIT3D/}}, a
fitting package described in \cite{sanchez06b} and \cite{sanchez11}. A
simple SSP template grid with 12 individual populations was adopted.
It comprises four stellar ages (0.09, 0.45, 1.00, and 17.78 Gyr), two
young and two old ones, and three metallicities (0.0004, 0.019, and
0.03, that is, subsolar, solar, or supersolar, respectively). The models were extracted from
the SSP template library provided by the MILES project
\citep{vazdekis10,falc11}. The use of different stellar ages and
metallicities or a larger set of templates does not qualitatively
affect the derived quantities that describe the stellar populations.
Indeed, we compared two of these parameters (luminosity weighted age
and metallicity), derived using this simple template and the more
elaborate one described in \cite{cid-fernandes13,cid-fernandes14} and
we found an agreement within $\sim$0.25 dex in both parameters.

After subtracting the underlying stellar population, the flux intensity
of the strong emission lines was extracted for each gas-pure spectrum
by fitting a single Gaussian model to each line,
resulting in a catalog of the emission line properties \citep{sanchez12b}. Finally,
the \HII regions were selected from this final catalog based on the
properties of the underlying stellar continuum. It was required that
they present a young stellar population, compatible with the
observed equivalent width of H$\alpha$  \citep{sanchez14}.

The final catalog comprises the strongest emission line and emission
line ratios from [\ion{O}{ii}]$\lambda$3727 to
[\ion{S}{ii}]$\lambda$6731 for 5229 \HII regions, together with their
equivalent widths and the luminosity weighted ages and metallicities
of the underlying stellar population. So far, this is the largest
catalog of \HII regions and agregations with spectroscopic
information. It is also one of the few catalogs derived for a
statistically well-defined sample of galaxies representative of the
entire population of galaxies in the local Universe (Walcher et al. 2014).

\section{Analysis and Results}\label{ana}

The main goal of this study is to understand how the ionization
conditions in \HII regions are related to the properties of the
galaxies, and if there is a connection between the ionization
conditions and the overall evolution and chemical enrichment of the
stellar populations.

To do so we use the catalog of emission-line fluxes and ratios,
and the corresponding properties of the stellar populations of the
\HII regions sample described before \citep{sanchez14}. 


\begin{figure*}
\centering
\includegraphics[width=7.5cm,angle=270,clip,trim=0 0 0 0]{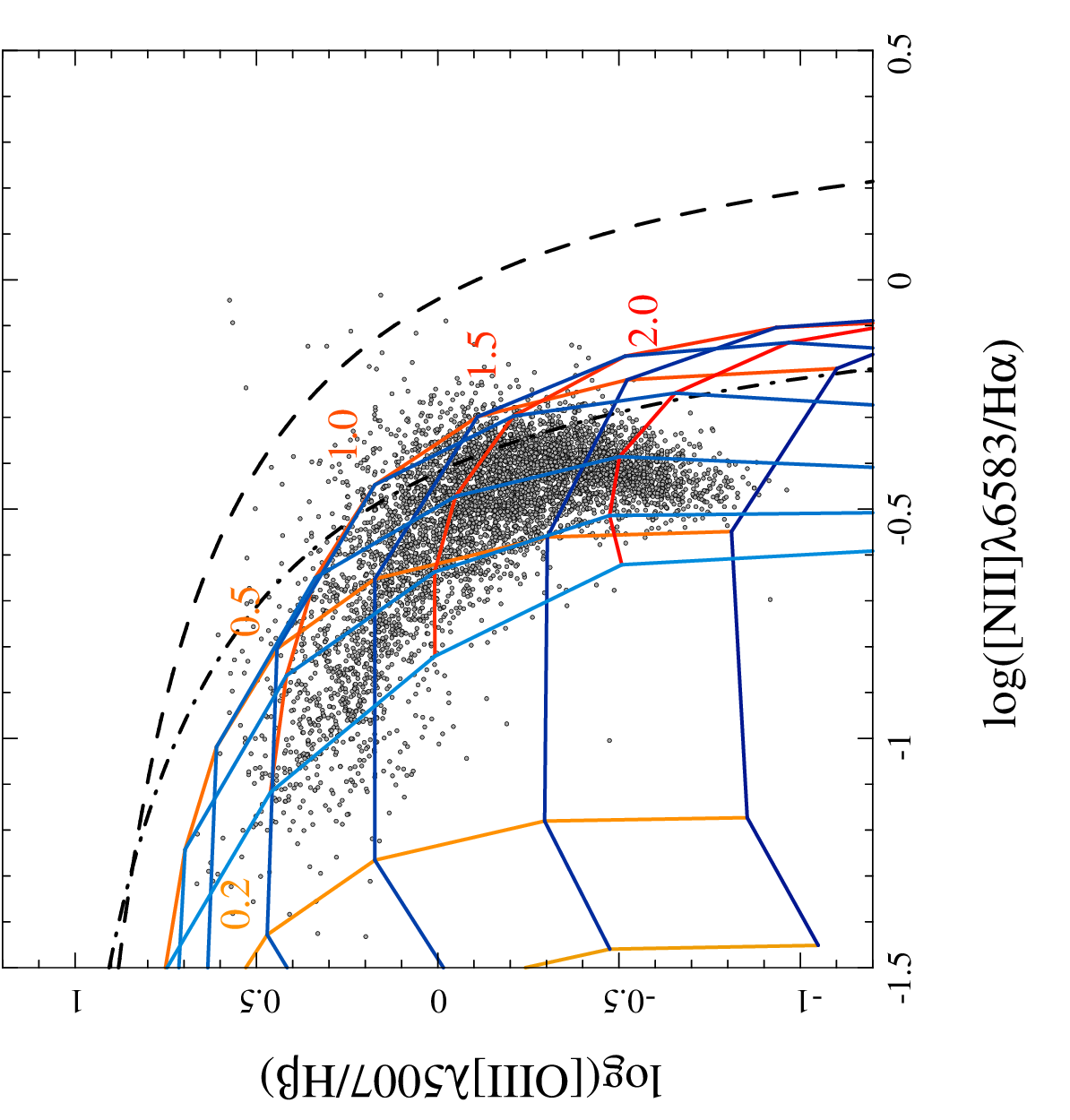}\includegraphics[width=7.5cm,angle=270,clip,trim=0 0 0 0]{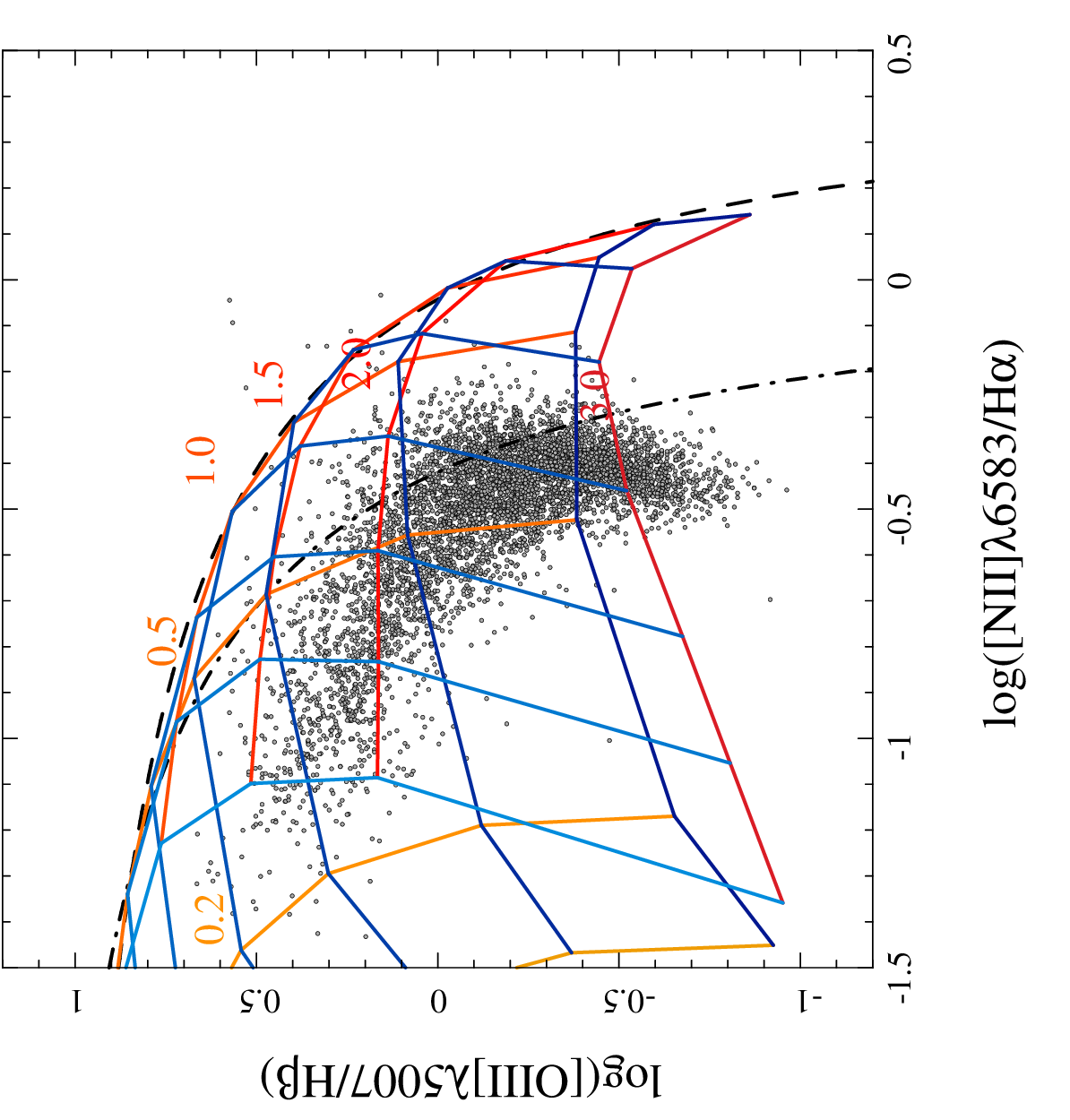}
\includegraphics[width=7.5cm,angle=270,clip,trim=0 0 0 0]{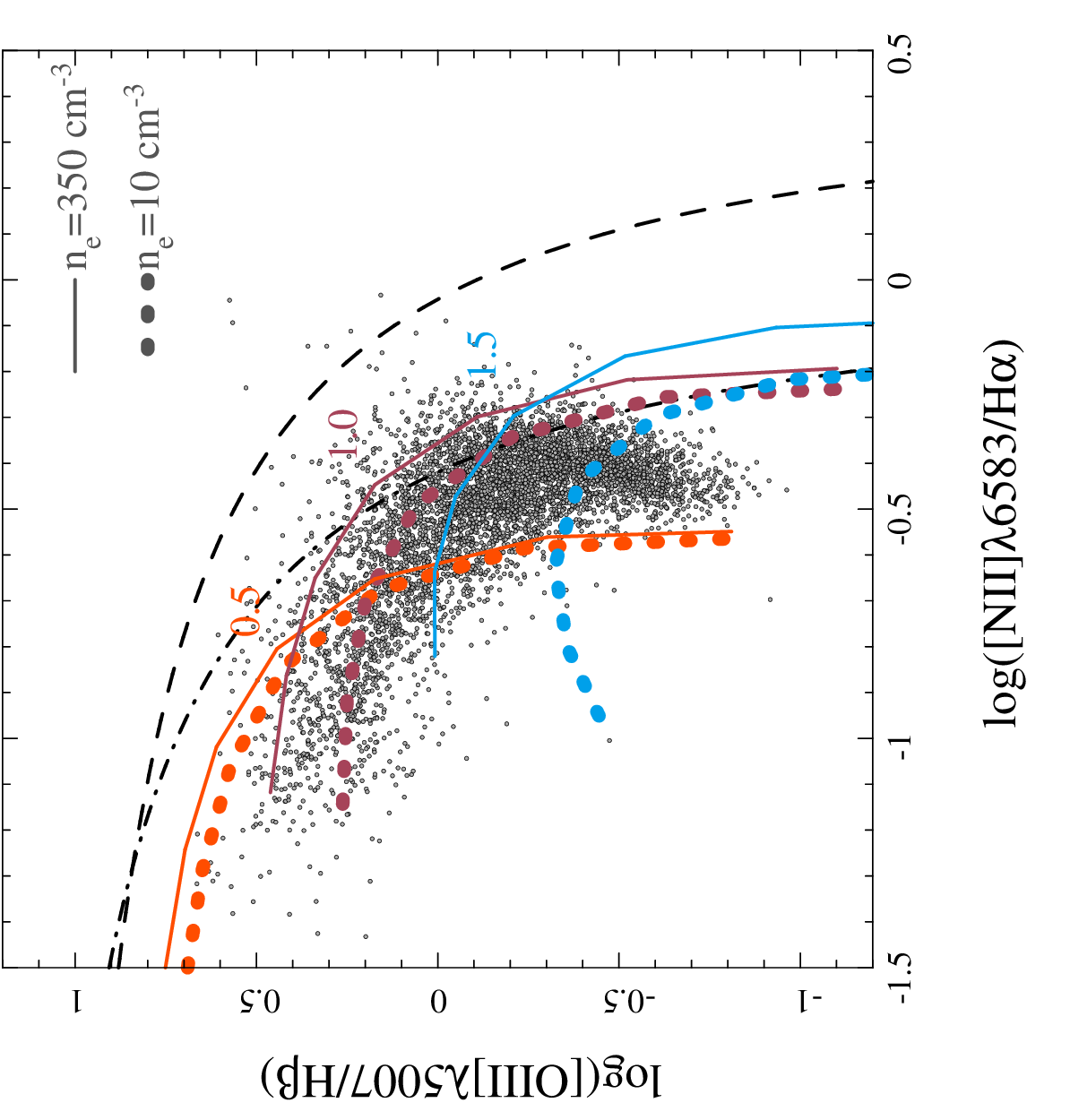}\includegraphics[width=7.5cm,angle=270,clip,trim=0 0 0 0]{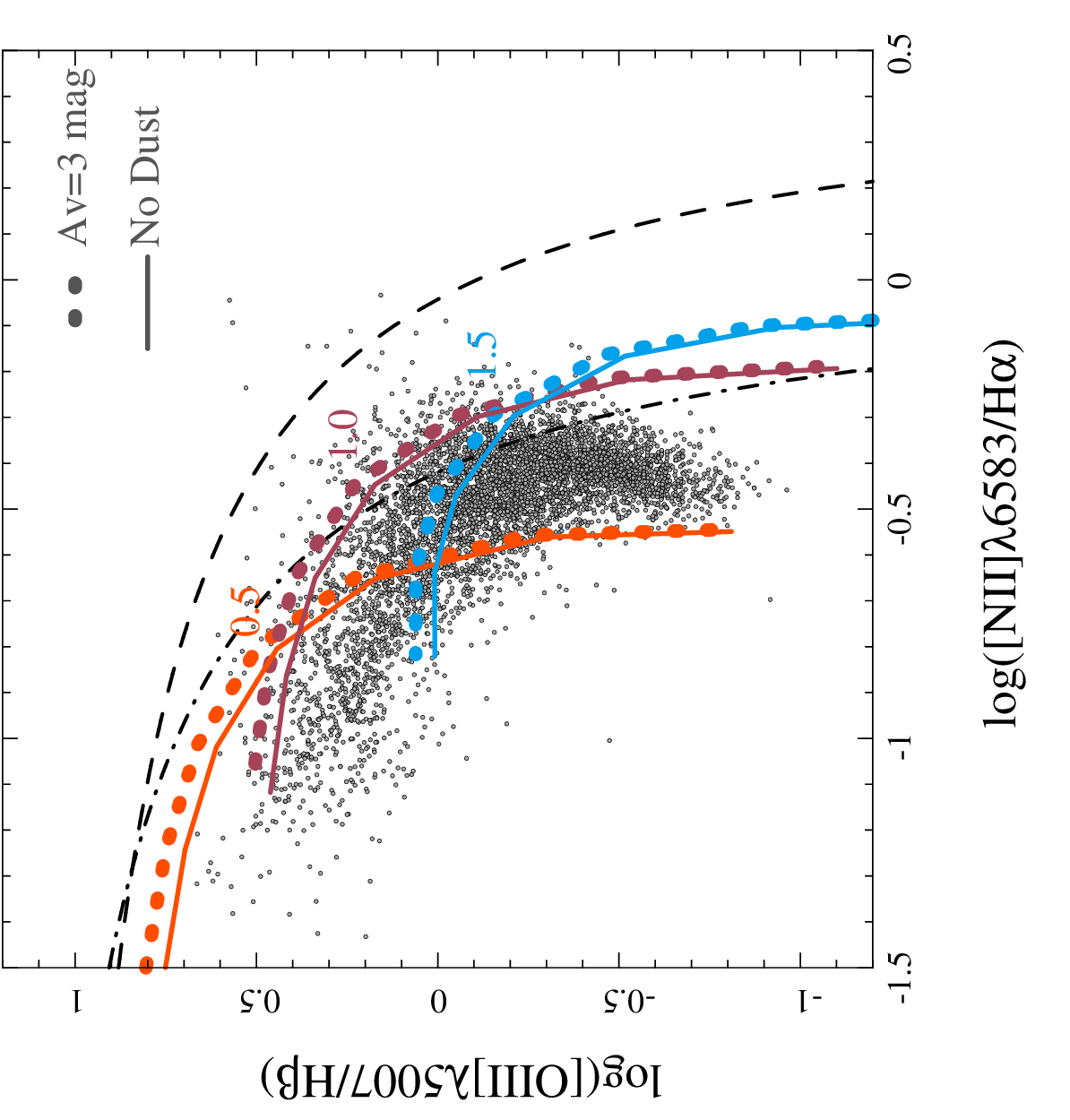}
\caption{\label{fig:grid} {\it Top-left panel:} [\ion{O}{iii}]~$\lambda$5007/H$\beta$ vs. [\ion{N}{ii}]~$\lambda$6583/H$\alpha$ diagnostic diagram for the $\sim$5000 \HII ionized regions included in our sample. The color solid-lines represent the expected line ratios derived from the MAPPINGS-III ionization models, based on an instantaneous zero-age starbust model based on the PEGASE spectral energy distribution \citep{kewley01,dopita00}, and a fixed electron density of 350 cm$^{-3}$. Each orange-to-red solid-line corresponds to a different metallicity, which is indicated with the corresponding label and color (ranging from 0.2 to 2.0 solar metallicities), and each blue-to-cyan solid-line corresponds to a different ionization parameter of the radiation field (ranging from log(u)=-3.7, for the dark blue, to log(u)=-2, at the cyan blue). {\it Top-right panel:} Similar plot, with the only difference that now the color solid-lines represent the expected line ratios for a continuous starburst model  4 Myr since its ignition. {\it Botton-left panel:} Similar plot, with the only difference being in this case that the color solid-lines and the dotted-lines correspond to the expected line ratios for a photoionization model based on an instantaneous zero-age starburst, with a constant metallicity (indicated with a label and represented with a fixed color), but for two different electron densities (10 cm$^{-3}$, for the dotted-line, and 350$cm^{-3}$, solid one).  {\it Botton-right panel:} Similar plot, with the only difference that in this case the color solid-lines and the dotted-lines correspond to the expected line ratios for a photoionization model based on an instantaneous zero-age starburst, with a constant metallicity, and a fixed electron density shown in top-left panel, but for two different dust attenuations, (no dust, for the solid-line, and $A_V=$3 mag, for the dotted one).  In all the panels, the dot-dashed- and dashed-line represent, respectively, the \cite{kauffmann03} and \cite{kewley01} demarcation curves. They are usually invoked to distinguish between classical star-forming objects (below the dashed-dotted line), and AGN powered sources (above the dashed-line). Regions between both lines are considered intermediate ones.}
\end{figure*}

\subsection{Distribution of \HII regions across the BPT diagram}\label{models}

Figure \ref{fig:grid} shows the distribution of our sample of \HII
regions across the BPT diagram, compared with different
photoionization models. The \cite{kauffmann03} and \cite{kewley01}
demarcation lines have been included as general references. The two
tophand panels include the expected location based on the MAPPINGS-III
photoionization models \citep{allen08}, adopting two different
ionizing populations \citep{kewley01}. The top left hand panel 
shows the grid corresponding to an instantaneous zero-age starbust
model based on the PEGASE spectral energy distribution
\citep{kewley01,dopita00}. Each red solid-line corresponds to a
different metallicity, which is indicated with the corresponding label
(ranging from 0.2 to 2.0 solar abundances), and each blue solid-line
corresponds to a different ionization parameter of the radiation field
(ranging from log(u)=-3.7, at the bottom, to log(u)=-2, at the top).
The top right hand panel shows a similar plot;  the only difference
being that here we assumed a continous starburst
model  4 Myr since its ignition. In all cases it is assumed an
average electron density of $n_e=$350 cm$^{-3}$.

As expected, most of the \HII regions are located below the envelope
defined by the instantaneous star-burst photoinization models, 
which was proposed by \cite{dopita00} as a demarcation line  to define the regions
that are clearly ionized by star-formation. This line is very similar
those defined by \cite{kauffmann03}, empirical, and \cite{stas06},
theoretical. We adopted here the \cite{kauffmann03} one, but the use of any of the
three will produce similar results. However, there is a non-negligible fraction of
$\sim$14\% of \HII regions located between this envelope and the one
defined by a a model with continuous star formation, as we already noticed in
\cite{sanchez13} and \cite{sanchez14}. \cite{kewley01} noticed that
starburst galaxies are preferentially located in this region. They
considered these galaxies to have experienced on-going star formation over
at least several Myr, and this prolonged star-formation activity is
reflected in the BPT diagrams. A similar interpretation can be applied
to the \HII regions above the \cite{dopita00} and/or
\cite{kauffmann03} demarcation lines.

In both cases, a single burst or a continuous star formation, the
distribution of \HII regions across the BPT diagram does not cover the
entire parameter space forseen by the considered photoionization
models.  Thus, {\it real} \HII regions cover a smaller parameter space
than is predicted by the models. On the other hand, there is a
degeneracy in the combination of ionization conditions that predict a
certain location in the diagram (abundance and ionization strength in
this particular grid). Therefore, the location of a particular \HII
could be explained by different ionization conditions. For example,
the curve described for a fixed solar abundance present values within
the BPT diagram that range from (-1,0.5), in the upper-left corner, to
(-0.2,-1), in the bottom-right corner. Indeed, it describes almost an
envelope of the distribution of \HII regions across the BPT diagram
(Fig. \ref{fig:grid}). Only when fixing one of the two parameters
involved in the grid it is possible to derive a trend with the other
parameter across the BPT diagram. Both results have two direct
consequences: (i) it is not possible to derive which ioniziation
conditions are associated with a particular \HII region based only on
their location in the BPT diagram, and (ii) there are ionization
conditions that are discarded by nature for reasons that cannot be
predicted based only on photoionization models. This result is well
known, and has limited our ability to recover the ionization
conditions of \HII regions for decades \citep[e.g.][]{epm14}.


We explore how the variation of other parameters affects the line
ratios predicted by photoionization models. The bottom left hand panel of Fig \ref{fig:grid}
 illustrates the change in the location across
the BPT diagram when different electron densities are assumed. We show the same grid of photoinization models presented in the
top left hand panel, but assuming two different electron densities
($n_e$=10 and 350 cm$^{-3}$). In general the photoionization models
predict that higher density regions of similar abundances and
ionization strengths should be located at higher values of
[\ion{O}{iii}]/H$\beta$ for very similar values of [\ion{N}{ii}]/H$\alpha$. Therefore,
  the regions located at the top right envelope of the distribution
  should be the densest ones.

Finally, we explore the variation of the line ratios when dust
attenuation is taken into account. The bottom right hand panel of Fig
\ref{fig:grid} shows the same grid of photoionization models presented
in top-lefthand panel, but assuming two different dust attenuations
($A_V$=0 and 3.5 mag).  The effects of the dust attenuation over the
line ratios were simulated adopting the \cite{cardelli89} extinction
law, with a specific dust attenuation of $R_{\rm V}=3.1$. The BPT
diagram is designed to be less affected by dust attenuation than other
ones frequently used \cite[e.g.][]{veil95}, since it involves emission
lines at very similar wavelengths for each ratio. Thus, even in the
case of very large dust attenuations the differences are very
small. Therefore, no trend is expected with the dust attenuation in
the distribution of \HII regions across the BPT diagram.

\begin{figure*}
\centering 
\includegraphics[width=7.5cm,angle=270,clip,trim=0  0 0 0]{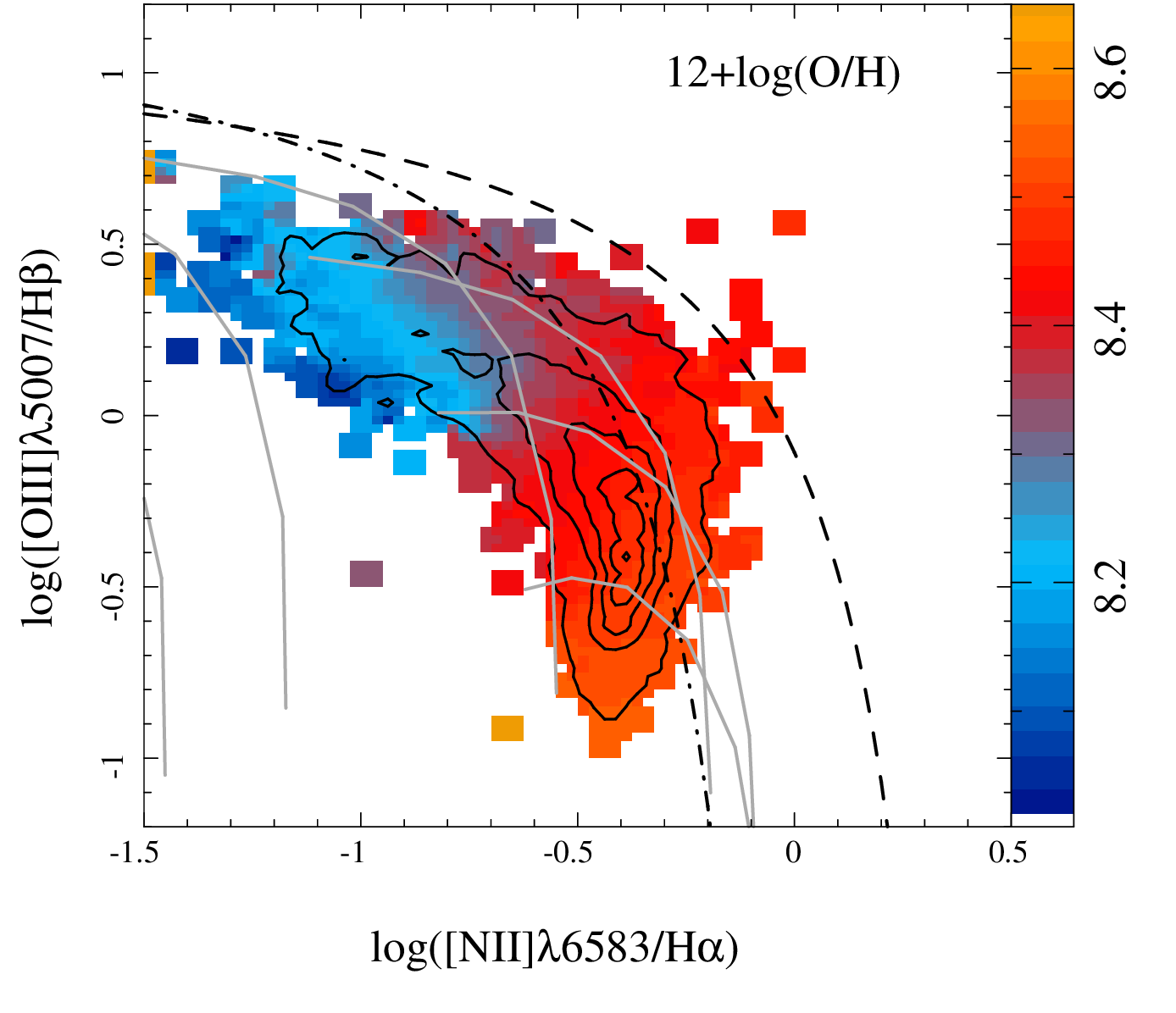}\includegraphics[width=7.5cm,angle=270,clip,trim=0 0 0 0]{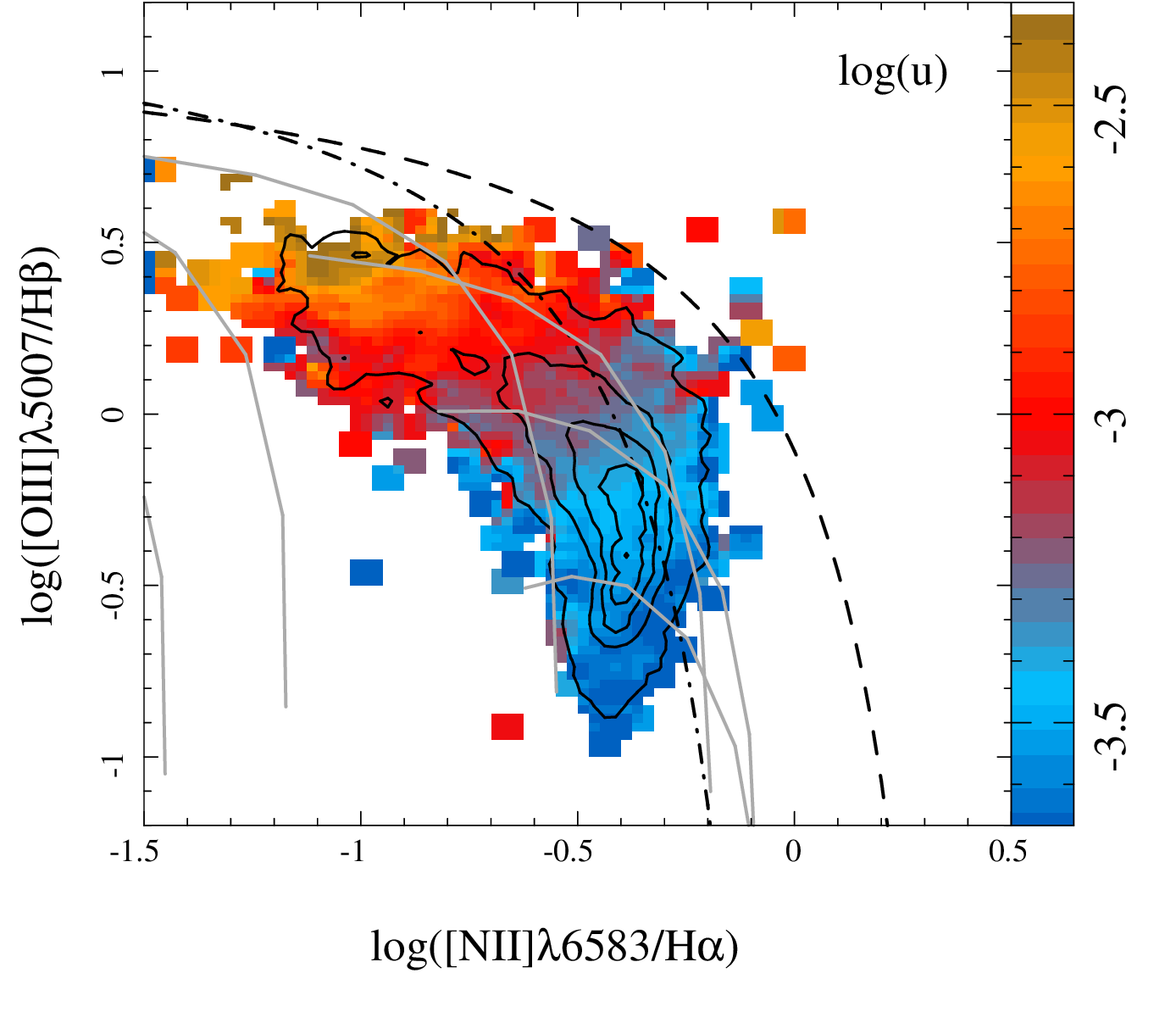}
\includegraphics[width=7.5cm,angle=270,clip,trim=0 0 0
  0]{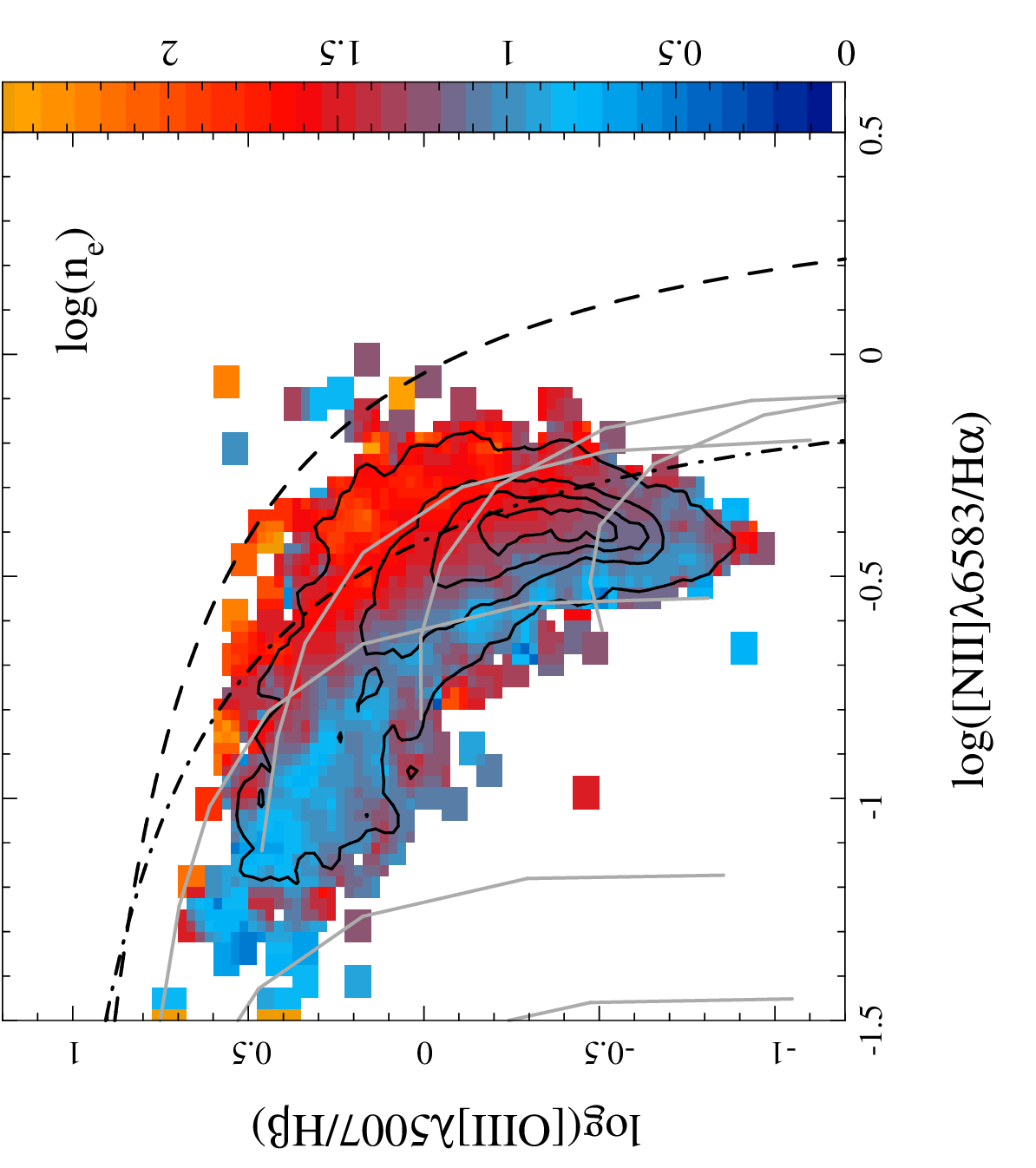}\includegraphics[width=7.5cm,angle=270,clip,trim=0
  0 0 0]{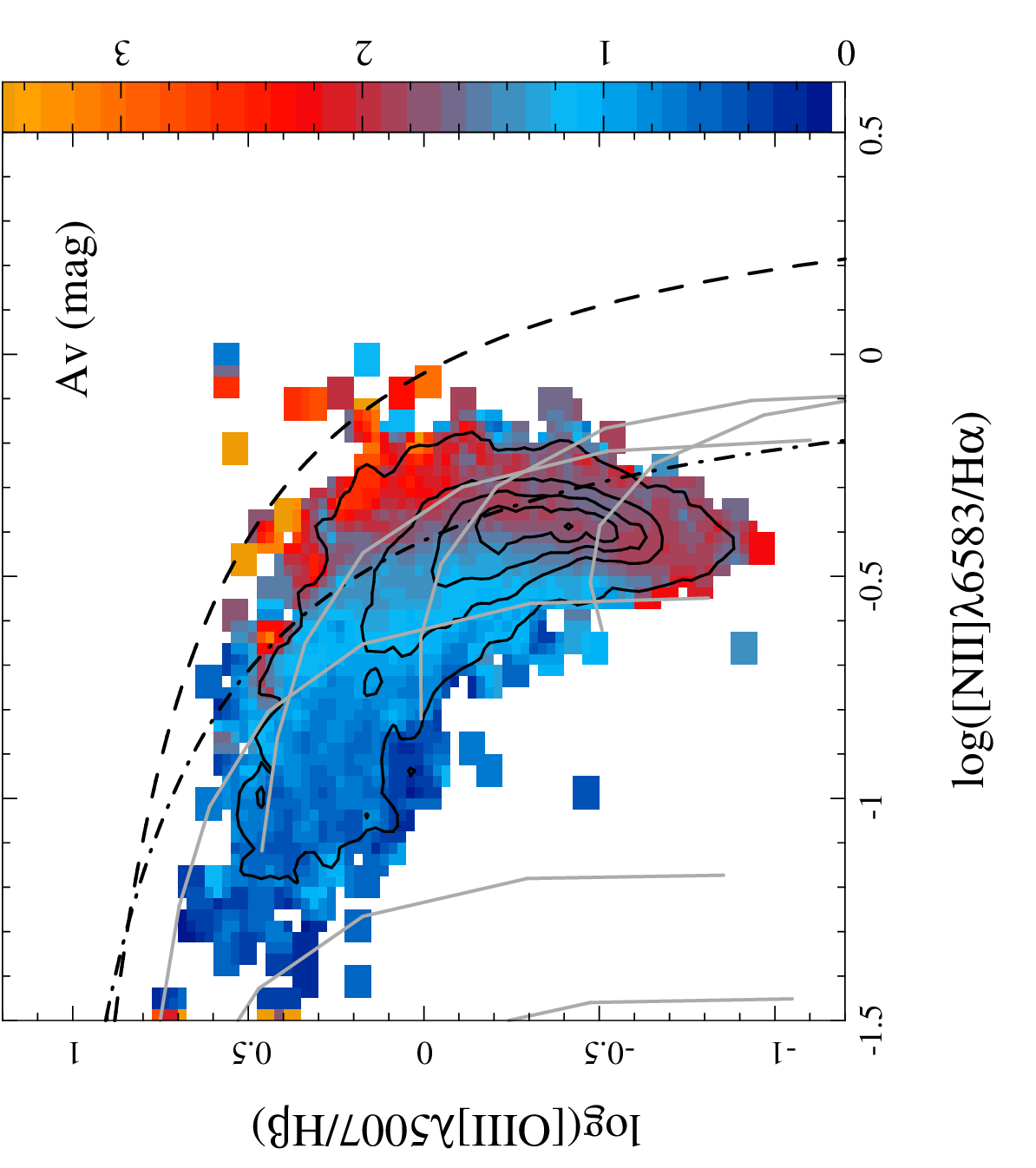}
\caption{\label{fig:BPT_prop} [\ion{O}{iii}]~$\lambda$5007/H$\beta$ vs. [\ion{N}{ii}]~$\lambda$6583/H$\alpha$ diagnostic diagram for the $\sim$5000 \HII regions included in our sample. The contours show the  density distribution of these regions within the diagram plane, with the  outermost contour enclosing 95\% of the regions, and each consecutive one  enclosing 20\% fewer regions. In each panel the color indicates the average value at the corresponding location in the diagram for one of the four parameters described in the text: {\it top-left panel:} the oxygen abundance; {\it top-right panel:} the ionization parameter; {\it botton-left panel:} the electron density; and {\it botton-right panel:} the dust attenuation. In all panels, the dot-dashed and dashed lines represent, respectively, the demarcation curves described in Fig.\ref{fig:grid}, and the grey solid-lines represent the expected line ratios derived from the photoionization models shown in Fig. \ref{fig:grid}, where each line corresponds to a different stellar metallicity (orage-to-red solid lines in Fig. \ref{fig:grid}). }
\end{figure*}

\subsection{Empirical estimation of the ionization parameters}\label{empirical}

We have seen in the previous section that the location in the BPT
diagram for a classical \HII region ionized by young stars of a
certain age is well defined by three main parameters (a) the
ionization parameter or fraction of Lyman continuum photons with respect to
total amount of gas, (b) the metallicity or chemical abundance of the ionized
gas, and (c) the electron density of the gas. In addition, we have seen
that the location should have no dependency on the dust attenuation
(for this particular diagram). However, the trends of the two first
parameters across the BPT diagram are not well defined based on
photoinization models. We investigate here how the oxygen abundance,  ionization
strength, electron density, and dust attenuation
actually change across the BPT diagram, based on empirical estimations
of these parameters.

The oxygen abundances were derived using the {\it counterpart}-method
described by \cite{P12}. This method uses the dust corrected emission line
ratios of a set of strong emission lines including
[\ion{O}{ii}]/H$\beta$, [\ion{O}{iii}]/H$\beta$, [\ion{N}{ii}]/H$\beta$ and
[\ion{S}{ii}]/H$\beta$, and compares the input values with a grid of line
ratios of \HII regions with known abundances derived using the direct
method (i.e., based on direct estimations of the electron temperature). This
calibrator attempts to be an improvement of the previously analytical
method proposed in \cite{pilyugin10}. The values derived using this
method were cross-checked with the ones provided using the O3N2
calibrator \citep{allo79,pettini04,stas06},

\begin{eqnarray}
 {\rm O3N2} &=& {\rm log } \left[\frac{I([\ion{O}{iii}]~\lambda5007)/I({\rm H}\beta)}{I([\ion{N}{ii}]~\lambda6584)/I({\rm H}\alpha)} \right] 
\end{eqnarray}
using the recently updated calibration presented by \cite{marino13},
that improves the abundance estimation by including many more \HII
regions than previous calibrations using this parameter
\citep[e.g.][]{pettini04}, in particular in the range of higher
metallicities.  The advantage of this method is that it depends on
{ strong, well deblended} emission lines, and in particular does not depend on
[\ion{O}{ii}]$\lambda$3727, a line that may be affected by vignetting
\citep{sanchez12a}, any inaccuracy in flux calibration in the blue-end
of the covered spectral range \citep{husemann13}, and uncertainties
related to the derivation of the dust attenuation. The values for
oxygen abundances match one another within a standard deviation of
$\sim$0.07 dex, a value of the order of the expected errors of the
individual measurements \citep{sanchez13} and the accuracy of the
calibrators \citep{marino13}.

For the ionization parameter, $u$, we adopted two different
calibrators described by \cite{diaz00}, based on
[\ion{S}{ii}]$\lambda$6717,31/H$\beta$, and
[\ion{O}{ii}]$\lambda$3727/H$\beta$, that require a previous knowledge
of the oxygen abundance, and a third one described by \cite{dors11},
that depends on the [\ion{O}{ii}]/[\ion{O}{iii}] line ratio. This
latter one is a recalibration of the classical one proposed by
\cite{diaz00}. We cross-checked the values obtained by the different
methods, and they seem to agree within the errors, although
[\ion{S}{ii}]/H$\beta$ yields slightly larger values than the other
two. We adopted the mean value of the three as our estimation of the
ionization parameter. { Recent results indicate that other line
  ratios, like [\ion{O}{iii}]/[\ion{S}{ii}] can be used as good
  tracers of the ionization parameter \citep{dopita13}. However, there
  is no published calibrator corresponding to this ratio to our
  knowledge, and therefore we could not compare the results in an
  independent way. } As a final sanity check we cross-checked that the
derived values follow the described relation between this parameter
and the equivalent width of H$\beta$ \citep{diaz89}.

The electron density, $n_e$, was derived based on the 
line ratio of the [\ion{S}{ii}] doublet \citep[e.g.,][]{osterbrock89}, by solving
the equation:

\begin{eqnarray}\label{eq2}
 \frac{I({\rm [\ion{S}{ii}]\lambda6717})}{I({\rm [\ion{S}{ii}]\lambda6731})} &=& 1.49 \frac{1+3.77{\rm x}}{1+12.8{\rm x}}
\end{eqnarray}
where $x$ is the density parameter, defined as $x=10^{-4} n_e
t^{-1/2}$ and $t$ is the electron temperature in units of $10^4$~K
\citep{McCa85}. For this calculation we assumed a typical electron
temperature of $T=10^4$K, an average value corresponding to the
expected conditions in \HII regions \cite{osterbrock89}. This equation
reflects the fact that the [\ion{S}{ii}] doublet ratio is sensitive to
changes in the electron density only for a limited range of
values. For high and low values it becomes asymptotic and the actual
derived value has to be treated with care, and should not be used for
quantitative statements. However, the value will be still valid for the
purposes of this study, allowing us to understand the global trends across the
BPT diagram with the electron density.

Finally, the dust attenuation, A$_{\rm V}$, was derived based on the
H$\alpha$/H$\beta$ Balmer line ratio. The extinction law by
\cite{cardelli89} was assumed, with a specific dust attenuation of
$R_{\rm V}=3.1$, and the theoretical value for the unobscured line
ratio for case B recombination of H$\alpha$/H$\beta=2.86$, for
$T_e$=10,000\,K and $n_e$=100\,cm$^{-3}$ \citep{osterbrock89}.
For this study we have assumed that the intrinsic H$\alpha$/H$\beta$ line ratio
does not vary significantly, although it is known that it presents a dependence
with the electron density and the temperature \citep[e.g.][]{sanchez05}.

\subsubsection{Distribution across the BPT diagram}\label{emp_dist}

The top left hand panel of Fig. \ref{fig:BPT_prop} shows the
distribution of \HII regions across the BPT diagram, with a color code
representing the oxygen abundance in each region. The more metal poor
\HII regions are located in the upper-left corner of the diagram, and
the more metal rich ones are located towards the lower-right
corner. None of these trends can be explained based only in the
expected values derived from photoionization models, as we have seen
in the previous section.  The opposite trend is found for the
ionization parameter.  The top left hand panel of Figure
\ref{fig:BPT_prop}, shows the same distribution with a color code
representing the value of the ionization parameter in each
region. There is a clear trend across the diagram. The \HII regions
with the largest ionization parameter are located in the upper-left
corner of the diagram, and those with the lowest values of this
parameter are located towards the lower-right corner.

The bottom left hand panel of Fig. \ref{fig:BPT_prop} shows the distribution
of \HII regions across the BPT diagram with a color code representing
the electron density. The denser regions are located towards the
upper-right envelope of the distribution, as expected from the
comparison with photoionization models, presented in the previous
section. Regions with higher electron density are expected in areas of
the galaxy that have suffered an over-pressure, in particular in the
leading front of a density wave generated by spiral arms or towards the
central regions of the galaxies. We will investigate that possibility
in the next sections.

Finally, we explore the relation between the distribution of \HII
regions across the BPT diagram and the amount of dust
attenuation. The bottom right hand panel of Fig. \ref{fig:BPT_prop} shows the
same distribution shown in previous panels with a color code
representing the dust attenuation derived from the Balmer
decrement. 
There is a clear trend that is not taken in to account by photoionization 
models: by construction the
location within the BPT diagram ought to be independent of dust
attenuation.  Regions with lower dust
attenuation are located in the upper-left region of the diagram, with
A$_{\rm V}$ increasing towards the lower-right, and then increasing
again at the upper-right end of the diagram. Indeed, the regions with
the largest dust attenuation are those located in the so-called
intermediate area, defined as the region between the
\cite{kauffmann03} and \cite{kewley01} curves.

We should note here that the derived parameters may present
cross-dependences due to the adopted formulae/calibration to derive
them. We tried to minimize or quantify those possible dependences to
exclude possible induced trends. For the oxygen abundance we adopted
the {\it counterpart}-method since it uses all the strong emission
lines in the considered wavelength range, and not only those involved
in the \HII diagram \cite[like O3N2 or N2 calibrators][]{marino13}. We
repeated the analysis using other abundance indicators, like R23 or
N2S2, without significant differences in the described trends. The
ionization parameter was derived using different calibrators, only one
of them is purely empirical (the one based on the
[\ion{O}{ii}]/[\ion{O}{iii}] line ratio). The other two calibrators
involve deprojections of photoionization models that try to remove the
degenerancy in the derivation of the oxygen abundance and the
ionization-parameter. We consider that this approach minimize the
possible dependencies between those two parameters induced by the
adopted estimators are minimized.

The derivation of the electron density is based on a formula derived
from basic principles of the quantum physic (equation \ref{eq2}). This
equation provides with an electron density that depends on the
electron temperature. We have fixed this second parameter, and
therefore we may have induced a possible secondary correlation with
any other parameter that depends on the temperature, like the oxygen
abundance or the ionization strength. However, the largest possible
bias introduced by fixing the electron temperature in the derivation
of $n_e$ is a of the order of a factor 2-3. This cannot explain the
range of derived electron densities, of two orders of magnitudes, or
their pattern along the BPT diagram.

Finally, in the derivation of the dust attenuation we assumed the a
fixed value for the intrinsic Balmer line ratios:
H$\alpha$/H$\beta=2.86$. However, it is well known that this line
ratio depends on the ionization conditions within the nebulae, and it
can range between 2.7 and 3.1 for usual \HII
regions \citep[e.g.][]{osterbrock89}. The possible effect of taking into
account this detail in the derivation of the dust attenuation has been
explored by previous studies and in general it has been found the 
differences in the derived dust attenuations is very small
\cite[e.g.][]{sanchez07c}. In general the possible trends of the
derived A$_{\rm V}$ with the electron temperature and electron density
are weaker than the observational errors of the considered line ratio,
or other effects (like assuming an uncorrect dust attenuation law or a
Milky-Way specific dust attenuation). In summary we consider that the
described trends are genuine, and not induced by possible correlations
implicit in the estimations used to derive each parameter.

\begin{figure}
\centering \includegraphics[width=6.5cm,angle=270,clip,trim=0 0 0
  0]{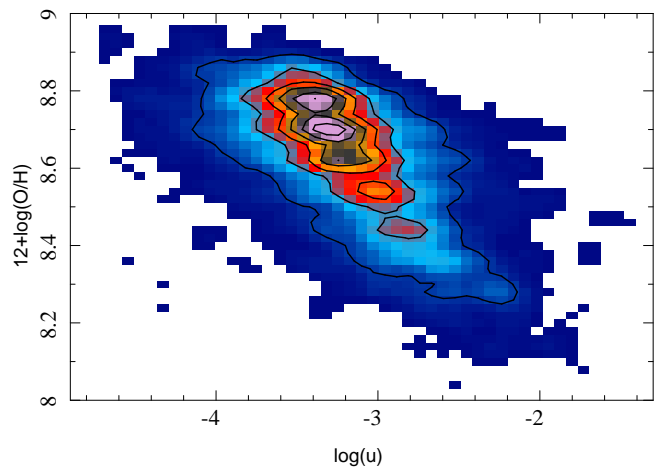}
\caption{\label{fig:para} Distribution of the $\sim$5000 \HII regions described in the text across the oxygen abudance vs. ionization parameter plane. The contours and color show the density distribution, with the outermost contour enclosing 95\% of the regions, and each consecutive one  enclosing 20\% fewer regions.}
\end{figure}

The described trends across the BPT diagram of the parameters that
define the ionization conditions imply that there are correlations
between those parameters that cannot be explained based purely on
photoionization models. Fig. \ref{fig:para} shows the distribution of
oxygen abundance along the ionization parameter. As expected, there is
a clear correlation. Those regions with higher oxygen abundance have
the lowest ionization strengths, and those regions with the lowest
oxygen abundances have the highest ionization strengths. This is a
well known relation, described for the first time by \cite{dopita86},
using strong-line indicators of both parameters. More recently,
\cite{epm14}, explored this relation based on a collection of \HII
regions with abundances derived using the direct method, ie., based on
the electron tempeture, described in \cite{marino13}.  The results of
\cite{epm14} clearly show that this relation is not induced by the
strong-line calibrator used to derive the abundance.

As a sanity check, we have repeated the figure using different
abundance and ionization parameter estimators, and we found the same
result for all of them. This trend cannot be explained based purely on
photoionization models, which assume that both parameters are totally
independent.

We should note here that these representations of the data are dominated
by the average properties and they highlight the dominant
trends. Secondary dependencies are mostly hidden or blurred when
adopting this scheme, and are appreciated only when the dominant trend
is removed.

\subsection{Distribution by galaxy morphology}\label{morph}

In the previous sections we have explored the distribution of the \HII
regions across the BPT diagram, comparing it with the predictions by
photoionization models (Sec. \ref{models}) and the empirical
estimation of a set of parameters that defines the ionization
conditions (Sec. \ref{empirical}). Now we explore the
dependence of the location of \HII regions across the BPT diagram on
the morphology of the host galaxy. The morphological classification
was performed by eye, based on independent analysis by five members of
the CALIFA collaboration, and will be described elsewhere in detail
(Walcher et al. 2014). Different tests indicate that our
morphological classification is fully compatible with pre-existing
ones, and the results agree with the expectations based on other
photometric/morphological parameters, such as the concentration index
\citep{sanchez13} or the Sersic index \citep{sersic68}.  

\begin{figure*}
\centering
\includegraphics[width=7.5cm,angle=270,clip,trim=0 0 0 0]{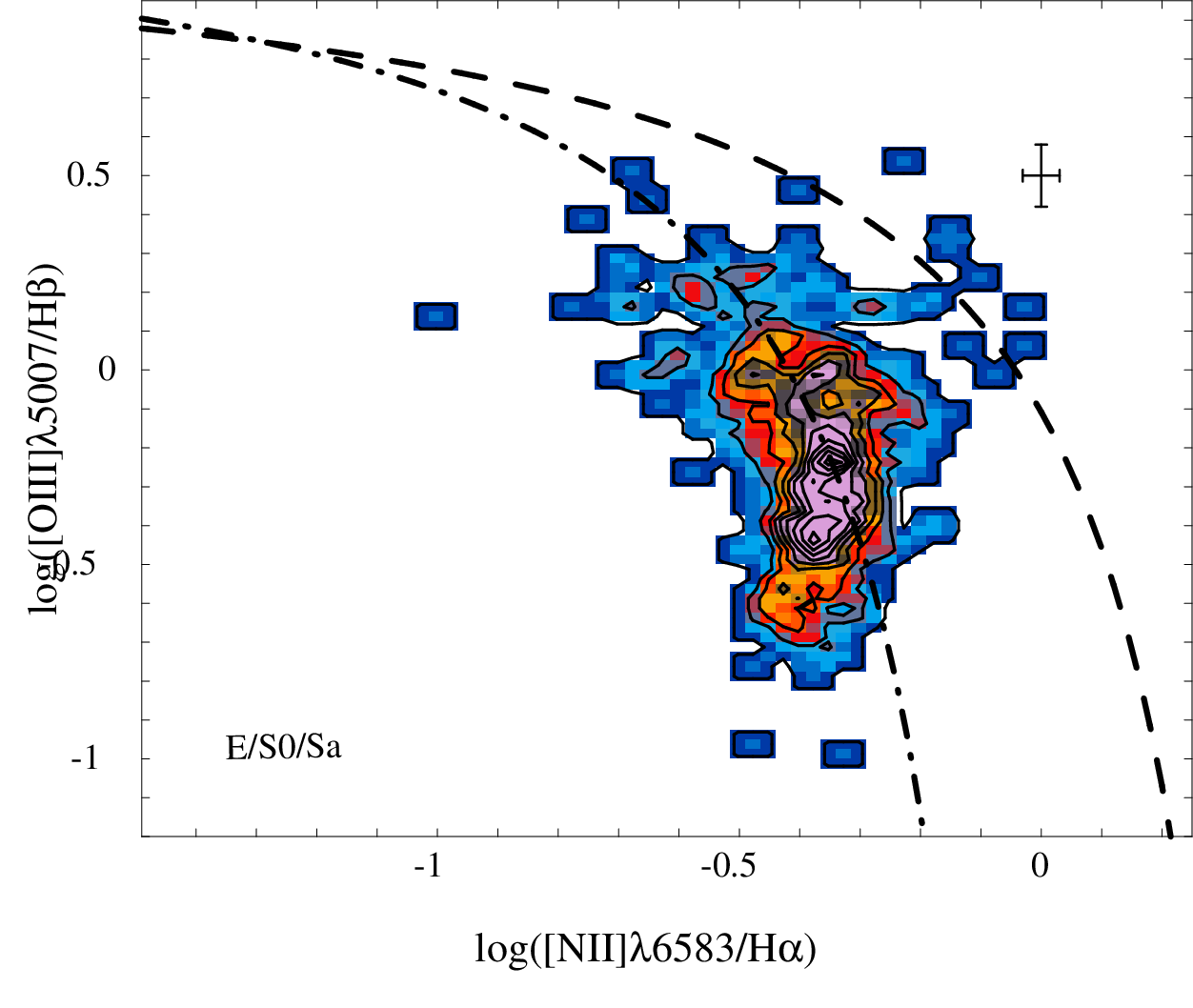}\includegraphics[width=7.5cm,angle=270,clip,trim=0 0 0 0]{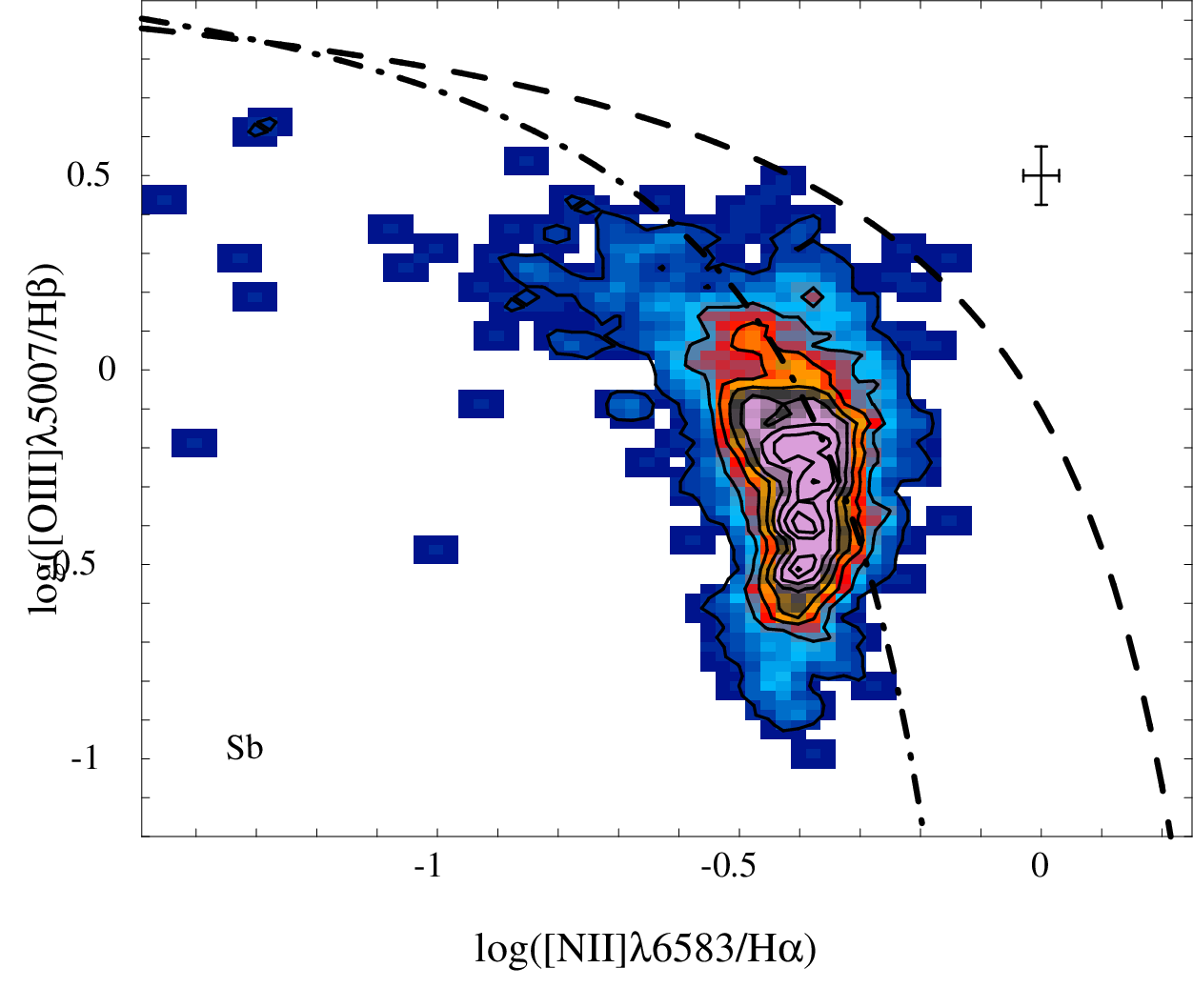}
\includegraphics[width=7.5cm,angle=270,clip,trim=0 0 0 0]{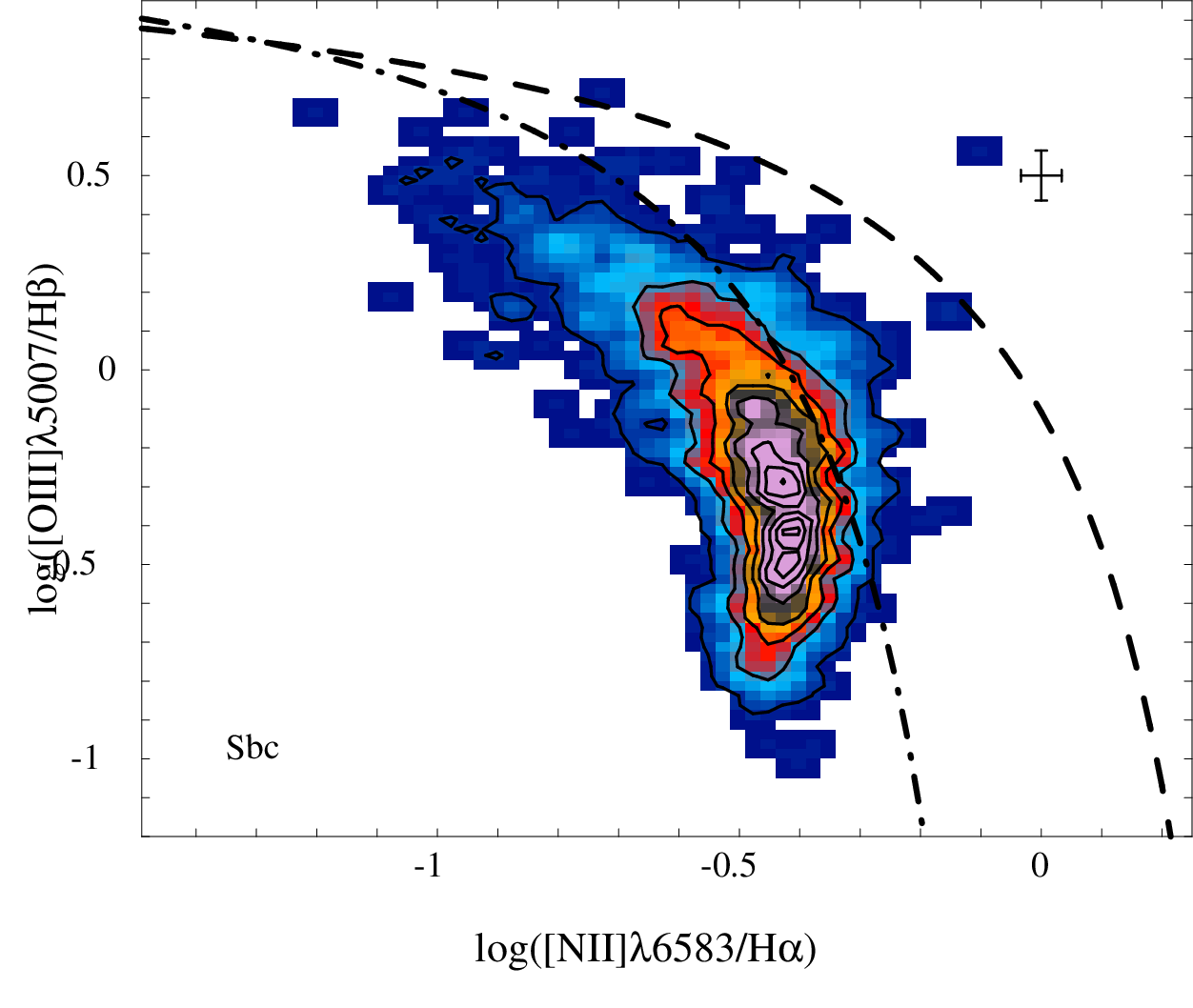}\includegraphics[width=7.5cm,angle=270,clip,trim=0 0 0 0]{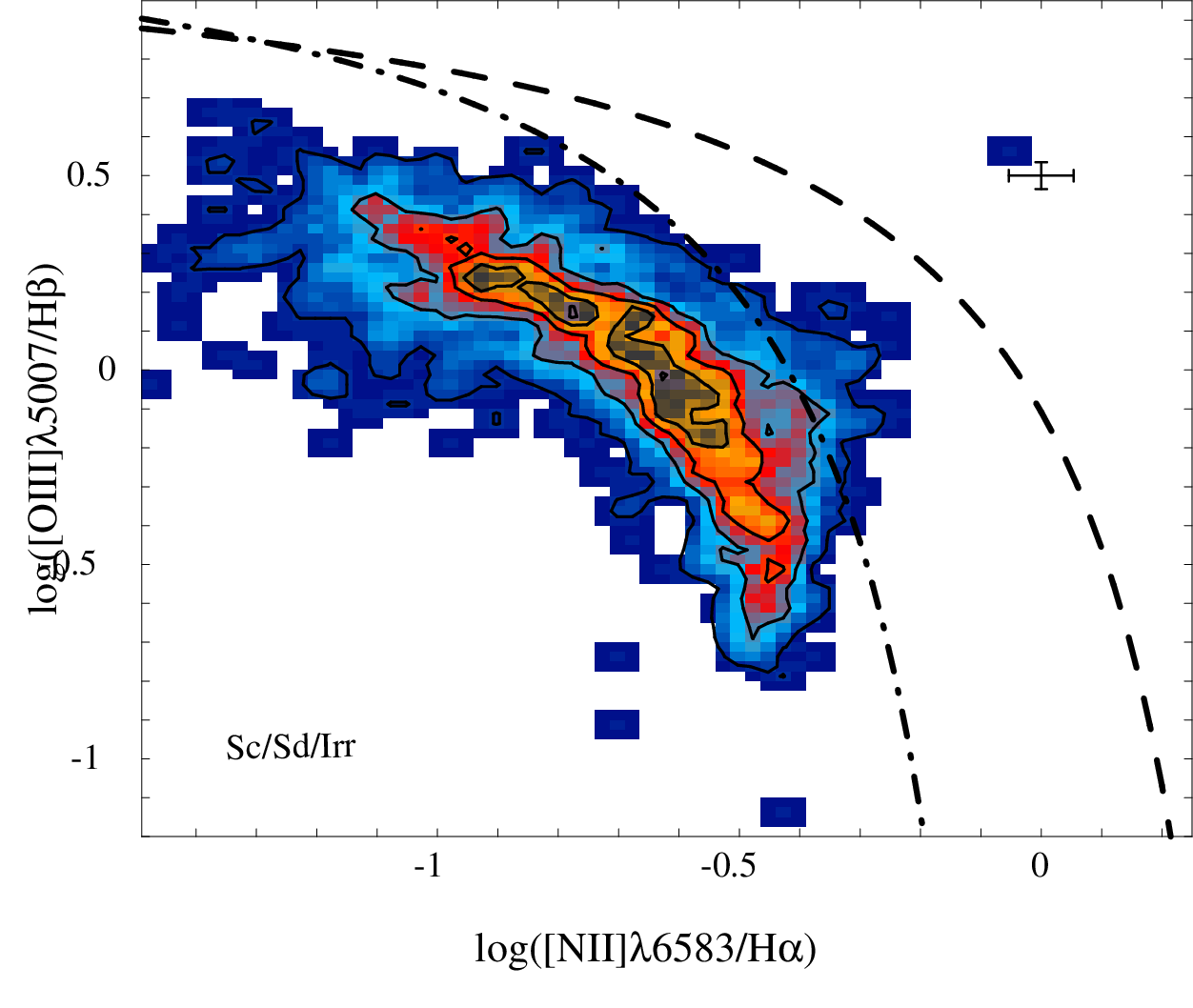}
\caption{\label{fig:BPT_morph} [\ion{O}{iii}]~$\lambda$5007/H$\beta$
  vs. [\ion{N}{ii}]~$\lambda$6583/H$\alpha$ diagnostic diagram for the
  $\sim$5000 \HII regions described in the text. The contours show the
  density distribution of these regions with the diagram plane, with the
  outermost contour enclosing 95\% of the regions, and each consecutive one
  enclosing 20\% fewer regions. In each panel, \HII regions corresponding
  to different galaxies with different morphological types are represented: (i) E/S0/Sa top left, (ii) Sb top right, (iii) Sbc bottom left, and (iv) Sc/Sd/Irr bottom right panel. In all panels, the dot-dashed and dashed lines represent, respectively, the demarcation curves described in Fig.\ref{fig:grid} }
\end{figure*}

Figure \ref{fig:BPT_morph} shows the distribution of \HII regions
across the BPT diagram segregated based on the morphology of their
host galaxies. Each panel shows the \HII regions corresponding to
galaxies from earlier to later morphological types, divided in four
groups with similar number of galaxies each one: (i) the top left hand
panel comprises those regions found in E/S0 and Sa galaxies (we found
\HII regions in only one galaxy classified as Elliptical, however, 2/3
of the classifiers selected them as S0 in Walcher et al. 2014); (ii)
the top right hand panel comprises those regions found in Sb galaxies;
(iii) the bottom left hand panel comprises those regions found in Sbc
galaxies; and finally, (iv) the bottom right hand panel comprises
those regions found in Sc/Sd and irregular galaxies.

There is a clear pattern in the distribution of \HII regions across
the BPT diagram depending on the morphology of the host
galaxies. Early-type galaxies have \HII regions located in the
right-end of the distribution.  For those galaxies the number of
regions within the {\it intermediate} area is much larger ($\sim$50\%
for E/S0 and Sa galaxies). Late-type galaxies have \HII regions
located along the classical sequence of these regions \cite[e.g.][]{osterbrock89}, with a clear
trend towards more regions in the upper-left end of the distribution
the later the type of the  galaxy.

{\it A priori}, we did not expect
a connection between the location in the BPT diagram that is
related to the ionization conditions of the nebulae, and the
morphological type of the host galaxies, since \HII regions are short-lived phenomena. This result indicates that it is not possible to {\it generate}
any kind of \HII region in any morphological type. I.e., there is
a connection between the {\it feasible} ionization conditions in a nebulae
and the morphological type of the host galaxy.

\begin{figure*}
\centering
\includegraphics[width=7.5cm,angle=270,clip,trim=0 0 0 0]{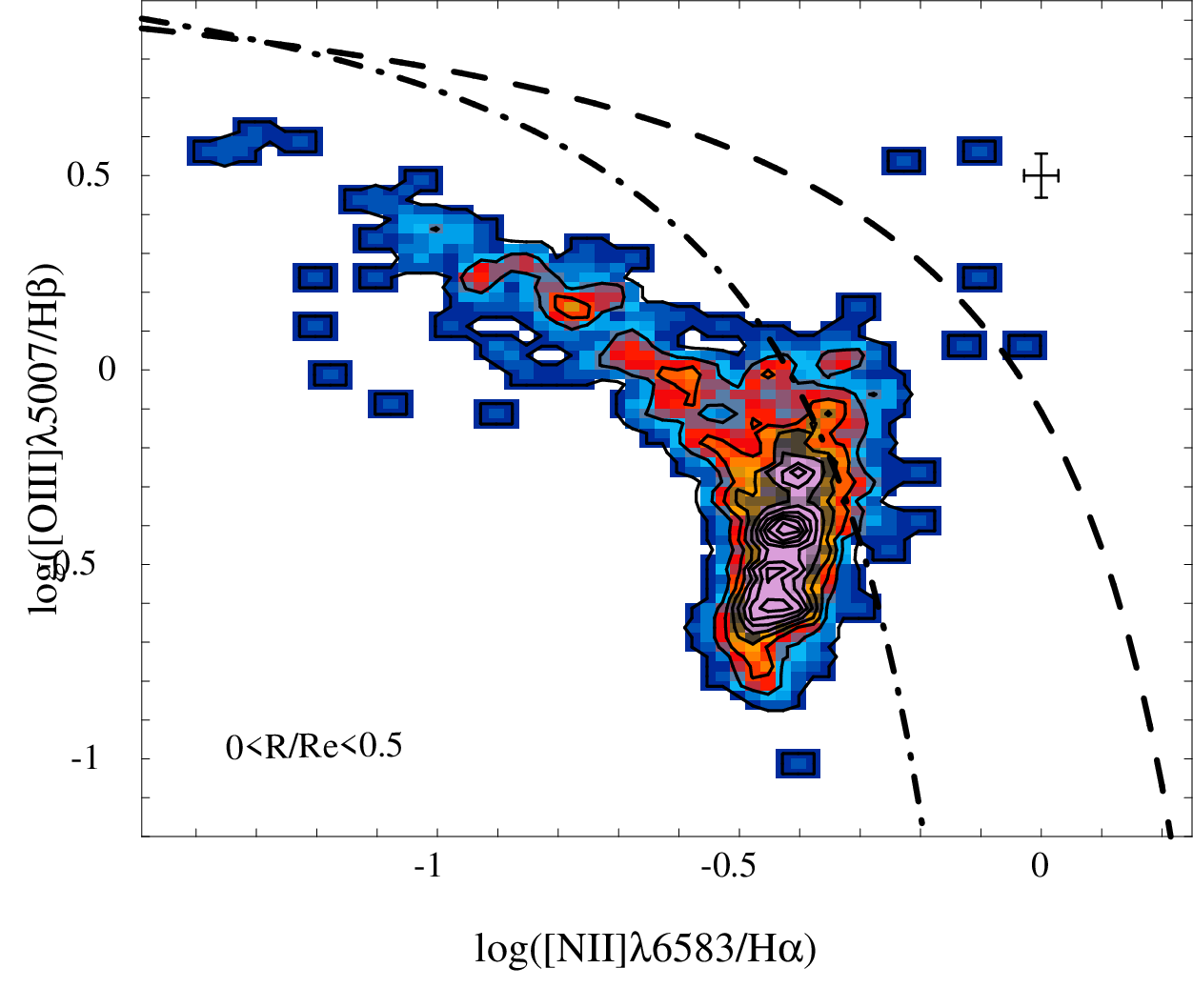}\includegraphics[width=7.5cm,angle=270,clip,trim=0 0 0 0]{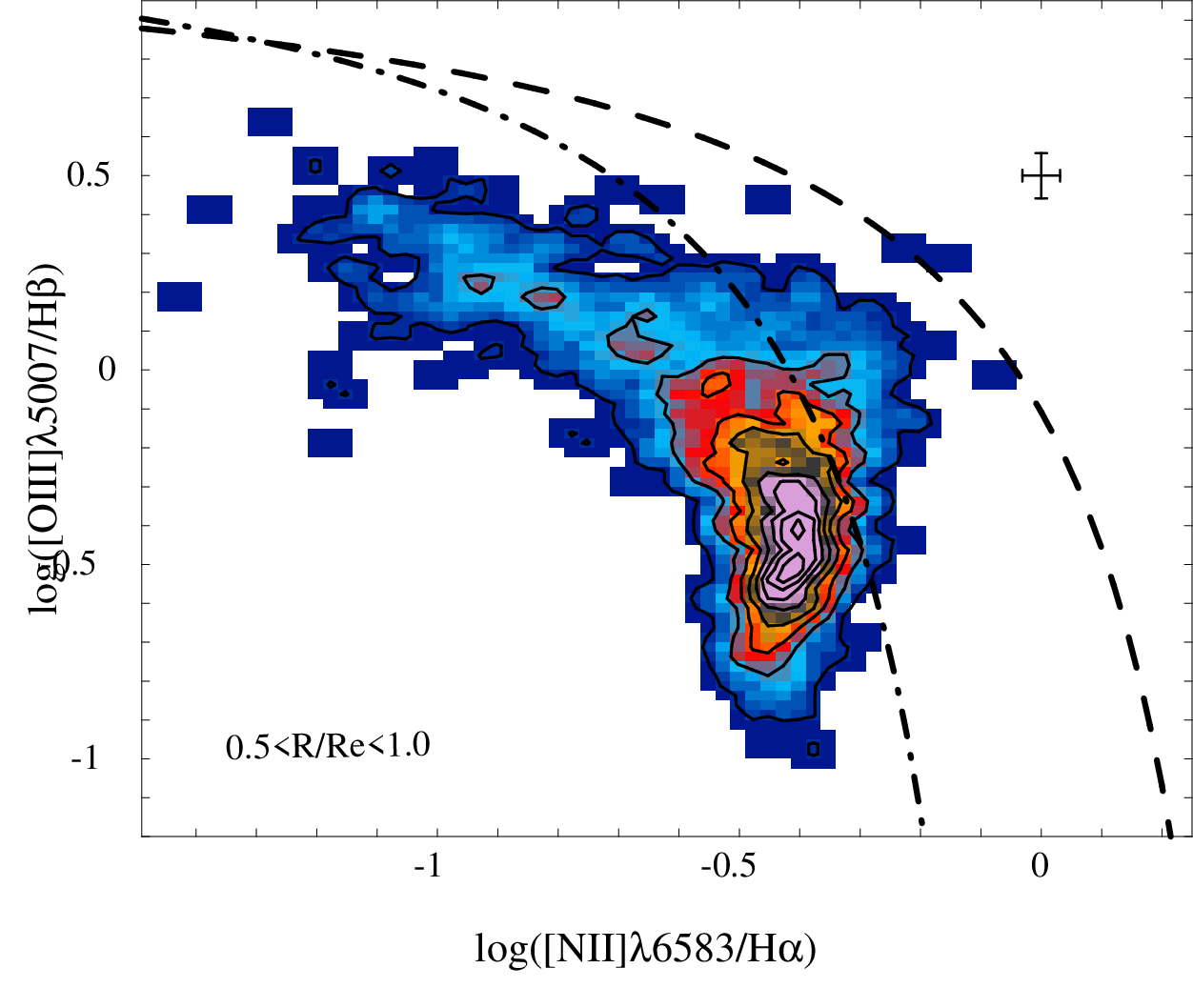}
\includegraphics[width=7.5cm,angle=270,clip,trim=0 0 0 0]{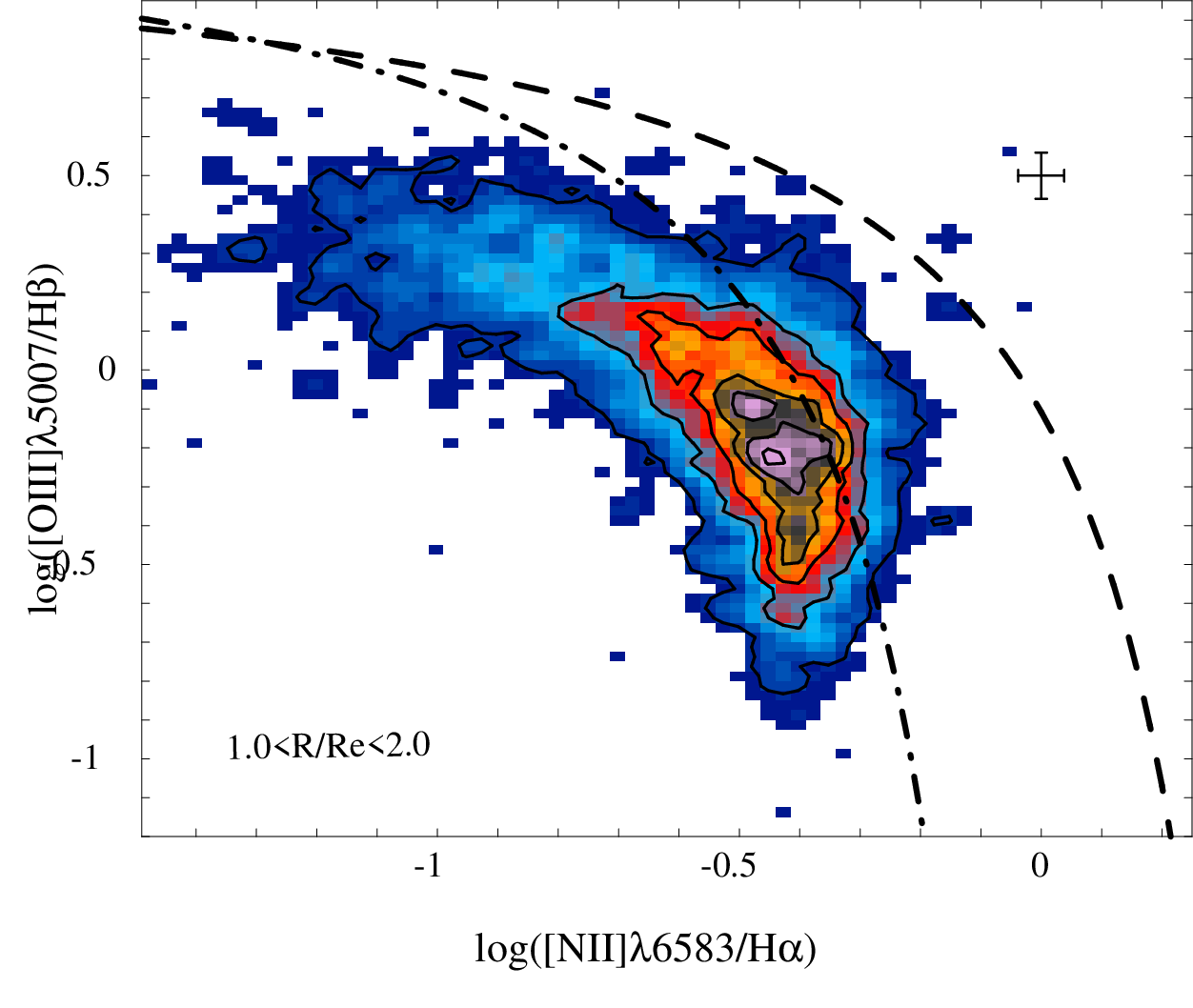}\includegraphics[width=7.5cm,angle=270,clip,trim=0 0 0 0]{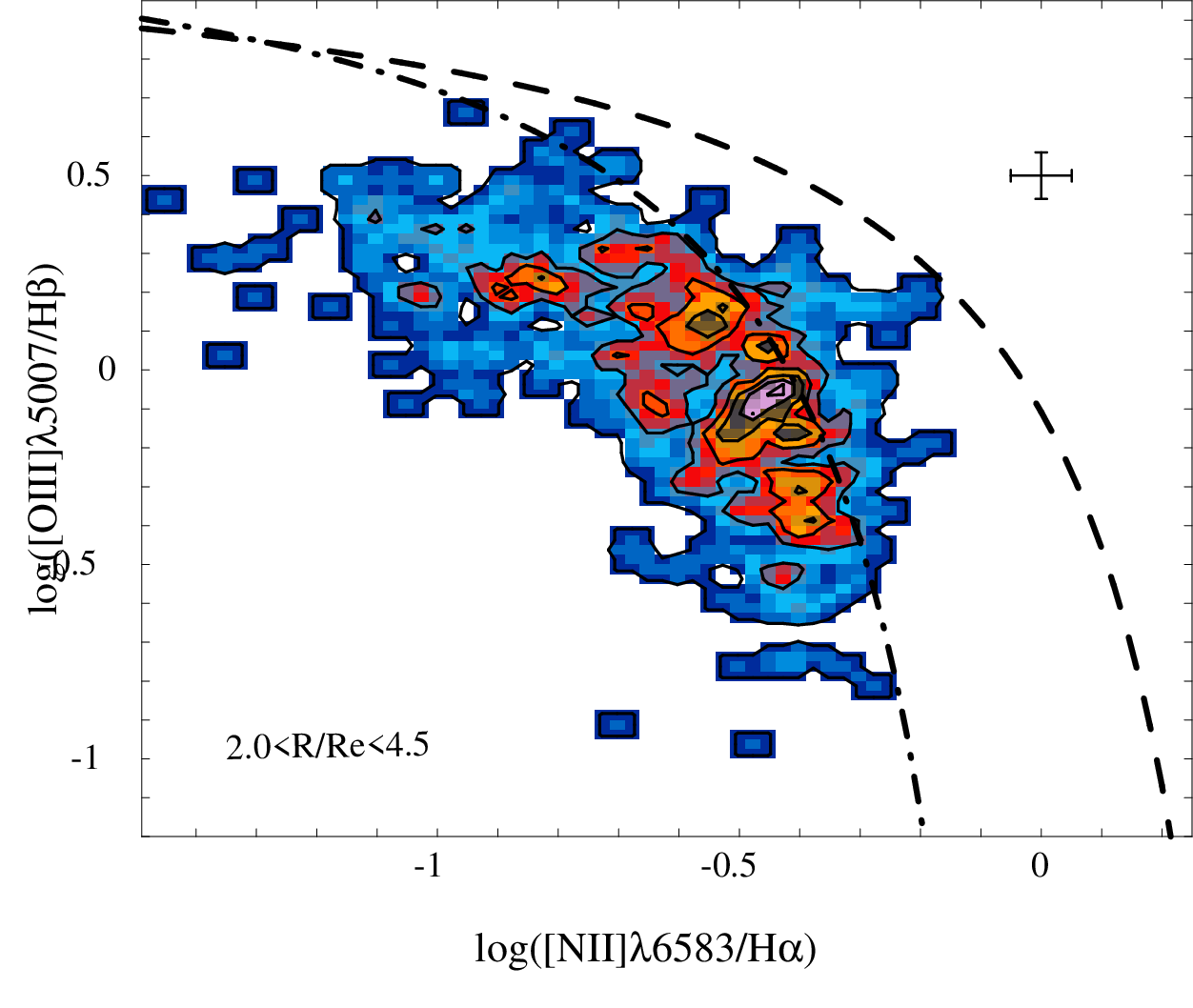}
\caption{\label{fig:BPT_dist} [\ion{O}{iii}]~$\lambda$5007/H$\beta$
  vs. [\ion{N}{ii}]~$\lambda$6583/H$\alpha$ diagnostic diagram for the
  $\sim$5000 \HII regions described in the text. The contours show the
  density distribution of these regions with the diagram plane, with the
  outermost contour enclosing 95\% of the regions, and each consecutive one
  enclosing 20\% fewer regions. In each panel shows the \HII regions corresponding
  to different galactocentric distances, normalized to the effective radius: (i) $R/Re<$0.5 top left, (ii) 0.5$<R/Re<$1.0 top right, (iii) 1.0$<R/Re<$2.0 bottom left, and (iv) $R/Re>$2.0 bottom right panel. In all panels, the dot-dashed and dashed-lines represent, respectively, the demarcation curves described in Fig.\ref{fig:grid}} 
\end{figure*}

\subsection{Distribution by galactocentric distance}\label{dist}

In the previous section we explored the distribution of the \HII
regions across the BPT diagram based on the morphological type of the
host galaxies. We found that galaxies of different morphologies show
different distributions of their star-forming regions across the BPT
diagram. In this section we explore the distribution of the \HII
regions based on their galactocentric distances. 

To do so, we use the galactocentric distances normalized to the
effective radius presented in \cite{sanchez13}. Those distances have been
derived from the catalog of \HII regions provided by {\sc HIIexplorer}
\citep{sanchez12b}, corrected for the estimated inclination of the
galaxy, and normalized to the disk effective radius \citep[see
  Appendix of][]{sanchez13}.

Figure \ref{fig:BPT_dist} shows the distribution of \HII regions
across the BPT diagram segregated based on their galactocentric
distances. Each panel shows the \HII regions corresponding to galaxies
from smaller to larger galactocentric distances, divided in to four
groups with similar number of regions each one: (i) the top left hand
panel comprises those regions at $R/R_e<$0.5; (ii) the top right hand
panel comprises those regions found between 0.5$<R/R_e<$1 ; (iii) the
bottom left hand panel comprises those regions found between
1$<R/R_e$<2; and finally, (iv) the bottom right hand panel comprises
those regions found at distance larger than $R/R_e>$2.

There is a trend similar to the one described for the morphological
type. The \HII regions in the central areas of the galaxies are more
frequently located in the right-end of the distribution, with a larger
fraction of them within the intermediate region. On the other hand,
the \HII regions in the outer regions follows a distribution more
consistent with the classical trend \citep[e.g.][]{osterbrock89}. This
is somewhat expected, since classical \HII regions were observed in the
disk of late-type spiral galaxies \citep[e.g.][]{diaz89}. { 
However, as indicated before, \cite{kennicutt89} and \cite{ho97}
recognized that \ion{H}{ii} regions in the center of galaxies
distinguish themselves spectroscopically from those in the disk by
their stronger low-ionization forbidden emission lines. Initially
it was thought that this may be due to contamination by an extra
source of ionization, such as diffuse emission or the presence of an
AGN. However, \cite{sanchez14} found that these regions are located
across the entire optical extension of the galaxies, and most probably
they are the consequence of a nitrogen enhancement
due to a natural aging process of \ion{H}{ii} regions and the surrounding ISM.}

\begin{figure*}
\centering
\includegraphics[width=7.5cm,angle=270,clip,trim=0 0 0 0]{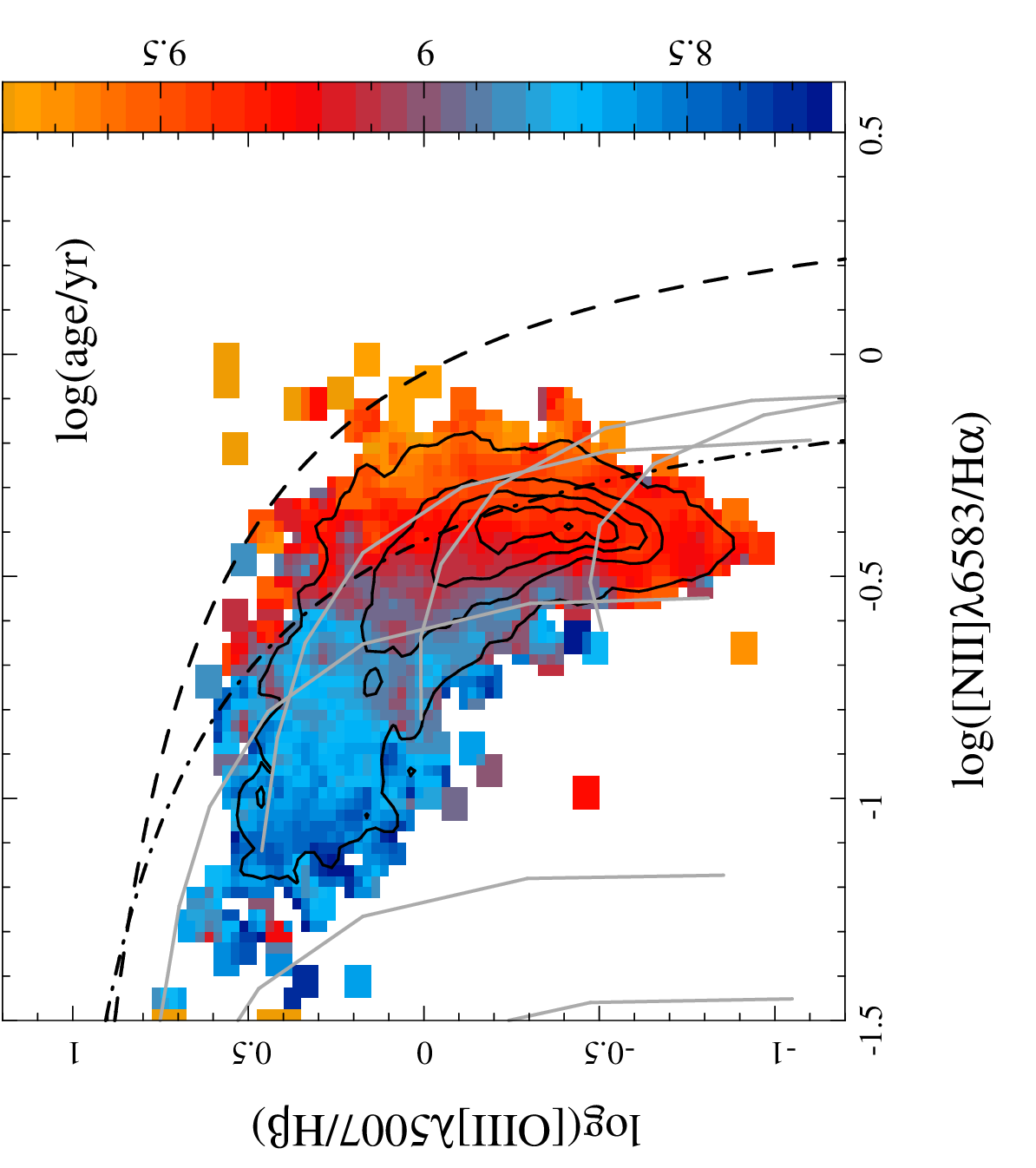}\includegraphics[width=7.5cm,angle=270,clip,trim=0 0 0 0]{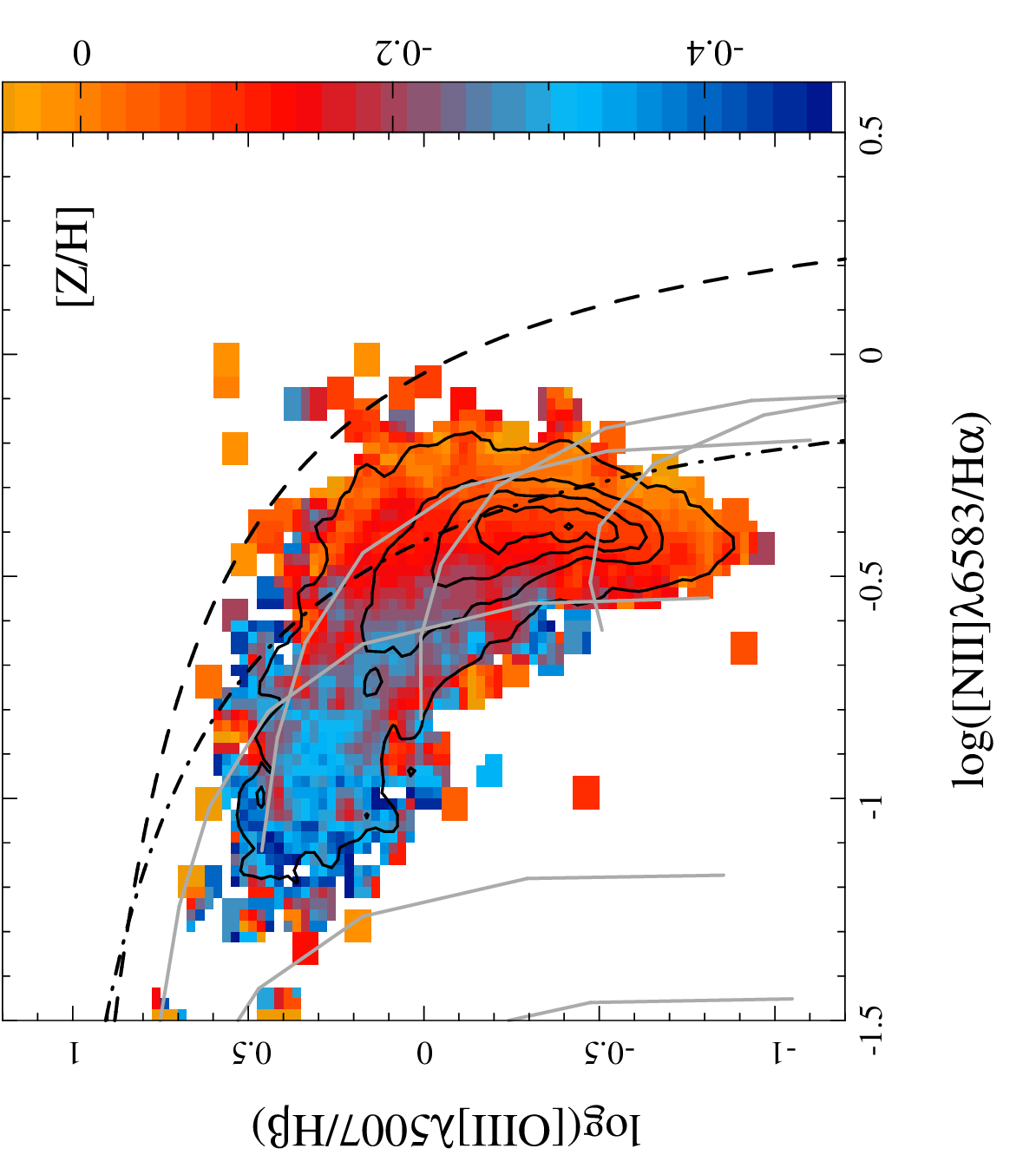}
\caption{\label{fig:BPT_AM} [\ion{O}{iii}]~$\lambda$5007/H$\beta$ vs. [\ion{N}{ii}]~$\lambda$6583/H$\alpha$ diagnostic diagram for the $\sim$5000 \HII regions included in our sample. The contours show the  density distribution of these regions within the diagram plane, with the  outermost contour enclosing 95\% of the regions, and each consecutive one  enclosing 20\% fewer regions.  In each panel the color indicates the average value at the corresponding location in the diagram of the luminosity weighted age ({\it left panel}) and  metallicity  ({\it right panel}) of the underlying stellar population, respectively. The grey solid-lines represent the expected line ratios derived from the photoionization models shown in Fig. \ref{fig:grid}, where each line corresponds to a different stellar metallicity (oragen-to-red solid lines in Fig. \ref{fig:grid}). In both panels, the dot-dashed and dashed line represent, respectively, the demarcation curves described in Fig.\ref{fig:grid}}
\end{figure*}

\subsection{Distribution by the properties of the underlying stellar population}\label{age_met}

In the two previous sections we explored the distribution of the \HII
regions across the BPT diagram based on the morphology of the host
galaxy and the galactocentric distance of the considered region. We
found that regions located in the central areas of galaxies, and in
earlier type galaxies are located in right/lower-right side of the
distribution, while those regions located at larger galactocentric
distances and in latter type galaxies are more frequently located
towards the upper-left side of the distribution. Since the stellar
population of the center of disk galaxies present similarities with
that of early type galaxies \citep[e.g.][]{rosa14}, the described
trends may indicate a connection between the location with the BPT
diagram and the properties of the underlying stellar population.

We derive the luminosity weighted ages and metallicities of the
underlying stellar population for each \HII region, adopting the
results from the analysis described in \cite{sanchez13} and \cite{sanchez14},
summarized before.

The left hand panel of Figure \ref{fig:BPT_AM} shows the distribution of \HII
regions across the BPT diagram with a color code representing the age
of the underlying stellar population. Those regions with older
underlying stellar population are located in the right/lower-right end
of the distribution, while those with younger underlying stellar
populations are located in the upper-left side one. We should note
here that we are not talking about the age of the ionizing population,
the derivation of which would require a more detailed analysis
\citep[e.g.][]{mira14}, but the complete  stellar population 
of each \HII region.

The right hand panel of Fig. \ref{fig:BPT_AM} shows the distribution of \HII
regions across the BPT diagram with a color code representing the
metallicity of the underlying stellar population, in units of [Z/H] (=log(Z/Z$_{\odot}$), where Z$_{\odot}=$0.02). Those regions with metal rich underlying
stellar populations are located in the right/lower-right end of the
distribution, while the metal poor ones are in the upper-left side
one. Despite the larger scatter this trend is very similar to the one
presented in Fig. \ref{fig:BPT_prop} for the oxygen abundance,
indicating a correspondence between the oxygen abundance and the
stellar metallicity of the underlying population. This correspondence
has been recently described by \cite{rosa14b}
and \cite{patri14}, in two independent analyses
of the same data.

We should remind again that these representations of the data are
dominated by the average properties and they highlight the dominant
trends. For example, the left-to-right gradient of the age appreciated
in the left hand panel of Figure \ref{fig:BPT_AM} should not be
interpreted as a lack of dependence of [\ion{O}{iii}] with this
parameter.  The correct interpretation is that the dependence of
[\ion{N}{ii}] with the age is much stronger, as we will see in the
next section.

\subsection{Correlations with the properties of the underlying stellar population}\label{corr}

The trends described in the previous section across the BPT diagram
with the properties of the underlying stellar population point towards
some kind of relationship between them and the line ratios shown in the
diagram. We explored these relations in the current section.

The left hand panel of Fig. \ref{fig:age} shows the distribution of the
[\ion{N}{ii}]/H$\alpha$ line ratio along the luminosity weighted age
of the underlying stellar population for the \HII regions shown in the
previous figures, with the stellar metallicity represented with a
color code.  There is a clear correlation between both parameters,
with a correlation coefficient of $r=0.58$. This correlation depends
on the abundance, being stronger for the higher metallicities
($r=0.70$ for the super-solar metallicities). In general, the older
and more metal rich  the underlying stellar population, the higher
 the value of the [\ion{N}{ii}]/H$\alpha$ line ratio. 

In order to illustrate more clearly the described correlation we have
split the data into bins of increasing metallicity, with a width of
0.1 dex in [Z/H] for each one, and derived the mean value of the line
ratio for different ages, separated by 0.25 dex in the logarithm of
the age.  The result is shown in Fig. \ref{fig:age}, where the
distribution of the mean line ratios (coloured circles) as a function
of the ages for each stellar metallicity are presented. The described
depedence of the line ratio on both parameters is clearly seen.

Thus, to first order the
[\ion{N}{ii}]/H$\alpha$ line ratio could be described as a linear
combination of both the luminosity weighted ages and metallicities of
the underlying stellar population, for which the best fitting linear regression is

\begin{eqnarray}
 {\rm log (}\frac{[\ion{N}{ii}]}{{\rm H}\alpha}{\rm )} &=& 0.24~{\rm log(age/yr)} +0.34~{\rm [Z/H]} -2.67
\end{eqnarray}

After correcting for this dependency, the standard deviation of the line ratio is reduced by  $\sim$70\%. 
Indeed, the
dispersion around the best fitted surface of $\sim$0.14 dex is similar
to the mean error of the considered line ratio $\sim$0.13 dex,
compared with the initial dispersion ($\sim$0.20 dex). This indicates
that the underlying stellar population dominates the definition of the
[\ion{N}{ii}]/H$\alpha$ line ratio, and any other physical conditions
of the nebulae seem to have little effect on it.

The right hand panel of Figure \ref{fig:age} shows the distribution of the
[\ion{O}{iii}]/H$\beta$ line ratio along the luminosity weighted age
of the underlying stellar population in our \HII region sample, with the stellar metallicity represented with a
color code.  The correlation between both parameters is weaker than
the one found for [\ion{N}{ii}]/H$\alpha$, with a correlation
coefficient of $r=0.42$. As in the previous case, there is a
dependency of this correlation with the abundance, being stronger for
the higher metallicities ($r=0.60$ for the super-solar
metallicities). The double dependence is more evident when it is
applied using the binning procedure for metallicity ranges and ages described
before for [\ion{N}{ii}]/H$\alpha$ (solid circles and lines in
Fig. \ref{fig:age}).

 In general, the younger and more metal poor  the underlying stellar
 population, the higher  the value of the [\ion{O}{iii}]/H$\beta$
 line ratio (contrary to the expected for the ionizing stellar
 population). The best fitting linear regression between this line ratio and
 the properties of the underlying stellar population is described by
 the formula

\begin{eqnarray}
 {\rm log (}\frac{[\ion{O}{iii}]}{{\rm H}\beta}{\rm )} &=& -0.25~{\rm log(age/yr)} -0.41~{\rm [Z/H]} +2.11
\end{eqnarray}

After correcting for this dependency the standard deviation of the
distribution of [\ion{O}{iii}]/H$\beta$ decreases only by $\sim$55\%.
For this line ratio the dispersion around the best fitted surface
($\sim$0.25 dex) is considerably larger than the mean error of the
considered line ratio ($\sim$0.13 dex).  In contrast with
[\ion{N}{ii}]/H$\alpha$, it seems that the value of
[\ion{O}{iii}]/H$\beta$ does not depend only on the underlying stellar
population. Rather, it seems to be sensitive to additional physical conditions in
the nebulae, such as the age of the ionizing population, the electron
density, the mass of the ionizing cluster, and the geometry of the
ionized gas \citep[e.g.][]{vargas95a}.

We repeated all the analysis using the luminosity weighted ages and
metallicities derived using STARLIGHT, as described in \cite{rosa14}
for the galaxies in common ($\sim$60\% of the sample). These values
are derived from a more detailed analysis of the underlying stellar
population, using a larger and more suitable SSP library. We derive
very similar results. The trend with the age of the underlying stellar
population is exactly the same, while the trend with metallicity seems
to be slighly weaker. However, this later result seems to be the
effect of the smaller statiscal number of points rather than a real
difference.

\section{Discussion}\label{dis}

So far explored the distribution of \HII regions
across the BPT diagram based on (i) the expected distributions derived from
photoionization models, (ii) 
an estimated set of parameters defining the ionization conditions, (iii) the morphology of the host
galaxies, (iv) the galactocentric distances of the \HII regions, and
(v) the main properties of the underlying stellar population. We found
that the location of an \HII region is not uniquely defined by a
single set of parameters for any set of photoionization models.  I.e.,
their location in the BPT diagram could be reproduced by the models
using different ionization conditions, with a clear degenerancy
between the ionization parameter, $\log(u)$ and the oxygen
abundance. On the other hand, not all the possible locations predicted
by photoionization models are covered by the observed \HII regions
(i.e., there are plausible combinations of physical parameters in
\HII regions which can be empirically excluded on the basis of the
derived BPT diagrams). In summary, from all the feasible hyper-space
of line ratios predicted by photoionization models, real \HII regions
cover a reduced region in this space, which  is not predicted by the
models.

Indeed, this result is not new. Other authors have previously noticed
that from all the line ratios predicted by photoionization models,
observed \HII regions cover only a particular set of them
\citep[e.g.][]{dopita00,kewley01}. However, our larger statistical
sample ensures that we can cover a much wider range of ionization
conditions, and the results still holds. { Thus, there are
  ionization conditions described by photoionization models that are
  not found in real \HII regions. This basically means that the
  observed \HII regions are restricted to a particular set of parameters
among those ones explored by photoionization models.}


Furthermore,  \HII regions do not only cover a particular region in
the hyper-space of possible ionization conditions. We found that there
are clear trends across the BPT diagram depending on the parameters
defining the ionization. \HII regions in the upper-left side of the
diagram are more metal poor, with higher ionization strengths, and
less dust attenuation that those regions in the right/lower-right end
of the diagram. Again, we are not the first to notice this
effect. Classical studies already found similar trends for more
reduced number of regions and/or a fewer number of galaxies
\cite{evans85,dopita86}. More recently, \citep[e.g.][]{bresolin12}
described the same trend with the ionization parameter across the BPT
diagram, for their sample of \HII regions. Indeed, these trends are
behind some of the well-known empirical calibrators between the oxygen
abundance and the strong-line ratios, like O3N2
\citep{allo79,pettini04,stas06,marino13} or N2
\citep{pettini04,marino13}. \cite{dopita86} already found that all the
abundance calibrators based on strong line indicators break the
degenerancy between the ionization parameter and the abundance
existing in photoionization models based on the existence of an
empirical correlation between both parameters. Indeed, \cite{epm14}
have  recently demonstrated that in the absence of this relation it is
not possible to derive a sensitive abundance based on any combination
of strong emission line ratios, even for the largest dataset of \HII
regions with direct abundance estimations. The correlation describred
by \cite{epm14} is very similar to the one presented in
Fig. \ref{fig:para}.

\begin{figure*}
\centering
\includegraphics[width=7.5cm,angle=270,clip,trim=0 0 0 0]{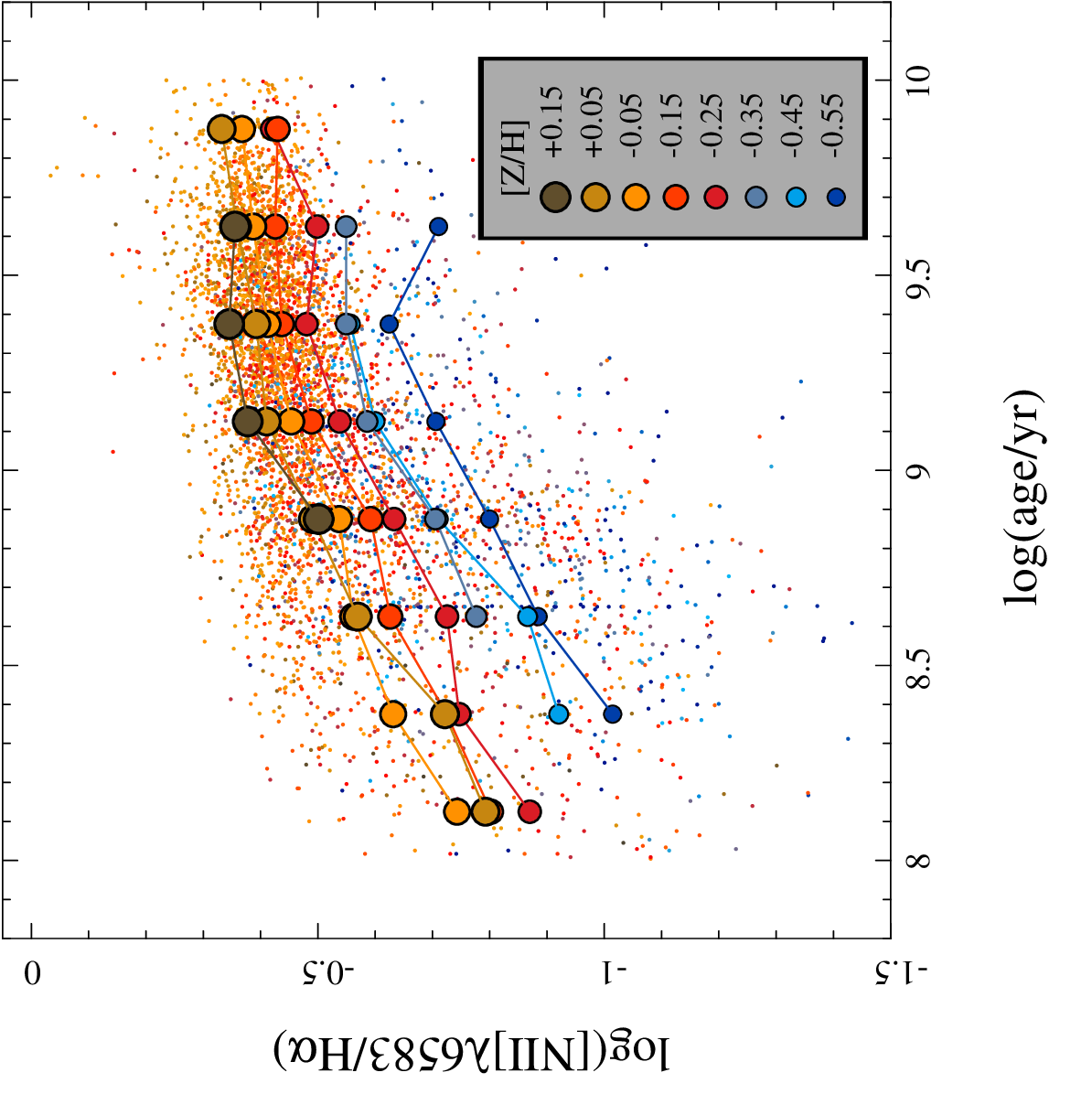}\includegraphics[width=7.5cm,angle=270,clip,trim=0 0 0 0]{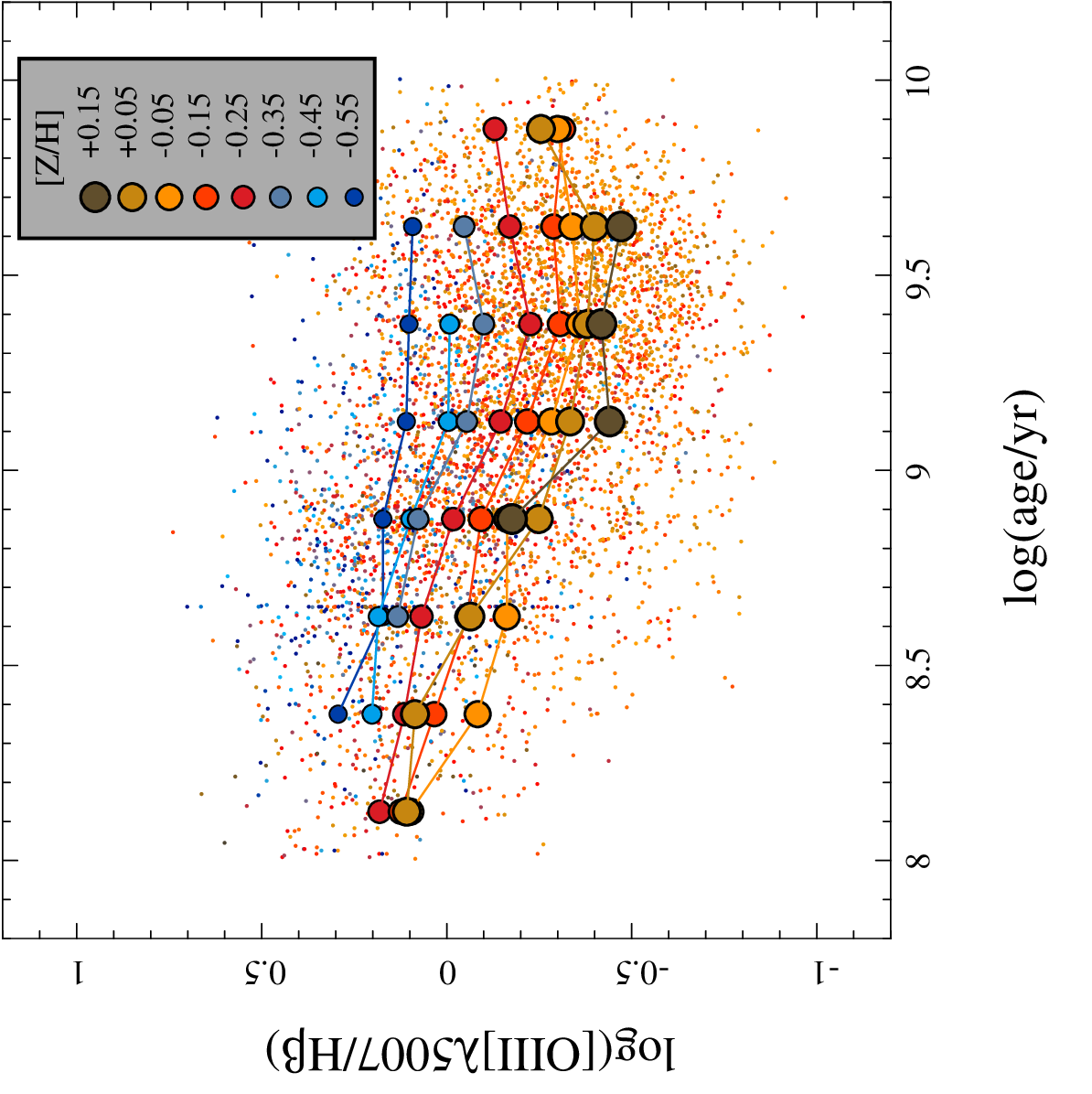}
\caption{\label{fig:age}  {\it Left panel:} Distribution of the [\ion{N}{ii}]~$\lambda$6583/H$\alpha$ line ratio along the luminosity weighted ages of the underlying stellar population for the  $\sim$5000 \HII regions included in our sample (individual dots), with a color code representing the luminosity weighted metallicity of the underlying stellar population. The solid circles and individual lines illustrate how the considered ratio change along the stellar population age for a certain metallicity range. Each circle represents the average line ratio within a bin of 0.25 dex in log(age/yr) for a fixed range of metallicities (within a range of 0.1 dex). Consecutive circles for a particular metallicity are linked with a solid line, and both of them are represented with a fixed color, different for each metallicity. The actual value of the metallicity corresponding to each color is indicated in the enhanced box. {\it Right panel:} Similar distribution for the [\ion{O}{iii}]~$\lambda$5007/H$\beta$ line. Symbols, solid lines and color codes are the same as the one presented in the previous panel.}
\end{figure*}

Several explanations have been proposed for the relation between the
oxygen abundance and the ionization parameter: (i) a correlation
between the IMF and the metallicity, { and/or the ionization
  parameter \citep{dopita06b}} (ii) the correlation is the result of
the dust absorption, and (iii) there are environmental effects which
change the geometry of the \HII regions at higher metallicities. The
first possibility was discussed by \cite{evans85}, without deriving a
conclusive result. The latter two were discussed by \cite{dopita86},
who considered that the dust extinction required to produce the
observed correlation should be larger than the observed one. Finally,
they considered that the third may be possible if the correlation is
related to a radial gradient in the density of the \HII regions, with
inner regions being more dense.  However, as we have show in
Fig. \ref{fig:BPT_prop} and Fig. \ref{fig:BPT_dist} the electron
density does not seem to present an obvious radial distribution and it
does not present a clear correlation with the other two parameters.

A possible alternative explanation to this relation could be that for
a given age, the ionizing stellar population becomes redder for higher
metallicities. First, as the metallicity of the stars increases, so
does the envelope opacity, causing a drop in their effective
temperatures and a shift of flux towards redder wavelengths
\citep[e.g.][]{pagel81}. Second, the ultraviolet emission is strongly
influenced by metal line blanketing in the stellar atmospheres
\citep[e.g.][]{lanz03}. Its effect is not as straight-forward to
understand as the one of the opacity. \cite{moki04} found that the
total number of Lyman continuum photons emitted is almost independent of line
blanketing effects and metallicity for a given effective temperature.
This is because the flux that is blocked by the forest of metal lines
at wavelengths shorter than 600 Angstrom is redistributed mainly
within the Lyman continuum. However, they also found that blanketing
produces a significant decrease in the effective temperature, as
previously reported by \cite{martins02}. Thus the net effect is a
decrease of the Lyman photons with the metallicity
\citep[e.g.][]{lanz03}.

Indeed, evolutionary models for giant \HII
regions describe a decrease of the ionization parameter with the
metallicity for a given age and mass of the ionizing cluster
qualitatively similar to the one described here
\citep{vargas95a,vargas95b}. The detailed dependency between both
parameters is different for each age and mass of the ionizing cluster,
as can be derived from the trends between emission lines shown in
\cite{vargas10}, and therefore a considerable scatter is expected in
the described relation (as observed in Fig. \ref{fig:para}).

{ Finally, the correlation may be induced by a totally different
  mechanism. \citep{dopita06a} proposed a scenario in which the
  expansion and internal pressure of the ionized cloud is driven by
  the net input of mechanical energy from the central cluster (due to
  winds of supernove events). Those models explain naturally the
  anticorrelation of ionization parameter with the oxygen abundance:
  (i) The stellar wind has a higher opacity for more metal rich
  clouds, and therefore absorbs a greater fraction of the ionizing
  photons, reducing $\log(u)$ in the surrounding \HII region; (ii) the
  atmosphere scatters the photons emitted from the photosphere more
  efficiently when the atmospheric abundances are higher, leading to a
  greater conversion efficiency from luminous energy flux to
  mechanical energy flux in the stellar wind base region. This also
  leads to a diminution of $\log(u)$ in the surrounding \HII region.}

The distribution shown in Fig. \ref{fig:BPT_morph} is connected to the
one shown in Fig. \ref{fig:BPT_prop}, upper left hand panel, via the
\mz\ relation \citep{tremonti04,sanchez13}. Early-type galaxies are,
in general, more massive, and therefore more metal rich than late-type
ones. Therefore their \HII regions are located towards the
right/lower-right end of the distribution. On the other hand, the \HII
regions of latter-type galaxies, more metal poor, are located towards
the left/upper-left side. The distribution shown in
Fig. \ref{fig:BPT_dist} can be explained based on the well known
inverse galactocentric abundance gradient, that seems to be present in
all galaxy types \citep[e.g.][ and references
  therein]{sanchez12b,sanchez14}. Indeed, both distributions can be
also explained based on \Sz\ relation \citep{rosales12,sanchez13},
since the central regions of galaxies present higher stellar mass
densities and therefore larger oxygen abundances. Thus, the location
of \HII regions within the BPT diagram changes with the morphology and
the galactocentric distance due to the described trend between this location
and the oxygen abundance.

The trend found with the dust attenuation was not a consequence of the
dependence of the line ratios with dust attenuation, and it was not
predicted by the adopted photoionization models.  By construction the
location within the BPT diagram should be indepedent of this parameter
(Fig. \ref{fig:BPT_prop}, bottom right hand panel). Based on the
previous results this trend indicates that there should be an inverse
gradient of the dust attenuation with the galactocentric distance and
a relation with the stellar mass density, similar to the ones
described for the oxygen abundance. To our knowledge these trends have
not been reported. In \cite{sanchez12b} we found only a very weak
inverse galactocentric gradient of the dust attenuation for a reduced
sample of face-on spiral galaxies. However, that sample was dominated
by Sc/Sd galaxies, which may have affected the results. Similar
results for individual objects were presented by \cite{ls04},
\cite{ls06} and \cite{monreal10}.  A possible explanation to this trend could arise from
the known relation between the stellar dust attenuation and the
\ion{H}{i} gas surface density \citep{bohl78}. Since gas and stellar attenuations present a tight relation \citep[e.g.][]{calz94,calz01},
there should a correlation between the dust attenuation in the gas and
the gas mass density, as recently highlighted by \cite{brin14}. Since
the gas surface density increases towards the center of spiral galaxies
\citep[e.g.][]{you91}, it is expected that there will be a
corresponding increase of $A_V,gas$ in the inner regions. On the other
hand, this trend could be a consequence of the relation between metals
and the dust grains, and therefore, with dust attenuation. 
{ Due to this relation the dust-to-gas ratio increases with metallicity, as nicely summarized by \cite{brinch13}. Indeed, for the Milky Way is $\sim$3 times larger than for the Large Magellanic Cloud and 
$\sim$1.5 times larger than for the Small Magellanic Cloud, as direct consequence
of the differences in metallicity between these three galaxies \citep[e.g][]{bohl78,fitz86,rodri97}}. In this
case the galactocentric decrease of the abundance may induce a similar
gradient for the dust attenuation.

Finally, the trend described for the electron density seems to be
explained naturally by photoionization models
(Fig. \ref{fig:BPT_prop}, bottom left hand panel). { Indeed,
  \cite{kewley13} has nicely illustrated the effect recently (e.g.,
  their Fig. 2).} Electron density is a proxy of the ionized gas
density, which is susceptible to modifications due to different
dynamical mechanisms { that modified the pressure}. { In an
  isobaric density distribution, the density is defined in terms of
  the ratio of the mean ISM pressure, $P$, and the mean electron
  temperature, $T_e$, through $n_e = P/T_e k$. For the typical
  electron temperature of ionized gas the density is simply determined
  by the ISM pressure \citep{kewley13}}. Some authors have described a
radial trend of the electron density in spiral galaxies
\citep[e.g.][]{guti10,beck13}, although it has not been tested in a
statistically significant large sample of galaxies. It is also
expected that the electron density increases in the wave front of
spiral arms \citep[e.g.][]{guti10}, and due to external effects, like
galaxy interaction \citep[e.g.][]{krabbe13}. { Highly active
  star-forming galaxies, like ULIRGs, and high redshift ones, both
  with dense ionized gas are known to be located in the upper-end of
  the classical location for \HII regions
  \citep[e.g.,][]{brinch08,lehnert09,yuan10,dopita14}, in agreement
  with our result.}  {Finally, the electron density could evolve with
  the age of the ionizing cluster, if we assume that \HII region is
  evolving as a mass-loss bubble pressurized by the combined ram
  pressure of the stellar wind and supernova explosions
  \citep{dopita14}, as decribed before.  In this scenario the density
  and the radius of the \HII regions are closely coupled, and both of
  them would depend on the age of the ionizing cluster. Under this
assumptions we expect a variation of the electron density across the optical 
extension of the galaxies. }

All these results indicate that the location within the BPT diagram is
tightly related to properties of the underlying stellar population.
Indeed, Fig. \ref{fig:age}, shows that both line ratios involved in
the BPT diagram present clear correlations with the age and
metallicity of the stellar populations. In particular, the
[\ion{N}{ii}]~$\lambda$6583/H$\alpha$ value is totally defined by both
parameters within the errors. This is somewhat expected since this line
ratio is a proxy of the oxygen abundance
\citep[e.g.][]{marino13}.  Since the gas is polluted by the last
population of stars, and assuming that the radial mixing is not too
strong $-$ which can be assumed based on the results by \cite{sanchez14}
and \cite{patri14} $-$, the oxygen abundance has to be correlated
with the metallicity of the younger population of stars.
Indeed, the gas abundance should be the envelope of the distribution of
metallicities within the underlying stellar population (as recently
found by Gonz\'alez-Delgado et al., 2014b). Therefore, a correlation between the abundance and the metallicity
is expected for two reasons: both the ionizing stars, the youngest ones, should
be  more metal rich, and the ionized gas should be polluted by the
last generation of metal rich stars. 

The correlation with the age is a consequence of the chemical
enrichment in galaxies. We know that disk-dominated galaxies grow
inside out, with the older and more metal rich stellar
populations living in the central regions, where the stellar mass density 
is larger \citep[e.g.][]{perez13,rosa14,patri14}. Therefore,
a correlation between the oxygen abundance of the \HII
regions and the age of the underlying stellar population is expected, and
therefore between this latter parameter and
[\ion{N}{ii}]~$\lambda$6583/H$\alpha$.

Finally, for [\ion{O}{iii}]~$\lambda$5007/H$\beta$, it is also
expected that the correlation is weaker. On one hand this line ratio
is less sensitive to the metallicity. The excitation of \ion{O}{$^{++}$}
requires a much higher energy than that of \ion{N}{$^{+}$}, and
therefore is more sensitive to variations in the geometry of the
nebulae, the gas density, and the properties of the ionizing
population.  Although this  also has an imprint of the local
chemical evolution of the galaxy, due to its metal content, its
evolution affects  this line ratio more than the former one.
Therefore, the location within the BPT diagram is largely
explained by the chemical enrichment and star-formation history, but
not totally.

Throughout this article we have explored the dependence of the location of
\HII regions within the BPT diagram with the properties of the
underlying stellar population. We selected the BPT diagram as the
archetype of the diagnostics diagrams, on one hand, and because the
line ratios involved are less sensitive to dust attenuation. However,
the main results described in this article would be similar if
other diagnostic diagrams are used. In Appendix \ref{ape}, we illustrate
how the distribution of \HII regions across other broadly used 
diagnostic diagrams depends on the properties of the underlying
stellar population. We prefer not to present them in the main text
for  clarity. 

\section{Summary and Conclusions}\label{sum}

In this study we have presented the distribution across the BPT
diagram for the largest catalog of \HII regions with spectroscopic
information currently available \citep{sanchez13,sanchez14}. We
compared their location in this diagram to that  predicted  by
photo-ionization models, finding that (i) \HII regions do not cover
 the entire range of parameters predicted by these models, and (ii) the
same location within the diagram can be reproduced by several
combinations of ionization parameters and photo-ionization models, in
many cases.

We have shown that these parameters (ionization parameter, oxygen
abundance, electron density and dust attenuation) follow well defined
trends across the BPT diagram. Only the trend described for the
electron density can be explained directly by photo-ionization models.
The remaining trends must have a different origin. Indeed, there is
clear correlation between the ionization parameter and the oxygen
abundance that cannot be explained based on photoinization models.
Different explanations for this correlation, extracted from the
literature, have been discussed. We propose an alternative one, in
which the relation is a consequence of the increase of the metal line
blanketing and the envelope opacity of the ionizing stars with the metallicity.

We found that the location of \HII regions within the BPT diagram
depends to first order on the morphology of the host galaxy, the
galactocentric distance, and ultimately, the properties of the
underlying stellar population (age and metallicity). The reason for
these empirical trends are related to the chemical evolution in galaxies.
Oxygen abundance is a tracer of stellar metallicity, and therefore a
by-product of the star-formation history and metal enrichment.  Due to
this, the metal sensitive line ratios are tighly correlated with the
properties of the underlying stellar population (i.e., the products of
the SFH). Finally, due to the correlation between the oxygen abundance
and the ionizing parameter, those line ratios more sensitive to this
latter physical quantity depend on the properties of the underlying
stellar population too.

Despite the fact that \HII regions are short lived phenomena, their
ionization conditions seem to be tightly related to the properties of
the underlying stellar population: i.e., they keep a {\it memory} of
the local star-formation history and metal enrichment at the location
where they are formed.

\begin{acknowledgements}

We thank Prf. Dr. M. Dopita for his comments and suggestions 
during the reviewing process that have help to improve significantly
the interpretation of our results and the ovarall manuscript.

SFS thanks the director of CEFCA, M. Moles, for his sincere support to
this project.

This study makes uses of the data provided by the Calar Alto Legacy
Integral Field Area (CALIFA) survey (http://califa.caha.es/).

CALIFA is the first legacy survey being performed at Calar Alto. The
CALIFA collaboration would like to thank the IAA-CSIC and MPIA-MPG as
major partners of the observatory, and CAHA itself, for the unique
access to telescope time and support in manpower and infrastructures.
The CALIFA collaboration also thanks the CAHA staff for the dedication
to this project.

Based on observations collected at the Centro Astron\'omico Hispano
Alem\'an (CAHA) at Calar Alto, operated jointly by the
Max-Planck-Institut f\"ur Astronomie and the Instituto de Astrof\'\i sica de
Andalucia (CSIC).

We thank the {\it Viabilidad , Dise\~no , Acceso y Mejora } funding program,
ICTS-2009-10, for supporting the initial developement of this project.

S.F.S, M.A.M and L.S.M. thank the {\it Plan Nacional de Investigaci\'on y Desarrollo} funding programs, AYA2012-31935, of the Spanish {\it Ministerio de Econom\'\i a y Competitividad}, for the support given to this project. 

S.F.S thanks the the {\it Ram\'on y Cajal} project RyC-2011-07590 of the spanish {\it Ministerio de Econom\'\i a y Competitividad}, for the support giving to this project.

PP is supported by a FCT Investigator 2013 consolidation grant.
He acknowledges support by the Funda\c{c}\~{a}o para a Ci\^{e}ncia
e a Tecnologia (FCT) under project FCOMP-01-0124-FEDER-029170 (Reference FCT
PTDC/FIS-AST/3214/2012), funded by FCT-MEC (PIDDAC) and FEDER (COMPETE).

R.G.D , E.P., and R.G.B. thank the {\it Plan Nacional de Investigaci\'on y
  Desarrollo} funding program AYA2010-15081.

Support for LG is provided by the Ministry of Economy, Development, and Tourism's Millennium Science Initiative through grant IC12009, awarded to The Millennium Institute of Astrophysics, MAS. LG acknowledges support by CONICYT through FONDECYT grant 3140566.

 R.A. Marino was also funded by the spanish programme of International Campus of Excellence Moncloa (CEI).

A. M.-I. acknowledges support from Agence Nationale de la Recherche through the STILISM project (ANR-12-BS05-0016-02).

\end{acknowledgements}

\bibliography{CALIFAI}

\begin{thebibliography}{117}
\expandafter\ifx\csname natexlab\endcsname\relax\def\natexlab#1{#1}\fi

\bibitem[{{Allen} {et~al.}(2008){Allen}, {Groves}, {Dopita}, {Sutherland}, \&
  {Kewley}}]{allen08}
{Allen}, M.~G., {Groves}, B.~A., {Dopita}, M.~A., {Sutherland}, R.~S., \&
  {Kewley}, L.~J. 2008, \apjs, 178, 20

\bibitem[{{Alloin} {et~al.}(1979){Alloin}, {Collin-Souffrin}, {Joly}, \&
  {Vigroux}}]{allo79}
{Alloin}, D., {Collin-Souffrin}, S., {Joly}, M., \& {Vigroux}, L. 1979, \aap,
  78, 200

\bibitem[{{Anderson}(2014)}]{ander14}
{Anderson}, L.~D. 2014, in American Astronomical Society Meeting Abstracts,
  Vol. 223, American Astronomical Society Meeting Abstracts 223, 312.01

\bibitem[{{Baldwin} {et~al.}(1981){Baldwin}, {Phillips}, \&
  {Terlevich}}]{baldwin81}
{Baldwin}, J.~A., {Phillips}, M.~M., \& {Terlevich}, R. 1981, \pasp, 93, 5

\bibitem[{{Beckman} {et~al.}(2013){Beckman}, {Cedr{\'e}s}, {Giammanco},
  {Bongiovanni}, \& {Cepa}}]{beck13}
{Beckman}, J.~E., {Cedr{\'e}s}, B., {Giammanco}, C., {Bongiovanni}, A., \&
  {Cepa}, J. 2013, in Revista Mexicana de Astronomia y Astrofisica Conference
  Series, Vol.~42, Revista Mexicana de Astronomia y Astrofisica Conference
  Series, 14--15

\bibitem[{{Binette} {et~al.}(2009){Binette}, {Flores-Fajardo}, {Raga},
  {Drissen}, \& {Morisset}}]{binn09}
{Binette}, L., {Flores-Fajardo}, N., {Raga}, A.~C., {Drissen}, L., \&
  {Morisset}, C. 2009, \apj, 695, 552

\bibitem[{{Binette} {et~al.}(1994){Binette}, {Magris}, {Stasi{\'n}ska}, \&
  {Bruzual}}]{binn94}
{Binette}, L., {Magris}, C.~G., {Stasi{\'n}ska}, G., \& {Bruzual}, A.~G. 1994,
  \aap, 292, 13

\bibitem[{{Bohlin} {et~al.}(1978){Bohlin}, {Savage}, \& {Drake}}]{bohl78}
{Bohlin}, R.~C., {Savage}, B.~D., \& {Drake}, J.~F. 1978, \apj, 224, 132

\bibitem[{{Bresolin} {et~al.}(2012){Bresolin}, {Kennicutt}, \&
  {Ryan-Weber}}]{bresolin12}
{Bresolin}, F., {Kennicutt}, R.~C., \& {Ryan-Weber}, E. 2012, ArXiv e-prints

\bibitem[{{Bresolin} {et~al.}(2009){Bresolin}, {Ryan-Weber}, {Kennicutt}, \&
  {Goddard}}]{bresolin09}
{Bresolin}, F., {Ryan-Weber}, E., {Kennicutt}, R.~C., \& {Goddard}, Q. 2009,
  \apj, 695, 580

\bibitem[{{Brinchmann} {et~al.}(2013{\natexlab{a}}){Brinchmann}, {Charlot},
  {Kauffmann}, {Heckman}, {White}, \& {Tremonti}}]{brin14}
{Brinchmann}, J., {Charlot}, S., {Kauffmann}, G., {et~al.} 2013{\natexlab{a}},
  \mnras, 432, 2112

\bibitem[{{Brinchmann} {et~al.}(2013{\natexlab{b}}){Brinchmann}, {Charlot},
  {Kauffmann}, {Heckman}, {White}, \& {Tremonti}}]{brinch13}
{Brinchmann}, J., {Charlot}, S., {Kauffmann}, G., {et~al.} 2013{\natexlab{b}},
  \mnras, 432, 2112

\bibitem[{{Brinchmann} {et~al.}(2008){Brinchmann}, {Pettini}, \&
  {Charlot}}]{brinch08}
{Brinchmann}, J., {Pettini}, M., \& {Charlot}, S. 2008, \mnras, 385, 769

\bibitem[{{Cair{\'o}s} {et~al.}(2012){Cair{\'o}s}, {Caon}, {Garc{\'{\i}}a
  Lorenzo}, {Kelz}, {Roth}, {Papaderos}, \& {Streicher}}]{cairos12}
{Cair{\'o}s}, L.~M., {Caon}, N., {Garc{\'{\i}}a Lorenzo}, B., {et~al.} 2012,
  \aap, 547, A24

\bibitem[{{Calzetti}(2001)}]{calz01}
{Calzetti}, D. 2001, \pasp, 113, 1449

\bibitem[{{Calzetti} {et~al.}(1994){Calzetti}, {Kinney}, \&
  {Storchi-Bergmann}}]{calz94}
{Calzetti}, D., {Kinney}, A.~L., \& {Storchi-Bergmann}, T. 1994, \apj, 429, 582

\bibitem[{{Cardelli} {et~al.}(1989){Cardelli}, {Clayton}, \&
  {Mathis}}]{cardelli89}
{Cardelli}, J.~A., {Clayton}, G.~C., \& {Mathis}, J.~S. 1989, \apj, 345, 245

\bibitem[{{Cid Fernandes} {et~al.}(2014){Cid Fernandes}, {Gonz{\'a}lez
  Delgado}, {Garc{\'{\i}}a Benito}, {P{\'e}rez}, {de Amorim}, {S{\'a}nchez},
  {Husemann}, {Falc{\'o}n Barroso}, {L{\'o}pez-Fern{\'a}ndez},
  {S{\'a}nchez-Bl{\'a}zquez}, {Vale Asari}, {Vazdekis}, {Walcher}, \&
  {Mast}}]{cid-fernandes14}
{Cid Fernandes}, R., {Gonz{\'a}lez Delgado}, R.~M., {Garc{\'{\i}}a Benito}, R.,
  {et~al.} 2014, \aap, 561, A130

\bibitem[{{Cid Fernandes} {et~al.}(2013){Cid Fernandes}, {Perez}, {Garcia
  Benito}, {Gonz\'alez Delgado}, {de Amorim}, {Sanchez}, {Husemann}, {Falcon
  Barroso}, {Sanchez-Blazquez}, {Walcher}, \& {Mast}}]{cid-fernandes13}
{Cid Fernandes}, R., {Perez}, E., {Garcia Benito}, R., {et~al.} 2013, ArXiv
  e-prints

\bibitem[{{Cid Fernandes} {et~al.}(2010){Cid Fernandes}, {Stasi{\'n}ska},
  {Schlickmann}, {Mateus}, {Vale Asari}, {Schoenell}, \&
  {Sodr{\'e}}}]{cid-fernandes10}
{Cid Fernandes}, R., {Stasi{\'n}ska}, G., {Schlickmann}, M.~S., {et~al.} 2010,
  \mnras, 403, 1036

\bibitem[{{Diaz}(1989)}]{diaz89}
{Diaz}, A.~I. 1989, in Evolutionary Phenomena in Galaxies, ed. {J.~E.~Beckman
  \& B.~E.~J.~Pagel}, 377--397

\bibitem[{{D{\'\i}az} {et~al.}(2000){D{\'\i}az}, {Castellanos}, {Terlevich}, \&
  {Luisa Garc{\'{\i}}a-Vargas}}]{diaz00}
{D{\'\i}az}, A.~I., {Castellanos}, M., {Terlevich}, E., \& {Luisa
  Garc{\'{\i}}a-Vargas}, M. 2000, \mnras, 318, 462

\bibitem[{{Dopita} \& {Evans}(1986)}]{dopita86}
{Dopita}, M.~A. \& {Evans}, I.~N. 1986, \apj, 307, 431

\bibitem[{{Dopita} {et~al.}(2006{\natexlab{a}}){Dopita}, {Fischera},
  {Sutherland}, {Kewley}, {Leitherer}, {Tuffs}, {Popescu}, {van Breugel}, \&
  {Groves}}]{dopita06a}
{Dopita}, M.~A., {Fischera}, J., {Sutherland}, R.~S., {et~al.}
  2006{\natexlab{a}}, \apjs, 167, 177

\bibitem[{{Dopita} {et~al.}(2006{\natexlab{b}}){Dopita}, {Fischera},
  {Sutherland}, {Kewley}, {Tuffs}, {Popescu}, {van Breugel}, {Groves}, \&
  {Leitherer}}]{dopita06b}
{Dopita}, M.~A., {Fischera}, J., {Sutherland}, R.~S., {et~al.}
  2006{\natexlab{b}}, \apj, 647, 244

\bibitem[{{Dopita} {et~al.}(2000){Dopita}, {Kewley}, {Heisler}, \&
  {Sutherland}}]{dopita00}
{Dopita}, M.~A., {Kewley}, L.~J., {Heisler}, C.~A., \& {Sutherland}, R.~S.
  2000, \apj, 542, 224

\bibitem[{{Dopita} {et~al.}(2014){Dopita}, {Rich}, {Vogt}, {Kewley}, {Ho},
  {Basurah}, {Ali}, \& {Amer}}]{dopita14}
{Dopita}, M.~A., {Rich}, J., {Vogt}, F.~P.~A., {et~al.} 2014, \apss, 350, 741

\bibitem[{{Dopita} \& {Sutherland}(1995)}]{dopita95}
{Dopita}, M.~A. \& {Sutherland}, R.~S. 1995, \apj, 455, 468

\bibitem[{{Dopita} {et~al.}(2013){Dopita}, {Sutherland}, {Nicholls}, {Kewley},
  \& {Vogt}}]{dopita13}
{Dopita}, M.~A., {Sutherland}, R.~S., {Nicholls}, D.~C., {Kewley}, L.~J., \&
  {Vogt}, F.~P.~A. 2013, \apjs, 208, 10

\bibitem[{{Dors} {et~al.}(2011){Dors}, {Krabbe}, {H{\"a}gele}, \&
  {P{\'e}rez-Montero}}]{dors11}
{Dors}, Jr., O.~L., {Krabbe}, A., {H{\"a}gele}, G.~F., \& {P{\'e}rez-Montero},
  E. 2011, \mnras, 415, 3616

\bibitem[{{Dottori}(1987)}]{dott87}
{Dottori}, H.~A. 1987, \rmxaa, 14, 463

\bibitem[{{Dottori} \& {Copetti}(1989)}]{dott89}
{Dottori}, H.~A. \& {Copetti}, M.~V.~F. 1989, \rmxaa, 18, 115

\bibitem[{{Evans} \& {Dopita}(1985)}]{evans85}
{Evans}, I.~N. \& {Dopita}, M.~A. 1985, \apjs, 58, 125

\bibitem[{{Falc{\'o}n-Barroso} {et~al.}(2011){Falc{\'o}n-Barroso},
  {S{\'a}nchez-Bl{\'a}zquez}, {Vazdekis}, {Ricciardelli}, {Cardiel}, {Cenarro},
  {Gorgas}, \& {Peletier}}]{falc11}
{Falc{\'o}n-Barroso}, J., {S{\'a}nchez-Bl{\'a}zquez}, P., {Vazdekis}, A.,
  {et~al.} 2011, \aap, 532, A95

\bibitem[{{Fitzpatrick}(1986)}]{fitz86}
{Fitzpatrick}, E.~L. 1986, \aj, 92, 1068

\bibitem[{{Flores-Fajardo} {et~al.}(2011){Flores-Fajardo}, {Morisset},
  {Stasi{\'n}ska}, \& {Binette}}]{flores11}
{Flores-Fajardo}, N., {Morisset}, C., {Stasi{\'n}ska}, G., \& {Binette}, L.
  2011, \mnras, 415, 2182

\bibitem[{{Garc{\'{\i}}a-Benito} {et~al.}(2011){Garc{\'{\i}}a-Benito},
  {P{\'e}rez}, {D{\'{\i}}az}, {Ma{\'{\i}}z Apell{\'a}niz}, \&
  {Cervi{\~n}o}}]{rgb2011}
{Garc{\'{\i}}a-Benito}, R., {P{\'e}rez}, E., {D{\'{\i}}az}, {\'A}.~I.,
  {Ma{\'{\i}}z Apell{\'a}niz}, J., \& {Cervi{\~n}o}, M. 2011, \aj, 141, 126

\bibitem[{{Garcia-Vargas} {et~al.}(1995{\natexlab{a}}){Garcia-Vargas},
  {Bressan}, \& {Diaz}}]{vargas95a}
{Garcia-Vargas}, M.~L., {Bressan}, A., \& {Diaz}, A.~I. 1995{\natexlab{a}},
  \aaps, 112, 13

\bibitem[{{Garcia-Vargas} {et~al.}(1995{\natexlab{b}}){Garcia-Vargas},
  {Bressan}, \& {Diaz}}]{vargas95b}
{Garcia-Vargas}, M.~L., {Bressan}, A., \& {Diaz}, A.~I. 1995{\natexlab{b}},
  \aaps, 112, 35

\bibitem[{{Garnett}(2002)}]{Garnett:2002p339}
{Garnett}, D.~R. 2002, \apj, 581, 1019

\bibitem[{{Gonz{\'a}lez Delgado} {et~al.}(2014{\natexlab{a}}){Gonz{\'a}lez
  Delgado}, {Cid Fernandes}, {Garc{\'{\i}}a-Benito}, {P{\'e}rez}, {de Amorim},
  {Cortijo-Ferrero}, {Lacerda}, {L{\'o}pez Fern{\'a}ndez}, {S{\'a}nchez}, {Vale
  Asari}, {Alves}, {Bland-Hawthorn}, {Galbany}, {Gallazzi}, {Husemann},
  {Bekeraite}, {Jungwiert}, {L{\'o}pez-S{\'a}nchez}, {de Lorenzo-C{\'a}ceres},
  {Marino}, {Mast}, {Moll{\'a}}, {del Olmo}, {S{\'a}nchez-Bl{\'a}zquez}, {van
  de Ven}, {V{\'{\i}}lchez}, {Walcher}, {Wisotzki}, {Ziegler}, \&
  {Collaboration920}}]{rosa14b}
{Gonz{\'a}lez Delgado}, R.~M., {Cid Fernandes}, R., {Garc{\'{\i}}a-Benito}, R.,
  {et~al.} 2014{\natexlab{a}}, \apjl, 791, L16

\bibitem[{{Gonz\'alez Delgado} \& {Perez}(1997)}]{rosa97}
{Gonz\'alez Delgado}, R.~M. \& {Perez}, E. 1997, \apjs, 108, 199

\bibitem[{{Gonz{\'a}lez Delgado} {et~al.}(2014{\natexlab{b}}){Gonz{\'a}lez
  Delgado}, {P{\'e}rez}, {Cid Fernandes}, {Garc{\'{\i}}a-Benito}, {de Amorim},
  {S{\'a}nchez}, {Husemann}, {Cortijo-Ferrero}, {L{\'o}pez Fern{\'a}ndez},
  {S{\'a}nchez-Bl{\'a}zquez}, {Bekeraite}, {Walcher}, {Falc{\'o}n-Barroso},
  {Gallazzi}, {van de Ven}, {Alves}, {Bland-Hawthorn}, {Kennicutt}, {Kupko},
  {Lyubenova}, {Mast}, {Moll{\'a}}, {Marino}, {Quirrenbach}, {V{\'{\i}}lchez},
  \& {Wisotzki}}]{rosa14}
{Gonz{\'a}lez Delgado}, R.~M., {P{\'e}rez}, E., {Cid Fernandes}, R., {et~al.}
  2014{\natexlab{b}}, \aap, 562, A47

\bibitem[{{Guti{\'e}rrez} \& {Beckman}(2010)}]{guti10}
{Guti{\'e}rrez}, L. \& {Beckman}, J.~E. 2010, \apjl, 710, L44

\bibitem[{{Ho} {et~al.}(1997){Ho}, {Filippenko}, \& {Sargent}}]{ho97}
{Ho}, L.~C., {Filippenko}, A.~V., \& {Sargent}, W.~L.~W. 1997, \apj, 487, 579

\bibitem[{{Hodge} \& {Kennicutt}(1983)}]{hodg83}
{Hodge}, P.~W. \& {Kennicutt}, Jr., R.~C. 1983, \apj, 267, 563

\bibitem[{{Husemann} {et~al.}(2013){Husemann}, {Jahnke}, {S{\'a}nchez},
  {Barrado}, {Bekerait*error*{\.e}}, {Bomans}, {Castillo-Morales},
  {Catal{\'a}n-Torrecilla}, {Cid Fernandes}, {Falc{\'o}n-Barroso},
  {Garc{\'{\i}}a-Benito}, {Gonz{\'a}lez Delgado}, {Iglesias-P{\'a}ramo},
  {Johnson}, {Kupko}, {L{\'o}pez-Fernandez}, {Lyubenova}, {Marino}, {Mast},
  {Miskolczi}, {Monreal-Ibero}, {Gil de Paz}, {P{\'e}rez}, {P{\'e}rez},
  {Rosales-Ortega}, {Ruiz-Lara}, {Schilling}, {van de Ven}, {Walcher}, {Alves},
  {de Amorim}, {Backsmann}, {Barrera-Ballesteros}, {Bland-Hawthorn}, {Cortijo},
  {Dettmar}, {Demleitner}, {D{\'{\i}}az}, {Enke}, {Florido}, {Flores},
  {Galbany}, {Gallazzi}, {Garc{\'{\i}}a-Lorenzo}, {Gomes}, {Gruel}, {Haines},
  {Holmes}, {Jungwiert}, {Kalinova}, {Kehrig}, {Kennicutt}, {Klar}, {Lehnert},
  {L{\'o}pez-S{\'a}nchez}, {de Lorenzo-C{\'a}ceres}, {M{\'a}rmol-Queralt{\'o}},
  {M{\'a}rquez}, {Mendez-Abreu}, {Moll{\'a}}, {del Olmo}, {Meidt}, {Papaderos},
  {Puschnig}, {Quirrenbach}, {Roth}, {S{\'a}nchez-Bl{\'a}zquez}, {Spekkens},
  {Singh}, {Stanishev}, {Trager}, {Vilchez}, {Wild}, {Wisotzki}, {Zibetti}, \&
  {Ziegler}}]{husemann13}
{Husemann}, B., {Jahnke}, K., {S{\'a}nchez}, S.~F., {et~al.} 2013, \aap, 549,
  A87

\bibitem[{{Kauffmann} {et~al.}(2003){Kauffmann}, {Heckman}, {Tremonti},
  {Brinchmann}, {Charlot}, {White}, {Ridgway}, {Brinkmann}, {Fukugita}, {Hall},
  {Ivezi{\'c}}, {Richards}, \& {Schneider}}]{kauffmann03}
{Kauffmann}, G., {Heckman}, T.~M., {Tremonti}, C., {et~al.} 2003, \mnras, 346,
  1055

\bibitem[{{Kehrig} {et~al.}(2012){Kehrig}, {Monreal-Ibero}, {Papaderos},
  {V{\'{\i}}lchez}, {Gomes}, {Masegosa}, {S{\'a}nchez}, {Lehnert}, {Cid
  Fernandes}, {Bland-Hawthorn}, {Bomans}, {Marquez}, {Mast}, {Aguerri},
  {L{\'o}pez-S{\'a}nchez}, {Marino}, {Pasquali}, {Perez}, {Roth},
  {S{\'a}nchez-Bl{\'a}zquez}, \& {Ziegler}}]{kehrig12}
{Kehrig}, C., {Monreal-Ibero}, A., {Papaderos}, P., {et~al.} 2012, \aap, 540,
  A11

\bibitem[{{Kehrig} {et~al.}(2008){Kehrig}, {V{\'{\i}}lchez}, {S{\'a}nchez},
  {Telles}, {P{\'e}rez-Montero}, \& {Mart{\'{\i}}n-Gord{\'o}n}}]{kehrig08}
{Kehrig}, C., {V{\'{\i}}lchez}, J.~M., {S{\'a}nchez}, S.~F., {et~al.} 2008,
  \aap, 477, 813

\bibitem[{{Kelz} {et~al.}(2006){Kelz}, {Verheijen}, {Roth}, {Bauer}, {Becker},
  {Paschke}, {Popow}, {S{\'a}nchez}, \& {Laux}}]{kelz06}
{Kelz}, A., {Verheijen}, M.~A.~W., {Roth}, M.~M., {et~al.} 2006, \pasp, 118,
  129

\bibitem[{{Kennicutt} {et~al.}(1989){Kennicutt}, {Keel}, \&
  {Blaha}}]{kennicutt89}
{Kennicutt}, Jr., R.~C., {Keel}, W.~C., \& {Blaha}, C.~A. 1989, \aj, 97, 1022

\bibitem[{{Kewley} {et~al.}(2013){Kewley}, {Dopita}, {Leitherer}, {Dav{\'e}},
  {Yuan}, {Allen}, {Groves}, \& {Sutherland}}]{kewley13}
{Kewley}, L.~J., {Dopita}, M.~A., {Leitherer}, C., {et~al.} 2013, \apj, 774,
  100

\bibitem[{{Kewley} {et~al.}(2001){Kewley}, {Dopita}, {Sutherland}, {Heisler},
  \& {Trevena}}]{kewley01}
{Kewley}, L.~J., {Dopita}, M.~A., {Sutherland}, R.~S., {Heisler}, C.~A., \&
  {Trevena}, J. 2001, \apj, 556, 121

\bibitem[{{Kewley} {et~al.}(2006){Kewley}, {Groves}, {Kauffmann}, \&
  {Heckman}}]{kewley06}
{Kewley}, L.~J., {Groves}, B., {Kauffmann}, G., \& {Heckman}, T. 2006, \mnras,
  372, 961

\bibitem[{{Knapen}(1998)}]{knap98}
{Knapen}, J.~H. 1998, \mnras, 297, 255

\bibitem[{{Krabbe} {et~al.}(2014){Krabbe}, {Rosa}, {Dors}, {Pastoriza},
  {Winge}, {H{\"a}gele}, {Cardaci}, \& {Rodrigues}}]{krabbe13}
{Krabbe}, A.~C., {Rosa}, D.~A., {Dors}, O.~L., {et~al.} 2014, \mnras, 437, 1155

\bibitem[{{Lanz} \& {Hubeny}(2003)}]{lanz03}
{Lanz}, T. \& {Hubeny}, I. 2003, \apjs, 146, 417

\bibitem[{{Lehnert} {et~al.}(2009){Lehnert}, {Nesvadba}, {Le Tiran}, {Di
  Matteo}, {van Driel}, {Douglas}, {Chemin}, \& {Bournaud}}]{lehnert09}
{Lehnert}, M.~D., {Nesvadba}, N.~P.~H., {Le Tiran}, L., {et~al.} 2009, \apj,
  699, 1660

\bibitem[{{Lequeux} {et~al.}(1979){Lequeux}, {Peimbert}, {Rayo}, {Serrano}, \&
  {Torres-Peimbert}}]{leque79}
{Lequeux}, J., {Peimbert}, M., {Rayo}, J.~F., {Serrano}, A., \&
  {Torres-Peimbert}, S. 1979, \aap, 80, 155

\bibitem[{{Lopez} {et~al.}(2011){Lopez}, {Krumholz}, {Bolatto}, {Prochaska}, \&
  {Ramirez-Ruiz}}]{lopez2011}
{Lopez}, L.~A., {Krumholz}, M.~R., {Bolatto}, A.~D., {Prochaska}, J.~X., \&
  {Ramirez-Ruiz}, E. 2011, \apj, 731, 91

\bibitem[{{L{\'o}pez-S{\'a}nchez} \& {Esteban}(2009)}]{angel09}
{L{\'o}pez-S{\'a}nchez}, A.~R. \& {Esteban}, C. 2009, \aap, 508, 615

\bibitem[{{L{\'o}pez-S{\'a}nchez} {et~al.}(2006){L{\'o}pez-S{\'a}nchez},
  {Esteban}, \& {Garc{\'{\i}}a-Rojas}}]{ls06}
{L{\'o}pez-S{\'a}nchez}, {\'A}.~R., {Esteban}, C., \& {Garc{\'{\i}}a-Rojas}, J.
  2006, \aap, 449, 997

\bibitem[{{L{\'o}pez-S{\'a}nchez} {et~al.}(2004){L{\'o}pez-S{\'a}nchez},
  {Esteban}, \& {Rodr{\'{\i}}guez}}]{ls04}
{L{\'o}pez-S{\'a}nchez}, {\'A}.~R., {Esteban}, C., \& {Rodr{\'{\i}}guez}, M.
  2004, \aap, 428, 425

\bibitem[{{Marino} {et~al.}(2012){Marino}, {Gil de Paz}, {Castillo-Morales},
  {Mu{\~n}oz-Mateos}, {S{\'a}nchez}, {P{\'e}rez-Gonz{\'a}lez}, {Gallego},
  {Zamorano}, {Alonso-Herrero}, \& {Boissier}}]{mari11}
{Marino}, R.~A., {Gil de Paz}, A., {Castillo-Morales}, A., {et~al.} 2012, \apj,
  754, 61

\bibitem[{{Marino} {et~al.}(2013){Marino}, {Rosales-Ortega}, {S{\'a}nchez},
  {Gil de Paz}, {V{\'{\i}}lchez}, {Miralles-Caballero}, {Kehrig},
  {P{\'e}rez-Montero}, {Stanishev}, {Iglesias-P{\'a}ramo}, {D{\'{\i}}az},
  {Castillo-Morales}, {Kennicutt}, {L{\'o}pez-S{\'a}nchez}, {Galbany},
  {Garc{\'{\i}}a-Benito}, {Mast}, {Mendez-Abreu}, {Monreal-Ibero}, {Husemann},
  {Walcher}, {Garc{\'{\i}}a-Lorenzo}, {Masegosa}, {Del Olmo Orozco},
  {Mour{\~a}o}, {Ziegler}, {Moll{\'a}}, {Papaderos},
  {S{\'a}nchez-Bl{\'a}zquez}, {Gonz{\'a}lez Delgado}, {Falc{\'o}n-Barroso},
  {Roth}, {van de Ven}, \& {Califa Team}}]{marino13}
{Marino}, R.~A., {Rosales-Ortega}, F.~F., {S{\'a}nchez}, S.~F., {et~al.} 2013,
  \aap, 559, A114

\bibitem[{{Martin} \& {Roy}(1994)}]{Martin:1994p1602}
{Martin}, P. \& {Roy}, J.-R. 1994, \apj, 424, 599

\bibitem[{{Mart{\'{\i}}n-Manj{\'o}n} {et~al.}(2010){Mart{\'{\i}}n-Manj{\'o}n},
  {Garc{\'{\i}}a-Vargas}, {Moll{\'a}}, \& {D{\'{\i}}az}}]{vargas10}
{Mart{\'{\i}}n-Manj{\'o}n}, M.~L., {Garc{\'{\i}}a-Vargas}, M.~L., {Moll{\'a}},
  M., \& {D{\'{\i}}az}, A.~I. 2010, \mnras, 403, 2012

\bibitem[{{Martins} {et~al.}(2002){Martins}, {Schaerer}, \&
  {Hillier}}]{martins02}
{Martins}, F., {Schaerer}, D., \& {Hillier}, D.~J. 2002, \aap, 382, 999

\bibitem[{{Mast} {et~al.}(2014){Mast}, {Rosales-Ortega}, {S{\'a}nchez},
  {V{\'{\i}}lchez}, {Iglesias-Paramo}, {Walcher}, {Husemann}, {M{\'a}rquez},
  {Marino}, {Kennicutt}, {Monreal-Ibero}, {Galbany}, {de Lorenzo-C{\'a}ceres},
  {Mendez-Abreu}, {Kehrig}, {del Olmo}, {Rela{\~n}o}, {Wisotzki},
  {M{\'a}rmol-Queralt{\'o}}, {Bekerait{\`e}}, {Papaderos}, {Wild}, {Aguerri},
  {Falc{\'o}n-Barroso}, {Bomans}, {Ziegler}, {Garc{\'{\i}}a-Lorenzo},
  {Bland-Hawthorn}, {L{\'o}pez-S{\'a}nchez}, \& {van de Ven}}]{mast14}
{Mast}, D., {Rosales-Ortega}, F.~F., {S{\'a}nchez}, S.~F., {et~al.} 2014, \aap,
  561, A129

\bibitem[{{McCall} {et~al.}(1985){McCall}, {Rybski}, \& {Shields}}]{McCa85}
{McCall}, M.~L., {Rybski}, P.~M., \& {Shields}, G.~A. 1985, \apjs, 57, 1

\bibitem[{{Miralles-Caballero} {et~al.}(2014){Miralles-Caballero},
  {D{\'{\i}}az}, {Rosales-Ortega}, {P{\'e}rez-Montero}, \&
  {S{\'a}nchez}}]{mira14}
{Miralles-Caballero}, D., {D{\'{\i}}az}, A.~I., {Rosales-Ortega}, F.~F.,
  {P{\'e}rez-Montero}, E., \& {S{\'a}nchez}, S.~F. 2014, \mnras, 440, 2265

\bibitem[{{Mokiem} {et~al.}(2004){Mokiem}, {Mart{\'{\i}}n-Hern{\'a}ndez},
  {Lenorzer}, {de Koter}, \& {Tielens}}]{moki04}
{Mokiem}, M.~R., {Mart{\'{\i}}n-Hern{\'a}ndez}, N.~L., {Lenorzer}, A., {de
  Koter}, A., \& {Tielens}, A.~G.~G.~M. 2004, \aap, 419, 319

\bibitem[{{Monreal-Ibero} {et~al.}(2011){Monreal-Ibero}, {Rela{\~n}o},
  {Kehrig}, {P{\'e}rez-Montero}, {V{\'{\i}}lchez}, {Kelz}, {Roth}, \&
  {Streicher}}]{monreal11}
{Monreal-Ibero}, A., {Rela{\~n}o}, M., {Kehrig}, C., {et~al.} 2011, \mnras,
  413, 2242

\bibitem[{{Monreal-Ibero} {et~al.}(2010){Monreal-Ibero}, {V{\'{\i}}lchez},
  {Walsh}, \& {Mu{\~n}oz-Tu{\~n}{\'o}n}}]{monreal10}
{Monreal-Ibero}, A., {V{\'{\i}}lchez}, J.~M., {Walsh}, J.~R., \&
  {Mu{\~n}oz-Tu{\~n}{\'o}n}, C. 2010, \aap, 517, A27

\bibitem[{{Morisset} \& {Georgiev}(2009)}]{mori09}
{Morisset}, C. \& {Georgiev}, L. 2009, \aap, 507, 1517

\bibitem[{{Moustakas} \& {Kennicutt}(2006)}]{moustakas06}
{Moustakas}, J. \& {Kennicutt}, Jr., R.~C. 2006, \apjs, 164, 81

\bibitem[{{Oey} {et~al.}(2003){Oey}, {Parker}, {Mikles}, \& {Zhang}}]{oey03}
{Oey}, M.~S., {Parker}, J.~S., {Mikles}, V.~J., \& {Zhang}, X. 2003, \aj, 126,
  2317

\bibitem[{{Osterbrock}(1989)}]{osterbrock89}
{Osterbrock}, D.~E. 1989, {Astrophysics of gaseous nebulae and active galactic
  nuclei} (University Science Books)

\bibitem[{{Pagel} \& {Edmunds}(1981)}]{pagel81}
{Pagel}, B.~E.~J. \& {Edmunds}, M.~G. 1981, \araa, 19, 77

\bibitem[{{Papaderos} {et~al.}(2013){Papaderos}, {Gomes}, {Vilchez}, {Kehrig},
  {Lehnert}, {Ziegler}, {Sanchez}, {Husemann}, {Monreal-Ibero},
  {Garcia-Benito}, {Bland-Hawthorn}, {Coritjo}, {de Lorenzo-Caceres}, {del
  Olmo}, {Falcon-Barroso}, {Galbany}, {Iglesias-Paramo}, {Lopez-Sanchez},
  {Marquez}, {Molla}, {Mast}, {van de Ven}, {Wisotzki}, \& {the CALIFA
  collaboration}}]{papa13}
{Papaderos}, P., {Gomes}, J.~M., {Vilchez}, J.~M., {et~al.} 2013, ArXiv
  e-prints

\bibitem[{{P{\'e}rez} {et~al.}(2013){P{\'e}rez}, {Cid Fernandes}, {Gonz{\'a}lez
  Delgado}, {Garc{\'{\i}}a-Benito}, {S{\'a}nchez}, {Husemann}, {Mast},
  {Rod{\'o}n}, {Kupko}, {Backsmann}, {de Amorim}, {van de Ven}, {Walcher},
  {Wisotzki}, {Cortijo-Ferrero}, \& {collaboration6}}]{perez13}
{P{\'e}rez}, E., {Cid Fernandes}, R., {Gonz{\'a}lez Delgado}, R.~M., {et~al.}
  2013, \apjl, 764, L1

\bibitem[{{P{\'e}rez-Montero}(2014)}]{epm14}
{P{\'e}rez-Montero}, E. 2014, ArXiv e-prints

\bibitem[{{Pettini} \& {Pagel}(2004)}]{pettini04}
{Pettini}, M. \& {Pagel}, B.~E.~J. 2004, \mnras, 348, L59

\bibitem[{{Pilyugin} {et~al.}(2012){Pilyugin}, {Grebel}, \& {Mattsson}}]{P12}
{Pilyugin}, L.~S., {Grebel}, E.~K., \& {Mattsson}, L. 2012, \mnras, 424, 2316

\bibitem[{{Pilyugin} {et~al.}(2010){Pilyugin}, {V{\'{\i}}lchez}, \&
  {Thuan}}]{pilyugin10}
{Pilyugin}, L.~S., {V{\'{\i}}lchez}, J.~M., \& {Thuan}, T.~X. 2010, \apj, 720,
  1738

\bibitem[{{Rela{\~n}o} {et~al.}(2010){Rela{\~n}o}, {Monreal-Ibero},
  {V{\'{\i}}lchez}, \& {Kennicutt}}]{rela10}
{Rela{\~n}o}, M., {Monreal-Ibero}, A., {V{\'{\i}}lchez}, J.~M., \& {Kennicutt},
  R.~C. 2010, \mnras, 402, 1635

\bibitem[{{Rodrigues} {et~al.}(1997){Rodrigues}, {Magalh{\~a}es}, {Coyne}, \&
  {Piirola}}]{rodri97}
{Rodrigues}, C.~V., {Magalh{\~a}es}, A.~M., {Coyne}, G.~V., \& {Piirola},
  S.~J.~V. 1997, \apj, 485, 618

\bibitem[{{Rosales-Ortega} {et~al.}(2011){Rosales-Ortega}, {D{\'{\i}}az},
  {Kennicutt}, \& {S{\'a}nchez}}]{rosales11}
{Rosales-Ortega}, F.~F., {D{\'{\i}}az}, A.~I., {Kennicutt}, R.~C., \&
  {S{\'a}nchez}, S.~F. 2011, \mnras, 415, 2439

\bibitem[{{Rosales-Ortega} {et~al.}(2010){Rosales-Ortega}, {Kennicutt},
  {S{\'a}nchez}, {D{\'{\i}}az}, {Pasquali}, {Johnson}, \&
  {Hao}}]{rosales-ortega10}
{Rosales-Ortega}, F.~F., {Kennicutt}, R.~C., {S{\'a}nchez}, S.~F., {et~al.}
  2010, \mnras, 405, 735

\bibitem[{{Rosales-Ortega} {et~al.}(2012){Rosales-Ortega}, {S{\'a}nchez},
  {Iglesias-P{\'a}ramo}, {D{\'{\i}}az}, {V{\'{\i}}lchez}, {Bland-Hawthorn},
  {Husemann}, \& {Mast}}]{rosales12}
{Rosales-Ortega}, F.~F., {S{\'a}nchez}, S.~F., {Iglesias-P{\'a}ramo}, J.,
  {et~al.} 2012, \apjl, 756, L31

\bibitem[{{Roth} {et~al.}(2005){Roth}, {Kelz}, {Fechner}, {Hahn}, {Bauer},
  {Becker}, {B{\"o}hm}, {Christensen}, {Dionies}, {Paschke}, {Popow}, {Wolter},
  {Schmoll}, {Laux}, \& {Altmann}}]{roth05}
{Roth}, M.~M., {Kelz}, A., {Fechner}, T., {et~al.} 2005, \pasp, 117, 620

\bibitem[{{S{\'a}nchez} {et~al.}(2005){S{\'a}nchez}, {Becker},
  {Garcia-Lorenzo}, {Benn}, {Christensen}, {Kelz}, {Jahnke}, \&
  {Roth}}]{sanchez05}
{S{\'a}nchez}, S.~F., {Becker}, T., {Garcia-Lorenzo}, B., {et~al.} 2005, \aap,
  429, L21

\bibitem[{{S{\'a}nchez} {et~al.}(2007c){S{\'a}nchez}, {Cardiel}, {Verheijen},
  {Mart{\'{\i}}n-Gord{\'o}n}, {Vilchez}, \& {Alves}}]{sanchez07c}
{S{\'a}nchez}, S.~F., {Cardiel}, N., {Verheijen}, M.~A.~W., {et~al.} 2007c,
  \aap, 465, 207

\bibitem[{{S{\'a}nchez} {et~al.}(2006b){S{\'a}nchez}, {Garc{\'{\i}}a-Lorenzo},
  {Jahnke}, {Mediavilla}, {Gonz{\'a}lez-Serrano}, {Christensen}, \&
  {Wisotzki}}]{sanchez06b}
{S{\'a}nchez}, S.~F., {Garc{\'{\i}}a-Lorenzo}, B., {Jahnke}, K., {et~al.}
  2006b, \nar, 49, 501

\bibitem[{{S{\'a}nchez} {et~al.}(2012{\natexlab{a}}){S{\'a}nchez}, {Kennicutt},
  {Gil de Paz}, {van de Ven}, {V{\'{\i}}lchez}, {Wisotzki}, {Walcher}, {Mast},
  {Aguerri}, {Albiol-P{\'e}rez}, {Alonso-Herrero}, {Alves}, {Bakos},
  {Bart{\'a}kov{\'a}}, {Bland-Hawthorn}, {Boselli}, {Bomans},
  {Castillo-Morales}, {Cortijo-Ferrero}, {de Lorenzo-C{\'a}ceres}, {Del Olmo},
  {Dettmar}, {D{\'{\i}}az}, {Ellis}, {Falc{\'o}n-Barroso}, {Flores},
  {Gallazzi}, {Garc{\'{\i}}a-Lorenzo}, {Gonz{\'a}lez Delgado}, {Gruel},
  {Haines}, {Hao}, {Husemann}, {Igl{\'e}sias-P{\'a}ramo}, {Jahnke}, {Johnson},
  {Jungwiert}, {Kalinova}, {Kehrig}, {Kupko}, {L{\'o}pez-S{\'a}nchez},
  {Lyubenova}, {Marino}, {M{\'a}rmol-Queralt{\'o}}, {M{\'a}rquez}, {Masegosa},
  {Meidt}, {Mendez-Abreu}, {Monreal-Ibero}, {Montijo}, {Mour{\~a}o},
  {Palacios-Navarro}, {Papaderos}, {Pasquali}, {Peletier}, {P{\'e}rez},
  {P{\'e}rez}, {Quirrenbach}, {Rela{\~n}o}, {Rosales-Ortega}, {Roth},
  {Ruiz-Lara}, {S{\'a}nchez-Bl{\'a}zquez}, {Sengupta}, {Singh}, {Stanishev},
  {Trager}, {Vazdekis}, {Viironen}, {Wild}, {Zibetti}, \&
  {Ziegler}}]{sanchez12a}
{S{\'a}nchez}, S.~F., {Kennicutt}, R.~C., {Gil de Paz}, A., {et~al.}
  2012{\natexlab{a}}, \aap, 538, A8

\bibitem[{{S{\'a}nchez} {et~al.}(2014){S{\'a}nchez}, {Rosales-Ortega},
  {Iglesias-P{\'a}ramo}, {Moll{\'a}}, {Barrera-Ballesteros}, {Marino},
  {P{\'e}rez}, {S{\'a}nchez-Blazquez}, {Gonz{\'a}lez Delgado}, {Cid Fernandes},
  {de Lorenzo-C{\'a}ceres}, {Mendez-Abreu}, {Galbany}, {Falcon-Barroso},
  {Miralles-Caballero}, {Husemann}, {Garc{\'{\i}}a-Benito}, {Mast}, {Walcher},
  {Gil de Paz}, {Garc{\'{\i}}a-Lorenzo}, {Jungwiert}, {V{\'{\i}}lchez},
  {J{\'{\i}}lkov{\'a}}, {Lyubenova}, {Cortijo-Ferrero}, {D{\'{\i}}az},
  {Wisotzki}, {M{\'a}rquez}, {Bland-Hawthorn}, {Ellis}, {van de Ven}, {Jahnke},
  {Papaderos}, {Gomes}, {Mendoza}, \& {L{\'o}pez-S{\'a}nchez}}]{sanchez14}
{S{\'a}nchez}, S.~F., {Rosales-Ortega}, F.~F., {Iglesias-P{\'a}ramo}, J.,
  {et~al.} 2014, \aap, 563, A49

\bibitem[{{S{\'a}nchez} {et~al.}(2013){S{\'a}nchez}, {Rosales-Ortega},
  {Jungwiert}, {Iglesias-P{\'a}ramo}, {V{\'{\i}}lchez}, {Marino}, {Walcher},
  {Husemann}, {Mast}, {Monreal-Ibero}, {Cid Fernandes}, {P{\'e}rez},
  {Gonz{\'a}lez Delgado}, {Garc{\'{\i}}a-Benito}, {Galbany}, {van de Ven},
  {Jahnke}, {Flores}, {Bland-Hawthorn}, {L{\'o}pez-S{\'a}nchez}, {Stanishev},
  {Miralles-Caballero}, {D{\'{\i}}az}, {S{\'a}nchez-Blazquez}, {Moll{\'a}},
  {Gallazzi}, {Papaderos}, {Gomes}, {Gruel}, {P{\'e}rez}, {Ruiz-Lara},
  {Florido}, {de Lorenzo-C{\'a}ceres}, {Mendez-Abreu}, {Kehrig}, {Roth},
  {Ziegler}, {Alves}, {Wisotzki}, {Kupko}, {Quirrenbach}, {Bomans}, \& {Califa
  Collaboration}}]{sanchez13}
{S{\'a}nchez}, S.~F., {Rosales-Ortega}, F.~F., {Jungwiert}, B., {et~al.} 2013,
  \aap, 554, A58

\bibitem[{{S{\'a}nchez} {et~al.}(2011){S{\'a}nchez}, {Rosales-Ortega},
  {Kennicutt}, {Johnson}, {Diaz}, {Pasquali}, \& {Hao}}]{sanchez11}
{S{\'a}nchez}, S.~F., {Rosales-Ortega}, F.~F., {Kennicutt}, R.~C., {et~al.}
  2011, \mnras, 410, 313

\bibitem[{{S{\'a}nchez} {et~al.}(2012{\natexlab{b}}){S{\'a}nchez},
  {Rosales-Ortega}, {Marino}, {Iglesias-P{\'a}ramo}, {V{\'{\i}}lchez},
  {Kennicutt}, {D{\'{\i}}az}, {Mast}, {Monreal-Ibero}, {Garc{\'{\i}}a-Benito},
  {Bland-Hawthorn}, {P{\'e}rez}, {Gonz{\'a}lez Delgado}, {Husemann},
  {L{\'o}pez-S{\'a}nchez}, {Cid Fernandes}, {Kehrig}, {Walcher}, {Gil de Paz},
  \& {Ellis}}]{sanchez12b}
{S{\'a}nchez}, S.~F., {Rosales-Ortega}, F.~F., {Marino}, R.~A., {et~al.}
  2012{\natexlab{b}}, \aap, 546, A2

\bibitem[{{Sanchez-Blazquez} {et~al.}(2014){Sanchez-Blazquez},
  {Rosales-Ortega}, {Mendez-Abreu}, {Perez}, {Sanchez}, {Zibetti}, {Aguerri},
  {Bland-Hawthorn}, {Catalan}, {Cid Fernandes}, {de Amorim}, {de
  Lorenzo-Caceres}, {Falcon-Barroso}, {Galazzi}, {Garcia Benito}, {Gil de Paz},
  {Gonzalez Delgado}, {Husemann}, {Iglesias-Paramo}, {Jungwiert}, {Marino},
  {Marquez}, {Mast}, {Mendoza}, {Molla}, {Papaderos}, {Ruiz-Lara}, {van de
  Ven}, {Walcher}, \& {Wisotzki}}]{patri14}
{Sanchez-Blazquez}, P., {Rosales-Ortega}, F., {Mendez-Abreu}, J., {et~al.}
  2014, ArXiv e-prints

\bibitem[{{Sanduleak}(1969)}]{sand69}
{Sanduleak}, N. 1969, \aj, 74, 47

\bibitem[{{Searle}(1971)}]{sear71}
{Searle}, L. 1971, \apj, 168, 327

\bibitem[{{Sersic}(1968)}]{sersic68}
{Sersic}, J.~L. 1968, {Atlas de galaxias australes}

\bibitem[{{Singh} {et~al.}(2013){Singh}, {van de Ven}, {Jahnke}, {Lyubenova},
  {Falc{\'o}n-Barroso}, {Alves}, {Cid Fernandes}, {Galbany},
  {Garc{\'{\i}}a-Benito}, {Husemann}, {Kennicutt}, {Marino}, {M{\'a}rquez},
  {Masegosa}, {Mast}, {Pasquali}, {S{\'a}nchez}, {Walcher}, {Wild}, {Wisotzki},
  \& {Ziegler}}]{sign13}
{Singh}, R., {van de Ven}, G., {Jahnke}, K., {et~al.} 2013, \aap, 558, A43

\bibitem[{{Skillman}(1989)}]{Skillman:1989p1592}
{Skillman}, E.~D. 1989, \apj, 347, 883

\bibitem[{{Stasi{\'n}ska} {et~al.}(2006){Stasi{\'n}ska}, {Cid Fernandes},
  {Mateus}, {Sodr{\'e}}, \& {Asari}}]{stas06}
{Stasi{\'n}ska}, G., {Cid Fernandes}, R., {Mateus}, A., {Sodr{\'e}}, L., \&
  {Asari}, N.~V. 2006, \mnras, 371, 972

\bibitem[{{Tremonti} {et~al.}(2004){Tremonti}, {Heckman}, {Kauffmann},
  {Brinchmann}, {Charlot}, {White}, {Seibert}, {Peng}, {Schlegel}, {Uomoto},
  {Fukugita}, \& {Brinkmann}}]{tremonti04}
{Tremonti}, C.~A., {Heckman}, T.~M., {Kauffmann}, G., {et~al.} 2004, \apj, 613,
  898

\bibitem[{{Vazdekis} {et~al.}(2010){Vazdekis}, {S{\'a}nchez-Bl{\'a}zquez},
  {Falc{\'o}n-Barroso}, {Cenarro}, {Beasley}, {Cardiel}, {Gorgas}, \&
  {Peletier}}]{vazdekis10}
{Vazdekis}, A., {S{\'a}nchez-Bl{\'a}zquez}, P., {Falc{\'o}n-Barroso}, J.,
  {et~al.} 2010, \mnras, 404, 1639

\bibitem[{{Veilleux} {et~al.}(1995){Veilleux}, {Kim}, {Sanders}, {Mazzarella},
  \& {Soifer}}]{veil95}
{Veilleux}, S., {Kim}, D.-C., {Sanders}, D.~B., {Mazzarella}, J.~M., \&
  {Soifer}, B.~T. 1995, \apjs, 98, 171

\bibitem[{{Veilleux} \& {Osterbrock}(1987)}]{veilleux87}
{Veilleux}, S. \& {Osterbrock}, D.~E. 1987, \apjs, 63, 295

\bibitem[{{Vila-Costas} \& {Edmunds}(1992)}]{VilaCostas:1992p322}
{Vila-Costas}, M.~B. \& {Edmunds}, M.~G. 1992, \mnras, 259, 121

\bibitem[{{Vogt} {et~al.}(2014){Vogt}, {Dopita}, {Kewley}, {Sutherland},
  {Scharwaechter}, {Basurah}, {Ali}, \& {Amer}}]{vogt14}
{Vogt}, F.~P.~A., {Dopita}, M.~A., {Kewley}, L.~J., {et~al.} 2014, ArXiv
  e-prints

\bibitem[{{Yoachim} {et~al.}(2010){Yoachim}, {Ro{\v s}kar}, \&
  {Debattista}}]{yoac10}
{Yoachim}, P., {Ro{\v s}kar}, R., \& {Debattista}, V.~P. 2010, \apjl, 716, L4

\bibitem[{{Young} \& {Scoville}(1991)}]{you91}
{Young}, J.~S. \& {Scoville}, N.~Z. 1991, \araa, 29, 581

\bibitem[{{Yuan} {et~al.}(2010){Yuan}, {Kewley}, \& {Sanders}}]{yuan10}
{Yuan}, T.-T., {Kewley}, L.~J., \& {Sanders}, D.~B. 2010, \apj, 709, 884

\bibitem[{{Zaritsky} {et~al.}(1994){Zaritsky}, {Kennicutt}, \&
  {Huchra}}]{zaritsky94}
{Zaritsky}, D., {Kennicutt}, Jr., R.~C., \& {Huchra}, J.~P. 1994, \apj, 420, 87

\end{thebibliography}
\bibliographystyle{aa}

\appendix

\section{Distribution of \HII regions across other diagnostic diagrams}\label{ape}

This study has been focused on the study of the distribution of
\HII regions across the most widely used diagnostic diagram, the
so-called BPT diagram (e.g., Fig. \ref{fig:grid}). In particular we
have studied the dependence of the location within this diagram and
the properties of the underlying stellar population (e.g.,
Fig. \ref{fig:BPT_AM}). The BPT diagram has the advantage that it
compares line ratios between ions that are very near in their
wavelengths, and therefore they are equally affected by dust
attenuation. Indeed, we have illustrated that even for large amount of
dust the location within the diagram is not significantly affected.

Throughout this article we have shown that indeed there is a tight correlation between
the location within the BPT diagram and the properties of the underlying stellar
population, and this dependence is most probably a consequence of the 
chemical evolution of the galaxies. 

\begin{figure*}
\centering
\includegraphics[width=6.5cm,angle=270,clip,trim=0 0 -20 0]{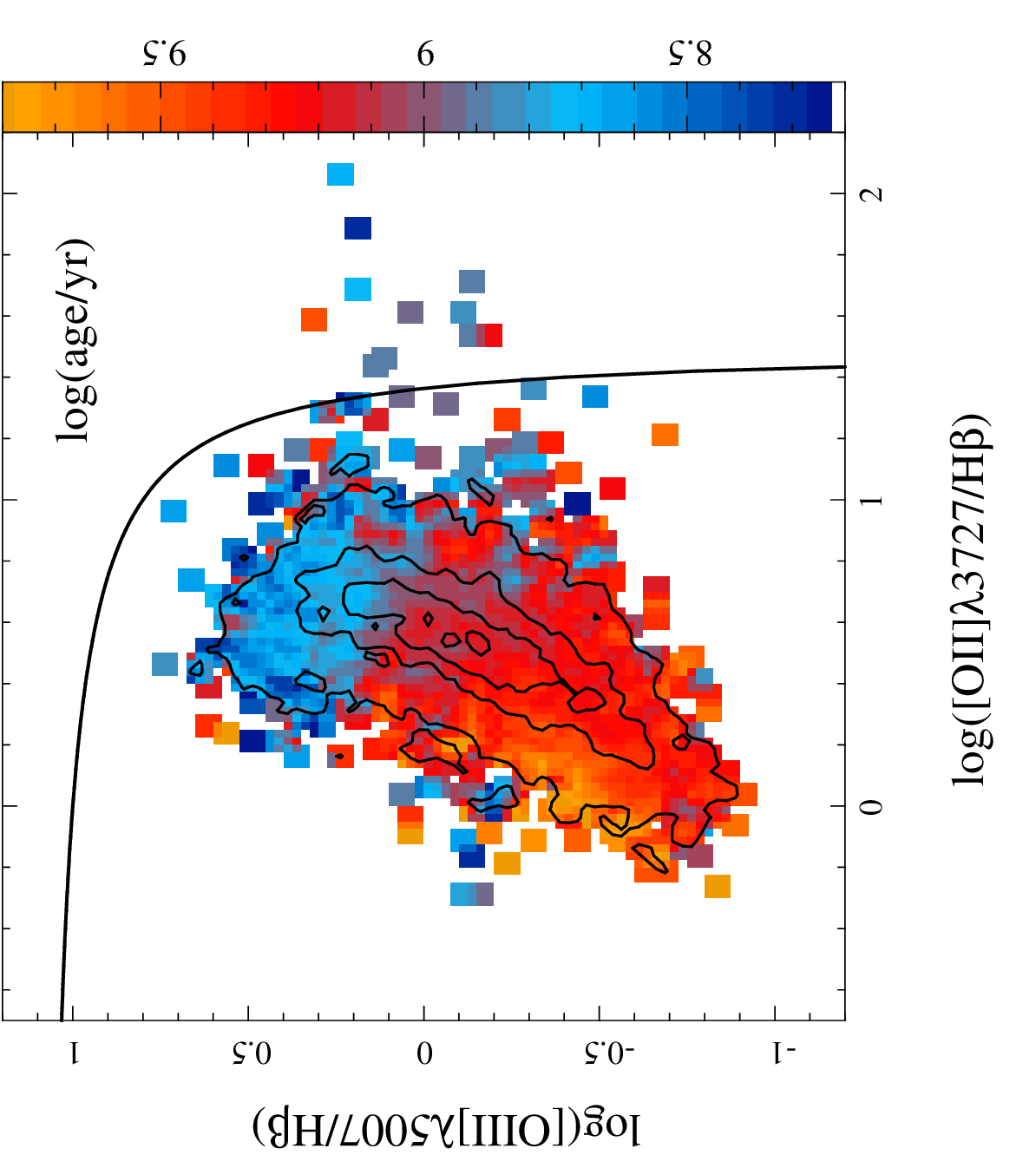}
\includegraphics[width=6.5cm,angle=270,clip,trim=0 0 -20 0]{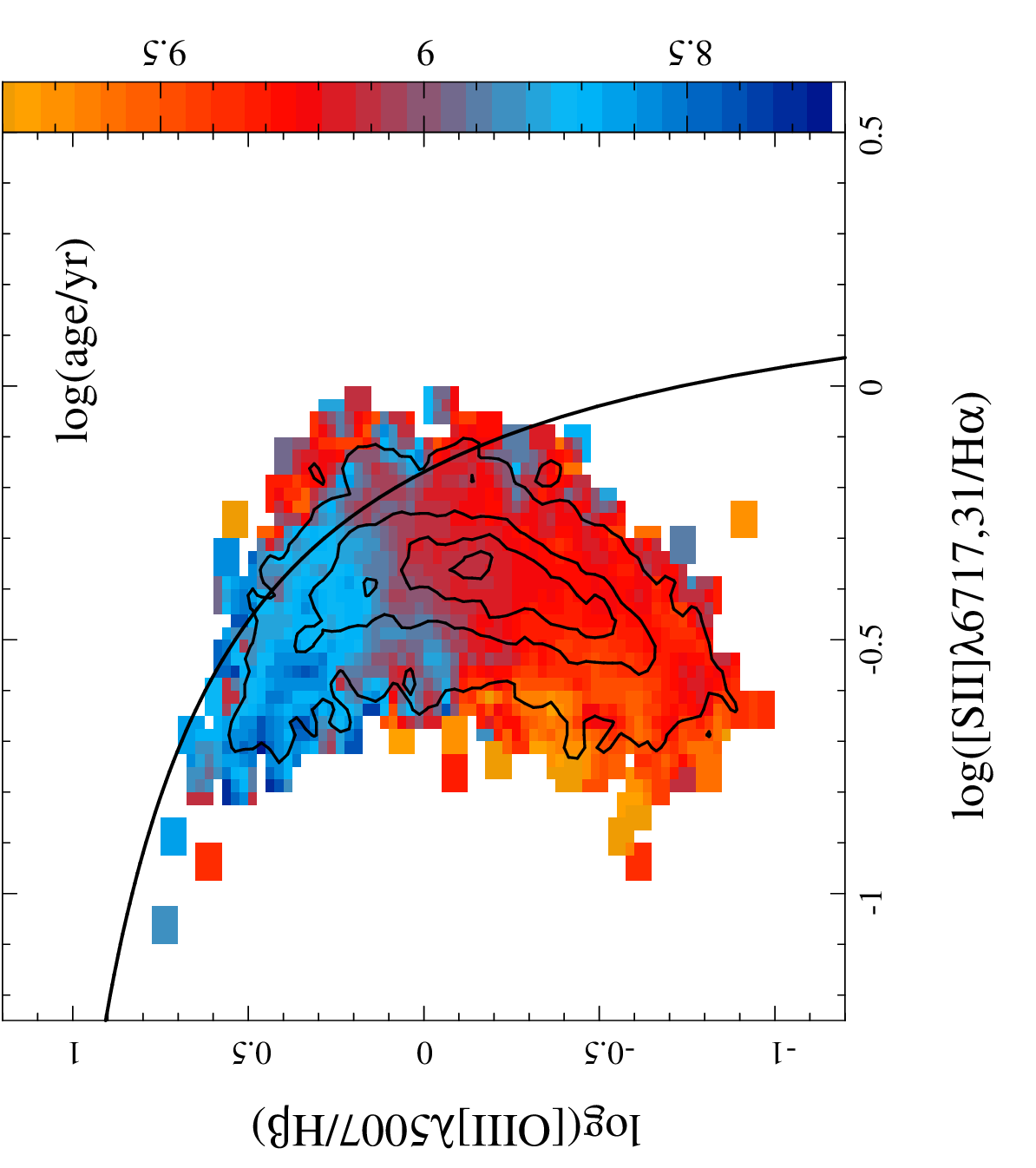}
\includegraphics[width=6.5cm,angle=270,clip,trim=0 0 -20 0]{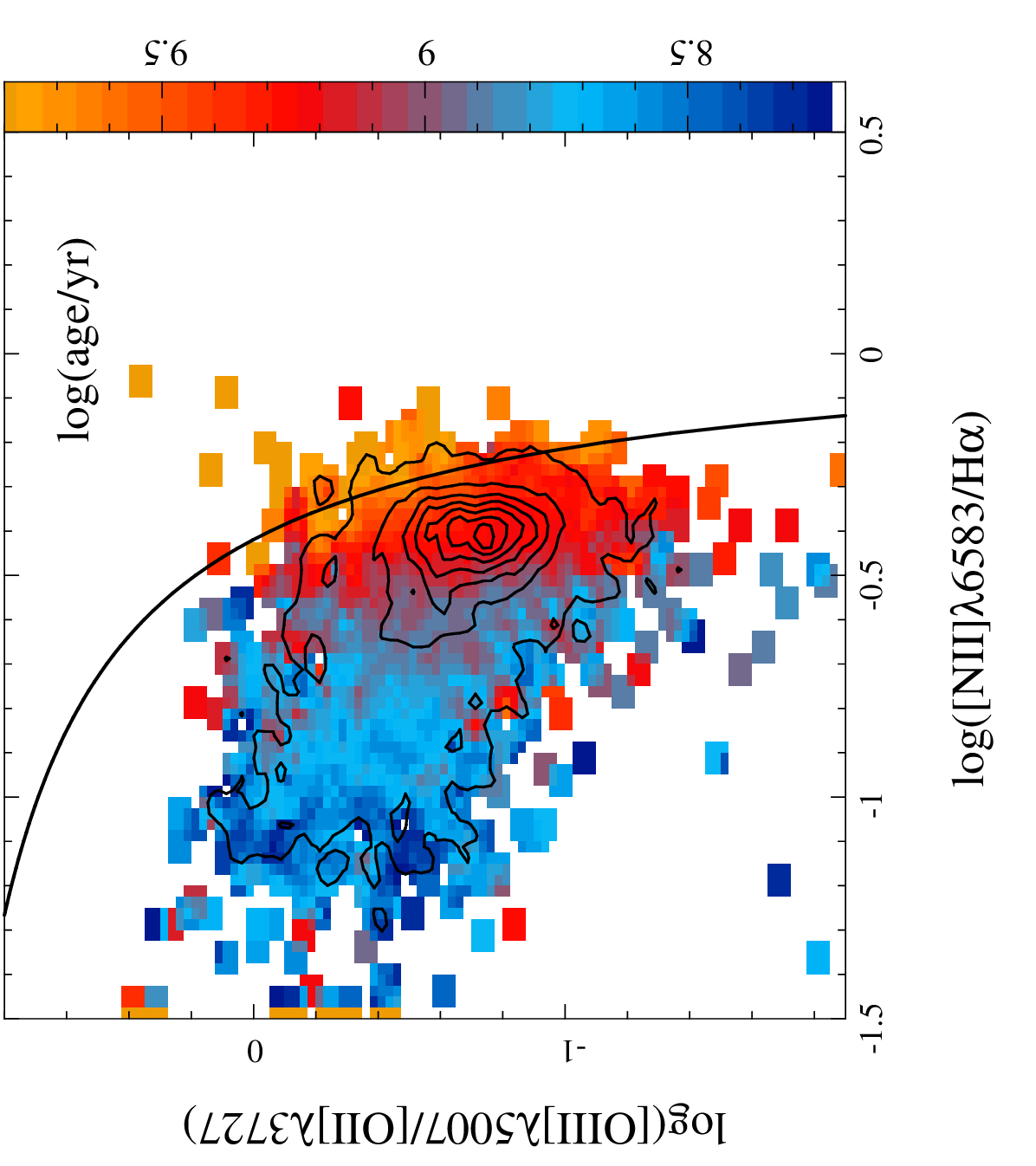}
\includegraphics[width=6.5cm,angle=270,clip,trim=0 0 -20 0]{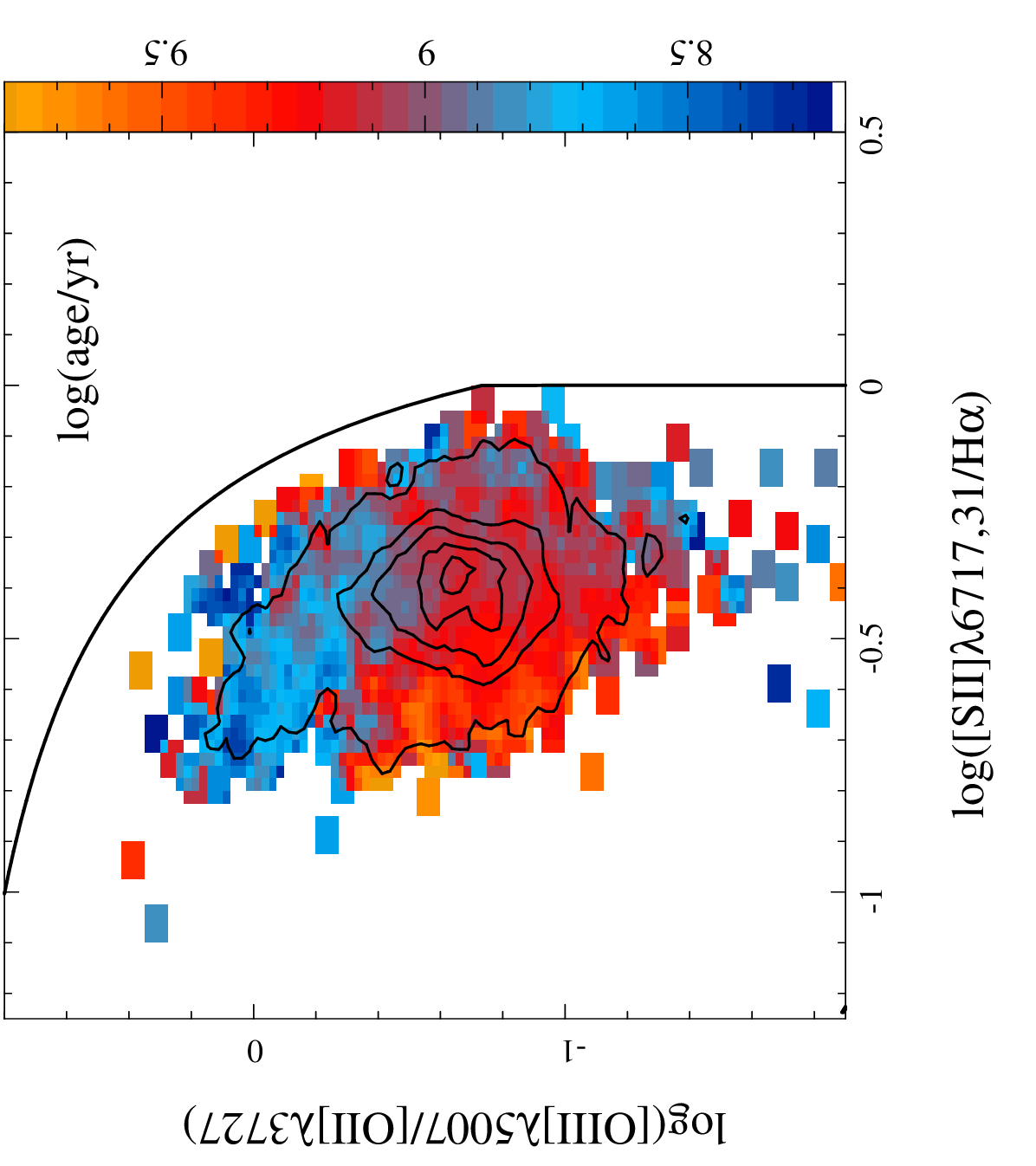}
\caption{\label{fig:ape} Distribution of the  $\sim$5000 \HII regions included in our sample across four
classical diagnostic diagrams: {\it top left panel,} [\ion{O}{iii}]~$\lambda$5007/H$\beta$ vs. [\ion{O}{ii}]~$\lambda$3727/H$\beta$; {\it top right panel,} [\ion{O}{iii}]~$\lambda$5007/H$\beta$ vs. [\ion{S}{ii}]~$\lambda$6717,31/H$\beta$; {\it bottom left panel,} [\ion{O}{iii}]~$\lambda$5007/[\ion{O}{ii}]~$\lambda$3727 vs. [\ion{N}{ii}]~$\lambda$6583/H$\beta$ and {\it bottom right panel,} [\ion{O}{iii}]~$\lambda$5007/[\ion{O}{ii}]~$\lambda$3727 vs.  [\ion{S}{ii}]~$\lambda$6717,31/H$\beta$. In each panel, the contours show the density distribution of the \HII regions within considered diagram, with the  outermost contour enclosing 95\% of the regions, and each consecutive one  enclosing 20\% less regions. The color indicates the average luminosity weighted age of the underlying stellar population at the corresponding location in the diagram. The solid line represent the classical demarcation line used to select \HII regions.}

\end{figure*}

In this Appendix we show that the imprint of the chemical evolution does
not only affect the location within the BPT diagram. Indeed, in
basically all the explored diagnostic diagrams we find  clear
trends related to the properties of the underlying stellar
population. Figure \ref{fig:ape} shows four additional diagnostic
diagrams, different than the BPT diagram, built based on the
comparison of the following emission line ratios:
[\ion{O}{iii}]~$\lambda$5007/H$\beta$,
[\ion{N}{ii}]~$\lambda$6583/H$\beta$,
[\ion{O}{ii}]~$\lambda$3727/H$\beta$,
[\ion{S}{ii}]~$\lambda$$\lambda$6717,31/H$\beta$ and
[\ion{O}{iii}]~$\lambda$5007/[\ion{O}{ii}]~$\lambda$3727. Each of
these line ratios are more sensitive to different properties of the
ionized gas and the ionizing sources. It is beyond  the scope of this
study to describe in detail the particular dependencies of each one,
which have been widely described in the literature
\citep[e.g.][]{veilleux87,dopita95,cid-fernandes10}.  However, regardless of which
physical properties are more sensible, Fig. \ref{fig:ape} shows that
it is possible to identify trends with the luminosity weighted age of
the underlying stellar population in the four diagrams. In some cases
the trend is stronger for one of the explored line ratios: e.g., in
the [\ion{O}{iii}]~$\lambda$5007/H$\beta$
vs. [\ion{S}{ii}]~$\lambda$$\lambda$6717,31/H$\beta$ diagram the stellar age is
more clearly associated with the first of the two parameters, while in
the [\ion{O}{iii}]~$\lambda$5007/[\ion{O}{ii}]~$\lambda$3727
vs. [\ion{N}{ii}]~$\lambda$6583/H$\beta$ one, it is more clearly
related to the second one. However, even in the case of the diagnostic
diagram comparing the two line ratios with the apparent weaker
dependence with the age of the stellar population
([\ion{O}{iii}]~$\lambda$5007/[\ion{O}{ii}]~$\lambda$3727 vs.
[\ion{S}{ii}]~$\lambda$6717,31/H$\beta$), there are clear trends that
depend on the two parameters at the same time.

In summary, we have shown that the main conclusion of the current study,
i.e., that there is a tight relation between the properties of the \HII regions
and those of the underlying stellar population, is independent of the diagnostic
diagram selected to explore this depedence.
 
\end{document}